\documentclass[a4paper,twoside]{article}

\usepackage{amssymb,amsmath}
\usepackage[dvips]{graphicx}

\numberwithin{equation}{section}
\allowdisplaybreaks[4]

\newcommand{\be}{\begin{equation}}
\newcommand{\ee}{\end{equation}}
\newcommand{\beq}{\begin{eqnarray}}
\newcommand{\eeq}{\end{eqnarray}}
\newcommand{\no}{\nonumber}
\newcommand{\bea}{\begin{array}}
\newcommand{\eea}{\end{array}}
\newcommand{\lb}{\label}
\newcommand{\mcal}{\mathcal}
\newcommand{\ve}{\varepsilon}
\newcommand{\ds}{\displaystyle}
\newcommand{\ts}{\textstyle}
\newcommand{\pp}{\partial}
\newcommand{\im}{\imath}
\newcommand{\ppr}{^{\prime}}
\newcommand{\pr}{\prime}
\newcommand{\dtot}{\mbox{d}}
\newcommand{\wtilde}{\widetilde}
\newcommand{\ovv}{\overline}

\newcommand{\trab}{\raisebox{-4pt}{$\mbox{tr}\atop {\scriptstyle a,b}$}}
\newcommand{\TRAB}{\raisebox{-4pt}{$\mbox{Tr}\atop {\scriptstyle p,q;a,b}$}}
\newcommand{\ph}{\phantom}
\newcommand{\scr}{\scriptstyle}

\begin{document}


\begin{center}
{\Large\bf Ensemble averaged coherent state path integral} \\
{\Large\bf for disordered bosons with a repulsive interaction} \\
{\Large\bf (Derivation of mean field equations)} \\
Bernhard Mieck\footnote{B. Mieck; {\sf E-mail : "Bernhard.Mieck@t-online.de",
phone: +49\,(0)203\,379\,4719, Fax: +49\,(0)203\,379\,4722}} \\
Department of Physics in Duisburg, University Duisburg-Essen, \\
Lotharstrasse 1, 47048 Duisburg, Germany\\
August 2007
\end{center}

\noindent {\bf Keywords:} Bose-Einstein condensation, ensemble averages for random potentials,
coherent state path integral, many-particle physics, Keldysh time contour.\newline
\noindent {\bf PACS} 03.75.Nt , 03.75.Kk , 03.75.Hh

\begin{abstract}
We consider bosonic atoms with a repulsive contact interaction in a trap potential for a Bose-Einstein condensation
(BEC) and additionally include a random potential. The ensemble averages for two models of static (I) and dynamic (II)
disorder are performed and investigated in parallel. The bosonic many body systems of the two disorder models are
represented by coherent state path integrals \(\ovv{Z_{I}[\mcal{J}]}\), \(\ovv{Z_{II}[\mcal{J}]}\) on the Keldysh time
contour which allow exact ensemble averages for zero and finite temperatures. These ensemble averages of coherent state
path integrals therefore present alternatives to replica field theories or super-symmetric averaging techniques.
Hubbard-Stratonovich transformations (HST) lead to two corresponding self-energies for
the hermitian repulsive interaction and for the non-hermitian disorder-interaction . The self-energy of the repulsive
interaction is absorbed by a shift into the disorder-self-energy which comprises as an element of a larger
symplectic Lie algebra $sp(4M)$ the self-energy of the repulsive interaction as a subalgebra (which is equivalent to
the direct product of $M$ times $sp(2)$; '$M$' is the number of discrete time intervals of the disorder-self-energy in
the generating function \(\ovv{Z_{I}[\mcal{J}]}\)). After removal of the remaining Gaussian integral for the
self-energy of the repulsive interaction, the first order variations of the coherent state path integrals
\(\ovv{Z_{I}[\mcal{J}]}\), \(\ovv{Z_{II}[\mcal{J}]}\) result in the exact mean field or saddle point equations, solely
depending on the disorder-self-energy matrix. These equations can be solved by continued fractions and are reminiscent
to the 'Nambu-Gorkov' Green function formalism in superconductivity because anomalous terms or pair condensates of the
bosonic atoms are also included into the selfenergies. The derived mean field equations of the models with static (I)
and dynamic (II) disorder are particularly applicable for BEC in $d=3$ spatial dimensions because of the singularity of
the density of states at vanishing wavevector. However, one usually starts out from restricted applicability of the
mean field approach for $d=2$; therefore, it is also pointed out that one should consider different HST's in $d=2$
spatial dimensions with the block diagonal densities as 'hinge' functions and that one has to introduce a coset
decomposition \(Sp(4M)\backslash U(2M)\) into densities and anomalous terms of the total disorder-self-energy
\(sp(4M)\) for deriving a nonlinear sigma model.
\end{abstract}
\vspace*{0.46cm}

\tableofcontents

\section{Introduction}\lb{s1}

The original Bose-Einstein condensation of atoms in traps has been
extended and performed in various manners \cite{And}-\cite{Kett2}. In this
article we investigate a BE-system of bosonic atoms with repulsive
interactions in a trap potential and include a random potential
which represents a model of ensemble averaged disorder. We
concentrate on the averaging procedure, derive the exact saddle
point equation and describe continued fractions as solutions of
the mean field equations which also comprise anomalous terms or
pair condensates in the coset parts of the selfenergies.
The exact saddle point
equation is obtained from first order variation of the self-energy of
the disorder in a coherent state path integral on the Keldysh time
contour \cite{Kel}-\cite{Ba2}. This equation finally results after Hubbard-Stratonovich
transformation (HST) of the repulsive interaction and integration over the
remaining bilinear condensate fields \(\psi^{*}(\vec{x},t)\ldots\psi(\vec{x},t)\) \cite{Strato,Neg}.
We introduce a 'Nambu'-doubling of the fields for the HST's so that the anomalous terms
are strictly incorporated into the selfenergies and the saddle point equations \cite{Nambu,Gold}.
The self-energy of the pure interaction (as the direct product of $M$ times \(sp(2)\) and as
subalgebra of the disorder-self-energy \(sp(4M)\))
can be absorbed by a shift into the 'larger' self-energy \(sp(4M)\) of
the ensemble averaged random potential ('\(M\)' denotes the number of discrete time intervals).
This disorder-self-energy \(sp(4M)\) is
represented by a larger group and also follows from a HST, but of a non-hermitian 'interaction' term
resulting from the average over the random disorder \cite{Strato,Neg}. After a
Gaussian integral of the self-energy of the repulsive interaction,
the self-energy of the ensemble averaged disorder only remains as a
matrix valued field depending on the two branches of the time
contour. The exact saddle point equation, obtained by a first order
variation, therefore only consists of the self-energy of the
disorder. Apart from the purely bosonic constituents of the disorder models the
presented approach is reminiscent of the Bogoliubov-de Gennes equations and
the Nambu-Gorkov Green function formalism in the theory of superconductivity \cite{Zago}-\cite{Mosk}.
Solutions of the saddle point equation can be achieved by
continued fraction of the disorder-self-energy starting from the free
Green function \cite{Groe,Viswa}. The iteration process for the continued fractions can
be simplified in the presence of spatial symmetries or time
independence. It is the imaginary increment \(-\im\;\ve_{p}\)
(\(\ve_{\pm}=\pm \ve\), \(\ve>0\)) in the Green function of the
disorder-self-energy which determines the solutions of the self-energy of the saddle point equation.
In fact the original coherent state
path integral, consisting only of the bosonic fields for the atoms,
is only defined by introducing this imaginary increment
\(-\im\;\ve_{p}\) on the time contour $t_{p}$. This imaginary
increment can be considered as a kind of regularization and is
necessary even without zero eigenvalues in the exponents of the time
development operators. This follows because Gaussian like
integrations (\ref{1}) as part of a time development operator only
lead to the absolute values of eigenvalues with respect to the sign
of the considered time intervals. Therefore, the important sign or
the information about bounded and unbounded states would be lost
under time reversal without the infinitesimal imaginary part
\(-\im\;\ve_{p}\)
\beq\lb{1}
\lefteqn{\int\dtot[\psi_{\vec{x}}(t_{p})]\;\;
\exp\bigg\{-\frac{\im}{\hbar}\int_{C}\dtot t_{p}\sum_{\vec{x}}
\psi_{\vec{x}}^{*}(t_{p})\;\;
\Big(-\im\;\ve_{p}+h_{\vec{x}}\Big)\;\;\psi_{\vec{x}}(t_{p})\bigg\}\propto } \\ \no &=&
\bigg[-\frac{\ve_{p}\;\;\Delta t_{p}}{\hbar}-\frac{\im}{\hbar}\;\Delta t_{p}\;\;h_{\vec{x}}\bigg]^{-1}\;\;
\hspace*{0.55cm}\ve_{p=\pm}=\pm\;\ve,\;\;(\ve>0);\hspace*{0.55cm}
\Delta t_{p=\pm}=\pm\Delta t,\;\;(\Delta t>0)\;\;\;.
\eeq
The imaginary increment has
therefore also to be taken into account in coherent state path
integrals without disorder in order to distinguish between advanced and
retarded Green functions which can appear after transformations and integrations
over field variables. One can improve the expansion of the actions
in the final path integral to quadratic and higher order in the
single self-energy matrix of the disorder for the inclusion of
fluctuation properties around the saddle point solution.

In section 2 we describe the average of a coherent state path integral on a time contour \cite{Bm3} and consider
the two cases of static and dynamic disorder. In section 3 the Hubbard Stratonovich transformations are given
and the exact saddle point equations are derived for static and dynamic disorder. We outline
the various steps for solving the saddle point equation via continued fractions and indicate
the reminiscence to many-body theory with anomalous terms
\cite{Zago}-\cite{Mosk},\cite{mbody1}-\cite{mbody3}.
Section 4 lists the relevant observables obtained by differentiation of the generating function
with respect to a source term. Section 5 contains a summary and points out
the extension of the mean field approach in sections \ref{s2}-\ref{s4}
for \(d=3\) spatial dimensions to the two dimensional case. Since the mean field approach is less
applicable in the case of \(d=2\), a nonlinear sigma model can be preferred, following from
spontaneous symmetry breaking and a gradient expansion of a determinant \cite{Bm1}
with additional 'Nambu'-doubled pair condensate terms in the self-energy.
However, in the \(d=2\) case different kinds of HST's have to be performed for a corresponding nonlinear
sigma model as for the \(d=3\) mean field approach (see section \ref{s5}).
The saddle point equations in \(d=3\) can lead to nonanalytic behaviour because of the
singularity of the density of states at vanishing wavevector.
In the spatial \(d=2\) case one has to apply the Weyl unitary trick for the parametrization
of the disorder-self-energy into densities and anomalous terms with a decomposition into a subgroup
\(U(2M)\) and coset part \(Sp(4M)\backslash U(2M)\) so that a nonlinear sigma model
can be achieved after a suitable HST and gradient expansion \cite{Weyl}-\cite{Corn},\cite{Bm2}.
In the strictly one dimensional case \(d=1\) one has to take into account large fluctuations
so that the transfer matrix approach of disorder-ensemble-averaged generating functions
can be chosen as the prevailing computational method for density and correlation functions
\cite{Bm4}-\cite{Bm6},\cite{Bm3}.

\section{Coherent state path integral} \lb{s2}

\subsection{Averaging methods for zero and finite temperatures}\lb{s21}

The Hamilton operators for the disordered bosonic systems
with the Bose field operators
(\(\big[\hat{\psi}_{\vec{x}}\:,\:\hat{\psi}_{\vec{x}\ppr}^{+}\big]=\delta_{\vec{x},\vec{x}\ppr}\);
\(\hat{\psi}_{\vec{x}}\), \(\hat{\psi}_{\vec{x}}^{+}\))
contain the trap potential $u(\vec{x})$, the quartic,
repulsive contact interaction with parameter \(V_{0}>0\) and the kinetic energy term with mass $m$.
Two kinds of random potentials \(V_{I}(\vec{x})\) (static disorder)
and \(V_{II}(\vec{x},t)\) (dynamic disorder) are introduced separately and result in the two different Hamilton operators
\(\hat{H}_{I}(\hat{\psi}_{\vec{x}}^{+},\hat{\psi}_{\vec{x}},V_{I})\) (\ref{2}) and
\(\hat{H}_{II}(\hat{\psi}_{\vec{x}}^{+},\hat{\psi}_{\vec{x}},V_{II})\) (\ref{3}) so that we consider two models
I and II in parallel. We examine these two models of disorder at zero temperature with coherent state path integrals and
incorporate a chemical potential or reference energy $\mu_{0}$. Furthermore, a $U(1)$ symmetry breaking, hermitian source
term with \(j_{\psi;\vec{x}}(t)\) is included for the creation of a coherent BE-condensate wavefunction.
Since we also consider pair condensates of bosonic atoms on the coset part
\(Sp(4M)\backslash U(2M)\) of the disorder-self-energy, we have to define a hermitian source term
\(j_{\psi\psi;\vec{x}}(t)\) for the creation of anomalous terms as
\(\langle\hat{\psi}_{\vec{x}}(t_{q}\ppr)\;\;\hat{\psi}_{\vec{x}}(t_{p})\rangle\) and
\(\langle\hat{\psi}_{\vec{x}}^{+}(t_{p})\;\;\hat{\psi}_{\vec{x}}^{+}(t_{q}\ppr)\rangle\) (for the contour
times \(t_{p},t_{q}\ppr\) see relations (\ref{11},\ref{12}))
\beq \lb{2}
\lefteqn{\hat{H}_{I}(\hat{\psi}^{+}_{\vec{x}},\hat{\psi}_{\vec{x}},V_{I})=\sum_{\vec{x}}\hat{\psi}_{\vec{x}}^{+}
\bigg(-\frac{\hbar^{2}}{2m}\Delta+u(\vec{x})-\mu_{0}+V_{I}(\vec{x})+
V_{0}\;\hat{\psi}_{\vec{x}}^{+}\hat{\psi}_{\vec{x}}\bigg)\hat{\psi}_{\vec{x}} +}
\\ \no &+&\sum_{\vec{x}}\Big(j_{\psi;\vec{x}}^{*}(t)\;\hat{\psi}_{\vec{x}}+
\hat{\psi}_{\vec{x}}^{+}\;j_{\psi;\vec{x}}(t)\Big)+\frac{1}{2}
\sum_{\vec{x}}\bigg(j_{\psi\psi;\vec{x}}^{*}(t)\;\;\hat{\psi}_{\vec{x}}\hat{\psi}_{\vec{x}}+
\hat{\psi}_{\vec{x}}^{+}\hat{\psi}_{\vec{x}}^{+}\;\;j_{\psi\psi;\vec{x}}(t)\bigg)
\eeq
\beq\lb{3}
\lefteqn{\hat{H}_{II}(\hat{\psi}^{+}_{\vec{x}},\hat{\psi}_{\vec{x}},V_{II})=\sum_{\vec{x}}\hat{\psi}_{\vec{x}}^{+}
\bigg(-\frac{\hbar^{2}}{2m}\Delta+u(\vec{x})-\mu_{0}+V_{II}(\vec{x},t)+
V_{0}\;\hat{\psi}_{\vec{x}}^{+}\hat{\psi}_{\vec{x}}\bigg)\hat{\psi}_{\vec{x}} + }
\\ \no &+&\sum_{\vec{x}}\Big(j_{\psi;\vec{x}}^{*}(t)\;\hat{\psi}_{\vec{x}}+
\hat{\psi}_{\vec{x}}^{+}\;j_{\psi;\vec{x}}(t)\Big)+\frac{1}{2}
\sum_{\vec{x}}\bigg(j_{\psi\psi;\vec{x}}^{*}(t)\;\;\hat{\psi}_{\vec{x}}\hat{\psi}_{\vec{x}}+
\hat{\psi}_{\vec{x}}^{+}\hat{\psi}_{\vec{x}}^{+}\;\;j_{\psi\psi;\vec{x}}(t)\bigg)  \\   \lb{4}    &&
\sum_{\vec{x}}\ldots=\sum_{\vec{x}_{i}}\bigg(\frac{\Delta x}{L}\bigg)^{d}\ldots
=\int_{L^{d}}\frac{d^{d}x}{L^{d}}\ldots\hspace*{1.0cm}
\Omega=\frac{1}{\Delta t}\hspace*{0.82cm}-\frac{T_{0}}{2}<t<+\frac{T_{0}}{2}_{\mbox{.}}
\eeq
Especially, in the case of disordered systems, the second moments of the Gaussian probability distributions of the
random potentials \(V_{I}(\vec{x})\) (\ref{5}) and \(V_{II}(\vec{x},t)\) (\ref{6})
have to be normalized in such a manner that the broadenings of the eigenvalue spectra
of the hermitian, ordered parts in \(\hat{H}_{I}(\hat{\psi}^{+}_{\vec{x}},\hat{\psi}_{\vec{x}},V_{I})\),
\(\hat{H}_{II}(\hat{\psi}^{+}_{\vec{x}},\hat{\psi}_{\vec{x}},V_{II})\) (\ref{2},\ref{3})
remain finite under the corresponding ensemble averages in model I and II. Therefore, the spatial integrals
\(\sum_{\vec{x}}\) in (\ref{2},\ref{3}) are given in a volume normalized kind as in (\ref{4})
for a system size of $L^{d}$, and the energy values
or frequencies of parameters and fields are related by the scale of the inverse time interval \(\Omega=1/\Delta t\).
We have to scale the actions in the generating functions at appropriate steps
for the derivation of the saddle point equations in model I and II
during a considered time development between times \(-T_{0}/2<t<+T_{0}/2\) (\ref{4}).
In the cases of model I and II with static and dynamic disorder, the suitable normalizations
of second moments for \(V_{I}(\vec{x})\), \(V_{II}(\vec{x},t)\) of Gaussian white-noise distributions are
obtained by the relations (\ref{5}) and (\ref{6}) with the parameters $R_{I}$ and $R_{II}$, respectively
(\(\mbox{dim}[R_{I}]=[\mbox{energy}\cdot\mbox{time}]\);
\(\mbox{dim}[R_{II}^{2}]=[(\mbox{energy})^{2}\cdot\mbox{time}]\)). The energy scale \(\hbar\Omega\)
follows from the discrete time steps, and the parameter \(\mcal{N}_{x}\) is the total number of discrete
points in the $d$-dimensional coordinate space of volume $L^{d}$ with discrete spatial intervals $\Delta x$
\beq \lb{5}
\ovv{V_{I}(\vec{x}_{1})\;V_{I}(\vec{x}_{2})}&=&\frac{R_{I}^{2}\;\Omega^{2}}{\mcal{N}_{x}}\;\;
\delta_{\vec{x}_{1},\vec{x}_{2}} \;;
\hspace*{0.5cm}\mcal{N}_{x}=\bigg(\frac{L}{\Delta x}\bigg)^{d}\hspace*{1.0cm}\mbox{static disorder}\\ \lb{6}
\ovv{V_{II}(\vec{x}_{1},t_{1})\;V_{II}(\vec{x}_{2},t_{2})}&=&R_{II}^{2}\;\;\delta_{\vec{x}_{1},\vec{x}_{2}}\;\;
\delta(t_{1}-t_{2})\hspace*{1.5cm}\mbox{dynamic disorder}\;\;\;.
\eeq
The generating functions \(Z[\mcal{J},V_{I}]\), \(Z[\mcal{J},V_{II}]\) of the disordered systems
in model I and II at zero temperature are represented by coherent state path integrals (\ref{7},\ref{8}) of the
unitary time development operators (\ref{9},\ref{10}) with
\(\hat{H}_{I}(\hat{\psi}^{+}_{\vec{x}},\hat{\psi}_{\vec{x}},V_{I})\) (\ref{2}) and
\(\hat{H}_{II}(\hat{\psi}^{+}_{\vec{x}},\hat{\psi}_{\vec{x}},V_{II})\) (\ref{3})
on the Keldysh time contour (\ref{11}) \cite{Neg}-\cite{Klau}. We have to take account of the negative sign
in the backward propagation of the Keldysh time contour (\ref{11}) and therefore introduce
the metric (\(\eta_{p}=p,\;p=\pm\)) (\ref{12}) of the contour time \(t_{p}=t_{\pm}\).
This metric (\(\eta_{p}=p,\;p=\pm\)) (\ref{12}) will frequently occur in the remainder
because the ensemble averages of the disordered
systems I, II couple the two '$\pm$' branches of the contour time in contrast to ordered systems where
the self-energy fields depend only on time arguments with a single branch of the contour, respectively \cite{Bm7}
\beq \lb{7}
Z[\mcal{J},V_{I}]&=&\langle 0|\hat{U}_{I}(-T_{0}/2,+T_{0}/2;V_{I};\mcal{J})
\;\;\hat{U}_{I}(+T_{0}/2,-T_{0}/2;V_{I};\mcal{J})|0\rangle \\ \lb{8}
Z[\mcal{J},V_{II}]&=&\langle 0|\hat{U}_{II}(-T_{0}/2,+T_{0}/2;V_{II};\mcal{J})
\;\;\hat{U}_{II}(+T_{0}/2,-T_{0}/2;V_{II};\mcal{J})|0\rangle
\eeq
\beq \lb{9}
\hat{U}_{I}(t,-T_{0}/2;V_{I};\mcal{J}) &=&\mcal{T}\exp\bigg\{-\frac{\im}{\hbar}\int_{-T_{0}/2}^{t}\dtot\tau\;\;
\hat{H}_{I}(\hat{\psi}^{+}_{\vec{x}},\hat{\psi}_{\vec{x}},V_{I};\mcal{J})\bigg\} \\ \lb{10}
\hat{U}_{II}(t,-T_{0}/2;V_{II};\mcal{J})
&=&\mcal{T}\exp\bigg\{-\frac{\im}{\hbar}\int_{-T_{0}/2}^{t}\dtot\tau\;\;
\hat{H}_{II}(\hat{\psi}^{+}_{\vec{x}},\hat{\psi}_{\vec{x}},V_{II};\mcal{J})\bigg\}
\eeq
\beq \lb{11}
\int_{C}\dtot t_{p}\ldots &=&\int_{-\infty}^{+\infty}\dtot t_{+}\ldots+\int_{+\infty}^{-\infty}\dtot t_{-}\ldots=
\int_{-\infty}^{+\infty}\dtot t_{+}\ldots-\int_{-\infty}^{+\infty}\dtot t_{-}\ldots \\ \lb{12}
\int_{C}\dtot t_{p}\ldots &=&\sum_{p=\pm}\int_{-\infty}^{+\infty}\dtot t_{p}\;\;\eta_{p}\ldots;\hspace{0.82cm}
\eta_{p}=\Big\{\underbrace{\eta_{+}=+1}_{p=+}\;;\;\underbrace{\eta_{-}=-1}_{p=-}\Big\}\;\;\;.
\eeq
The coherent state path integrals \(Z[\mcal{J},V_{I}]\), \(Z[\mcal{J},V_{II}]\) (\ref{7},\ref{8})
at zero temperature are normalized in the case of vanishing 'exterior' source term \(\mcal{J}\)
which allows to obtain observables from differentiating
\(Z[\mcal{J},V_{I}]\), \(Z[\mcal{J},V_{II}]\)
(\ref{7},\ref{8}) by \(\mcal{J}\) (compare section \ref{s4}).
The property of normalization of \(Z[\mcal{J},V_{I}]\), \(Z[\mcal{J},V_{II}]\) (\ref{13},\ref{14})
is guaranteed by the unitary time development $\hat{U}_{I}$, $\hat{U}_{II}$ (\ref{9},\ref{10})
in forward '$t_{+}$' and backward '$t_{-}$' direction on the time contour (\ref{11},\ref{12}).
The presence of the source fields \(j_{\psi;\vec{x}}(t_{p})\) and \(j_{\psi\psi;\vec{x}}(t_{p})\)
creates Bose particles from the vacuum states \(|0\rangle\), \(\langle 0|\) with the corresponding coherent
state fields \(\psi_{\vec{x}}(t_{p})\), \(\psi_{\vec{x}}^{*}(t_{p})\) and nonvanishing anomalous terms
\(\langle\psi_{\vec{x}}(t_{p})\;\psi_{\vec{x}}(t_{p})\rangle\), \(\langle\psi_{\vec{x}}^{*}(t_{p})\;\psi_{\vec{x}}^{*}(t_{p})\rangle\).
However, we have also to require in final relations for observables that the source terms
\(j_{\psi;\vec{x}}(t_{p})\), \(j_{\psi\psi;\vec{x}}(t_{p})\) have the same values on the two branches of
the time contour. This is defined by relation (\ref{15}) with the vertical line and has to be added to the generating functions.
Therefore, the required normalization property of ensemble averaged disordered systems is fulfilled and
possible problems with limits in replica field theories or super-symmetric extensions are circumvented \cite{Lern1,Lern2}.
According to the property of normalization at zero temperature (\ref{13},\ref{14}), the Gaussian ensemble
averages in model I and II (\ref{16},\ref{17}) are well defined and can be transferred to other physical problems
with disordered parts for generalized coherent states (as e.g. $SU(2)$-coherent states \cite{Gil1})
\beq \lb{13}
Z[\mcal{J}\equiv 0,V_{I}]\Big|_{\{j_{\psi},j_{\psi\psi}\}} &\equiv & 1 \\ \lb{14}
Z[\mcal{J}\equiv 0,V_{II}]\Big|_{\{j_{\psi},j_{\psi\psi}\}} & \equiv & 1  \\ \lb{15}
\ldots\Big|_{\{j_{\psi},j_{\psi\psi}\}} &:=& \Big\{j_{\psi;\vec{x}}(t_{+})=j_{\psi;\vec{x}}(t_{-})\;;\;
j_{\psi\psi;\vec{x}}(t_{+})=j_{\psi\psi;\vec{x}}(t_{-})\Big\}
\eeq
\beq \lb{16}
\ovv{Z_{I}[\mcal{J}]}&=&\ovv{\langle 0|\hat{U}_{I}(-T_{0}/2,+T_{0}/2;V_{I};\mcal{J})
\;\;\hat{U}_{I}(+T_{0}/2,-T_{0}/2;V_{I};\mcal{J})|0\rangle}\Big|_{\{j_{\psi},j_{\psi\psi}\}}  \\ \lb{17}
\ovv{Z_{II}[\mcal{J}]}&=&\ovv{\langle 0|\hat{U}_{II}(-T_{0}/2,+T_{0}/2;V_{II};\mcal{J})
\;\;\hat{U}_{II}(+T_{0}/2,-T_{0}/2;V_{II};\mcal{J})|0\rangle}\Big|_{\{j_{\psi},j_{\psi\psi}\}\mbox{.}}
\eeq
The normalized unitary time development at zero temperature (\ref{7}-\ref{17}) has to be modified
in the case of a finite temperature. We briefly describe the suitably normalized generating function for finite
temperature in the case of model I (static disorder, compare with the Hamilton and unitary time development operators
(\ref{2},\ref{9}) for model I). The inclusion of the grand
canonical statistical operator \(\exp\{-\beta(\hat{h}_{I}-\mu\;\hat{N})\}\) with $\hat{h}_{I}(\hat{\psi}^{+}_{\vec{x}},\hat{\psi}_{\vec{x}},V_{I})$ (\ref{19}) as part of
$\hat{H}_{I}(\hat{\psi}^{+}_{\vec{x}},\hat{\psi}_{\vec{x}},V_{I})$ (\ref{18},\ref{2}) and its appearance in the
denominator with the trace $Z_{\beta}[V_{I}]$ (\ref{21})
of the total generating function \(Z[\mcal{J},\beta,V_{I}]\) (\ref{20})
lead to an expansion of a large (\(n\rightarrow\infty\), \(n\in \mbox{\sf N}_{0}\geq0\))
limit with $Z_{\beta}[V_{I}]$ (\ref{22}-\ref{24}).
This follows from the representation of the
inverse of $Z_{\beta}[V_{I}]$ (\ref{22}) by an exponential integral
with auxiliary integration variable $x$ and the Taylor
expansion (\ref{22}) of the exponential \(\exp\{-x\;Z_{\beta}[V_{I}]\}\) in the integrand with \(x\in [0,\infty)\)
\beq \lb{18}
\hat{H}_{I}(\hat{\psi}^{+}_{\vec{x}},\hat{\psi}_{\vec{x}},V_{I})&=&
\hat{h}_{I}(\hat{\psi}^{+}_{\vec{x}},\hat{\psi}_{\vec{x}},V_{I}) +
\sum_{\vec{x}}\Big(j_{\psi;\vec{x}}^{*}(t)\;\hat{\psi}_{\vec{x}}+
\hat{\psi}_{\vec{x}}^{+}\;j_{\psi;\vec{x}}(t)\Big)+ \\ \no &+&\frac{1}{2}
\sum_{\vec{x}}\bigg(j_{\psi\psi;\vec{x}}^{*}(t)\;\;\hat{\psi}_{\vec{x}}\hat{\psi}_{\vec{x}}+
\hat{\psi}_{\vec{x}}^{+}\hat{\psi}_{\vec{x}}^{+}\;\;j_{\psi\psi;\vec{x}}(t)\bigg) \\ \lb{19}
\hat{h}_{I}(\hat{\psi}^{+}_{\vec{x}},\hat{\psi}_{\vec{x}},V_{I})  &=&
\sum_{\vec{x}}\hat{\psi}_{\vec{x}}^{+}
\bigg(-\frac{\hbar^{2}}{2m}\Delta+u(\vec{x})-\mu_{0}+V_{I}(\vec{x})+
V_{0}\;\hat{\psi}_{\vec{x}}^{+}\hat{\psi}_{\vec{x}}\bigg)\hat{\psi}_{\vec{x}}
\eeq
\be \lb{20}
Z[\mcal{J},\beta,V_{I}]=
\frac{\mbox{Tr}\Big[\exp\{-\beta(\hat{h}_{I}-\mu\hat{N})\}\;
\hat{U}_{I}(-T_{0}/2,+T_{0}/2;V_{I};\mcal{J})\;\;
\hat{U}_{I}(+T_{0}/2,-T_{0}/2;V_{I};\mcal{J})\Big]}{Z_{\beta}[V_{I}]}
\ee
\beq \lb{21}
Z_{\beta}[V_{I}]&=&\mbox{Tr}\Big[\exp\{-\beta(\hat{h}_{I}-\mu\hat{N})\}\Big]\hspace*{1.54cm}
Z[\mcal{J}\equiv0,\beta,V_{I}]\Big|_{\{j_{\psi},j_{\psi\psi}\}} \equiv 1   \\ \lb{22}
\frac{1}{Z_{\beta}[V_{I}]}&=&\int_{0}^{\infty}\dtot x\;\;\exp\{-x\;Z_{\beta}[V_{I}]\}=
\int_{0}^{\infty}\dtot x\sum_{n=0}^{\infty}
\frac{(-x)^{n}}{n!}\;\;\big(Z_{\beta}[V_{I}]\big)^{n}\;\;\;.
\eeq
The ensemble averaged generating function \(\ovv{Z_{I}[\mcal{J},\beta]}\) (\ref{23}) of
\(Z[\mcal{J},\beta,V_{I}]\) (\ref{20})
for finite temperatures is given by the sum of averaged generating functions
\(\ovv{Z_{I,n}[\mcal{J},\beta]}\) (\ref{24}) with increasing number of fields
in $\big(Z_{\beta}[V_{I}]\big)^{n}$ so that the large (\(n\rightarrow \infty\), \(n\in \mbox{\sf N}_{0}\geq0\))
limit of field theories has to be considered.
The unitary time development operator $\hat{U}_{I}$ is determined by the relation (\ref{9}) with
the operator \(\hat{H}_{I}(\hat{\psi}^{+}_{\vec{x}},\hat{\psi}_{\vec{x}},V_{I})\) (\ref{18},\ref{2})
which includes the symmetry breaking source terms and additionally the 'exterior' source variable
\(\mcal{J}\) for the observables
\beq \lb{23}
\ovv{Z_{I}[\mcal{J},\beta]}&=&\int_{0}^{\infty}\dtot x\sum_{n=0}^{\infty}\frac{(-x)^{n}}{n!}\;\;
\ovv{Z_{I,n}[\mcal{J},\beta]} \\ \lb{24}
\ovv{Z_{I,n}[\mcal{J},\beta]}&=&\ovv{\mbox{Tr}\Big[\exp\{-\beta(\hat{h}_{I}-\mu\hat{N})\}\;
\hat{U}_{I}(-T_{0}/2,+T_{0}/2;V_{I};\mcal{J})\;\times\ldots}
\\ \no &&\ovv{\ldots\times\hat{U}_{I}(+T_{0}/2,-T_{0}/2;V_{I};\mcal{J})\Big]\;\;\;\;
\Big(\mbox{Tr}\Big[\exp\{-\beta(\hat{h}_{I}-\mu\hat{N})\}\Big]\Big)^{n} } \Big|_{\{j_{\psi},j_{\psi\psi}\}}.
\eeq
However, we restrict in this paper to zero temperature for the models I, II.
A coset decomposition into densities and pair condensate terms is preferable in \(d=2\) spatial dimensions
with a gradient expansion of a determinant leading to the Goldstone modes of a spontaneous symmetry breaking
within a nonlinear sigma model \cite{Bm7,Bm2}.

\subsection{Ensemble averages in model I and II}\lb{s22}

We list in relations (\ref{25},\ref{26}) the coherent state path integral representation
of the unitary time development operators $\hat{U}_{I}$, $\hat{U}_{II}$ (\ref{9},\ref{10})
in \(Z[\mcal{J},V_{I}]\), \(Z[\mcal{J},V_{II}]\) (\ref{7},\ref{8})
at zero temperature for the disorder models I and II
\beq \lb{25}
\lefteqn{Z[\mcal{J},V_{I}]=
\int\dtot[\psi_{\vec{x}}(t_{p})]\;
\exp\bigg\{-\frac{\im}{2\hbar}\int_{C}\dtot t_{p}\;\dtot t\ppr_{q}\sum_{\vec{x},\vec{x}\ppr}
\Psi_{\vec{x}\ppr}^{+b}(t_{q}\ppr)\;\;\mcal{J}_{\vec{x}\ppr,\vec{x}}^{ba}(t_{q}\ppr,t_{p})\;\;
\Psi_{\vec{x}}^{a}(t_{p})\bigg\} }
\\ \no &\times&
\exp\bigg\{-\frac{\im}{\hbar}\int_{C}\dtot t_{p}\sum_{\vec{x}}
\psi_{\vec{x}}^{*}(t_{p})\;\Big[\hat{h}_{p}(t_{p})+V_{I}(\vec{x})+
V_{0}\;\psi_{\vec{x}}^{*}(t_{p})\;\psi_{\vec{x}}(t_{p})\Big]\;
\psi_{\vec{x}}(t_{p})\bigg\}
\\ \no &\times& \exp\bigg\{-\frac{\im}{\hbar}\int_{C}\dtot t_{p}\sum_{\vec{x}}\Big[
j_{\psi;\vec{x}}^{*}(t_{p})\;\psi_{\vec{x}}(t_{p})+\psi_{\vec{x}}^{*}(t_{p})\;j_{\psi;\vec{x}}(t_{p})\Big]\bigg\}
\\ \no
&\times& \exp\bigg\{-\frac{\im}{2\hbar} \int_{C}\dtot t_{p}\sum_{\vec{x}}\Big[
j_{\psi\psi;\vec{x}}^{*}(t_{p})\;\psi_{\vec{x}}(t_{p})\;\psi_{\vec{x}}(t_{p})+\psi_{\vec{x}}^{*}(t_{p})\;
\psi_{\vec{x}}^{*}(t_{p})\;j_{\psi\psi;\vec{x}}(t_{p}) \Big]\bigg\}
\eeq
\beq \lb{26}
\lefteqn{Z[\mcal{J},V_{II}]=
\int\dtot[\psi_{\vec{x}}(t_{p})]\;
\exp\bigg\{-\frac{\im}{2\hbar}\int_{C}\dtot t_{p}\;\dtot t\ppr_{q}\sum_{\vec{x},\vec{x}\ppr}
\Psi_{\vec{x}\ppr}^{+b}(t_{q}\ppr)\;\;\mcal{J}_{\vec{x}\ppr,\vec{x}}^{ba}(t_{q}\ppr,t_{p})\;\;
\Psi_{\vec{x}}^{a}(t_{p})\bigg\} } \\ \no &\times&
\exp\bigg\{-\frac{\im}{\hbar}\int_{C}\dtot t_{p}\sum_{\vec{x}}
\psi_{\vec{x}}^{*}(t_{p})\;\Big[\hat{h}_{p}(t_{p})+V_{II}(\vec{x},t)+
V_{0}\;\psi_{\vec{x}}^{*}(t_{p})\;\psi_{\vec{x}}(t_{p})\Big]\;
\psi_{\vec{x}}(t_{p})\bigg\}
\\ \no &\times& \exp\bigg\{-\frac{\im}{\hbar}\int_{C}\dtot t_{p}\sum_{\vec{x}}\Big[
j_{\psi;\vec{x}}^{*}(t_{p})\;\psi_{\vec{x}}(t_{p})+\psi_{\vec{x}}^{*}(t_{p})\;j_{\psi;\vec{x}}(t_{p})\Big]\bigg\}
\\ \no &\times& \exp\bigg\{-\frac{\im}{2\hbar} \int_{C}\dtot t_{p}\sum_{\vec{x}}\Big[
j_{\psi\psi;\vec{x}}^{*}(t_{p})\;\psi_{\vec{x}}(t_{p})\;\psi_{\vec{x}}(t_{p})+\psi_{\vec{x}}^{*}(t_{p})\;
\psi_{\vec{x}}^{*}(t_{p})\;j_{\psi\psi;\vec{x}}(t_{p}) \Big]\bigg\}\;\;\;.
\eeq
The one-particle parts are given by $\hat{h}_{p}(t_{p})+V_{I}(\vec{x})$,
$\hat{h}_{p}(t_{p})+V_{II}(\vec{x},t)$
which consist of the common Hamilton operator $\hat{h}_{p}(t_{p})$ (\ref{27},\ref{28}) with the kinetic energy,
the trap potential $u(\vec{x})$, the chemical potential $\mu_{0}$,
the energy operator \(-\im\hbar\;\pp/\pp t_{p}\) of the corresponding branch of the contour time
and the random potentials $V_{I}(\vec{x})$, $V_{II}(\vec{x},t)$.
Note the inclusion of the small imaginary energy increment \(-\im\;\ve_{p}\) (\ref{27})
on both branches of the time contour
which allows the selection between advanced and retarded Green functions.
This imaginary increment \(-\im\;\ve_{p}\) determines a direction for the time development
so that the coherent state path integrals (\ref{25},\ref{26}) are well defined
for all kinds of effective energies whether vanishing, bounded or unbounded
\beq \lb{27}\hspace*{-1.0cm}
\hat{h}_{p}(t_{p})&=&-\im\hbar\frac{\pp}{\pp t_{p}}-\im\;\ve_{p}-\frac{\hbar^{2}}{2m}\Delta+
u(\vec{x})-\mu_{0} \\ \lb{28}
\hat{h}_{\vec{x},\vec{x}\ppr}(t_{p},t_{q}\ppr)&=&\delta_{p,q}\;\eta_{p}\;\delta(t_{p}-t_{q}\ppr)\;
\delta_{\vec{x},\vec{x}\ppr}\;\;\hat{h}_{p}(t_{p}) \;;
\hspace*{0.46cm}\ve_{p}=\eta_{p}\;\ve\;\;;(\ve>0;\;\;\eta_{\pm}=\pm 1)\\ \lb{29}
j_{\psi;\vec{x}}(t_{+})&=&j_{\psi;\vec{x}}(t_{-})\hspace*{1.54cm}
j_{\psi\psi;\vec{x}}(t_{+})=j_{\psi\psi;\vec{x}}(t_{-})\;\;\;.
\eeq
The symmetry breaking source terms $j_{\psi;\vec{x}}(t_{\pm})$, $j_{\psi\psi;\vec{x}}(t_{\pm})$ (\ref{29},\ref{15})
for the creation of a coherent BE-condensate wavefunction and pair condensates
have to be set to the same values on the two branches of the contour time
in the final relations for observables with vanishing 'exterior' source
\(\mcal{J}_{\vec{x}\ppr,\vec{x}}^{ba}(t_{q}\ppr,t_{p})\).
We perform a 'Nambu'-doubling of the coherent state fields $\psi_{\vec{x}}(t_{p})$
on the time contour with its complex conjugated fields $\psi_{\vec{x}}^{*}(t_{p})$
in order to obtain also anomalous terms as \(\langle\psi_{\vec{x}}(t_{p})\;\psi_{\vec{x}}(t_{p})\rangle\)
by a single differentiation with \(\mcal{J}_{\vec{x}\ppr,\vec{x}}^{ba}(t_{q}\ppr,t_{p})\).
We denote this 'Nambu'-doubled field by \(\Psi_{\vec{x}}^{a(=1/2)}(t_{p(=\pm)})\) (\ref{30},\ref{31})
(with capital '$\Psi$' instead of the lower-case letter '$\psi$')
and introduce the additional indices (\(a,b=1,2\))
for referring to $\psi_{\vec{x}}(t_{p})$ (\(a=1\)) or to the complex conjugated part
\(\psi_{\vec{x}}^{*}(t_{p})\) (\(a=2\)).
There are two possible orders of the four component 'Nambu'-doubled field
\(\Psi_{\vec{x}}^{a(=1/2)}(t_{p(=\pm)})\).
In the listing (\ref{30}) one gives priority with respect to the two branches of the contour time
so that the first two components of \(\Psi_{\vec{x}}^{a}(t_{p})\) are on the plus branch of $t_{p=+}$
whereas the listing (\ref{31}) prefers ordering with respect to the anomalous pair condensates
\beq\lb{30}
\mbox{contour time ordering} &:&  \\ \no
\Psi_{\vec{x}}^{a(=1/2)}(t_{p(=\pm)})&=&\left(
\bea{c}
\psi_{\vec{x}}(t_{p})\;(a=1) \\ \psi_{\vec{x}}^{*}(t_{p})\;(a=2)
\eea\right)=\left(
\bea{c}
\psi_{\vec{x}}(t_{+}) \;\;(a=1)\\ \psi_{\vec{x}}^{*}(t_{+})\;\;(a=2) \\ \psi_{\vec{x}}(t_{-})\;\;(a=1) \\
\psi_{\vec{x}}^{*}(t_{-})\;\;(a=2)
\eea\right)  \\ \lb{31}  \mbox{ordering for anomalous terms} & : & \\ \no
\Psi_{\vec{x}}^{a(=1/2)}(t_{p(=\pm)})&=&\left(
\bea{c}
\psi_{\vec{x}}(t_{p})\;(a=1) \\ \psi_{\vec{x}}^{*}(t_{p})\;(a=2)
\eea\right)=\left(
\bea{c}
\psi_{\vec{x}}(t_{+})\;\;(a=1) \\ \psi_{\vec{x}}(t_{-})\;\;(a=1) \\ \psi_{\vec{x}}^{*}(t_{+})\;\;(a=2) \\
\psi_{\vec{x}}^{*}(t_{-})\;\;(a=2)
\eea\right)_{\mbox{.}}
\eeq
In the remainder we partially use a different notation
for the coherent state field variables \(\psi_{\vec{x}}(t\raisebox{1pt}{\mbox{$_{p(=\pm)}$}})\),
\(\psi_{\vec{x}\ppr}^{*}(t\raisebox{2pt}{\mbox{$_{q(=\pm)}\ppr$}})\)
in the generating functions of models I, II in order to emphasize the complete independence of the fields
\(\psi_{\vec{x}}(t_{+}):=\psi_{\vec{x},+}(t)\), \(\psi_{\vec{x}}(t_{-}):=\psi_{\vec{x},-}(t)\) and also
\(\psi_{\vec{x}\ppr}^{*}(t_{+}):=\psi_{\vec{x}\ppr,+}^{*}(t)\),
\(\psi_{\vec{x}\ppr}^{*}(t_{-}):=\psi_{\vec{x}\ppr,-}^{*}(t)\) concerning the two branches of the time contour
(\ref{32},\ref{33}). However, if classical approximations are implemented
in the coherent state path integrals of disorder models I
and II, the fields \(\psi_{\vec{x}}(t\raisebox{1pt}{\mbox{$_{p(=\pm)}$}})\),
\(\psi_{\vec{x}\ppr}^{*}(t\raisebox{2pt}{\mbox{$_{q(=\pm)}\ppr$}})\)
may take exactly the same values on both contour time branches
\(\psi_{\vec{x}}(t_{+})\raisebox{-2pt}{\mbox{$\stackrel{!}{=}$}}\psi_{\vec{x}}(t_{-})\),
\(\psi_{\vec{x}\ppr}^{*}(t_{+}\ppr)\raisebox{-2pt}{\mbox{$\stackrel{!}{=}$}}\psi_{\vec{x}\ppr}^{*}(t_{-}\ppr)\)
(for classical approximations following from variations with respect to contour fields)
\beq\lb{32}
\mbox{contour time ordering} & : & \\ \no
\Psi_{\vec{x}}^{a(=1/2)}(t_{p(=\pm)})&=&\Psi_{\vec{x},p(=\pm)}^{a(=1/2)}(t)=
\left(
\bea{c}
\psi_{\vec{x}}(t_{p})\;(a=1) \\ \psi_{\vec{x}}^{*}(t_{p})\;(a=2)
\eea\right)=\left(
\bea{c}
\psi_{\vec{x},p}(t)\;(a=1) \\ \psi_{\vec{x},p}^{*}(t)\;(a=2)
\eea\right)   \\ \no &=&
\left(
\bea{c}
\psi_{\vec{x}}(t_{+}) \;\;(a=1)\\ \psi_{\vec{x}}^{*}(t_{+})\;\;(a=2) \\ \psi_{\vec{x}}(t_{-})\;\;(a=1) \\
\psi_{\vec{x}}^{*}(t_{-})\;\;(a=2)
\eea\right)=\left(
\bea{c}
\psi_{\vec{x},+}(t) \;\;(a=1)\\ \psi_{\vec{x},+}^{*}(t)\;\;(a=2) \\ \psi_{\vec{x},-}(t)\;\;(a=1) \\
\psi_{\vec{x},-}^{*}(t)\;\;(a=2)
\eea\right)  \\ \no \\  \lb{33}
\mbox{ordering for anomalous terms} & : & \\ \no
\Psi_{\vec{x}}^{a(=1/2)}(t_{p(=\pm)})&=&\Psi_{\vec{x},p(=\pm)}^{a(=1/2)}(t)=\left(
\bea{c}
\psi_{\vec{x}}(t_{p})\;(a=1) \\ \psi_{\vec{x}}^{*}(t_{p})\;(a=2)
\eea\right)=\left(
\bea{c}
\psi_{\vec{x},p}(t)\;(a=1) \\ \psi_{\vec{x},p}^{*}(t)\;(a=2)
\eea\right) \\ \no &=& \left(
\bea{c}
\psi_{\vec{x}}(t_{+})\;\;(a=1) \\ \psi_{\vec{x}}(t_{-})\;\;(a=1) \\ \psi_{\vec{x}}^{*}(t_{+})\;\;(a=2) \\
\psi_{\vec{x}}^{*}(t_{-})\;\;(a=2)
\eea\right)=\left(
\bea{c}
\psi_{\vec{x},+}(t)\;\;(a=1) \\ \psi_{\vec{x},-}(t)\;\;(a=1) \\ \psi_{\vec{x},+}^{*}(t)\;\;(a=2) \\
\psi_{\vec{x},-}^{*}(t)\;\;(a=2)
\eea\right)_{\mbox{.}}
\eeq
Therefore, one can also rewrite the 'Nambu'-doubled bilinear term (\ref{34}) with a matrix
\(\hat{M}_{\vec{x}\ppr,\vec{x}}^{ba}(t_{q}\ppr,t_{p}):=\hat{M}_{\vec{x}\ppr,q;\vec{x},p}^{ba}(t\ppr,t)\)
and its integrations over contour times $t_{q}\ppr$, $t_{p}$ as follows (by using the metric $\eta_{p}$, $\eta_{q}$
(\ref{11},\ref{12}))
\beq \lb{34}
\lefteqn{\int_{C}\dtot t_{p}\;\dtot t_{q}\ppr\sum_{\vec{x},\vec{x}\ppr}\sum_{a,b=1,2}
\Psi_{\vec{x}\ppr}^{+b}(t_{q}\ppr)\;\;\hat{M}_{\vec{x}\ppr,\vec{x}}^{ba}(t_{q}\ppr,t_{p})\;\;
\Psi_{\vec{x}}^{a}(t_{p}) = } \\ \no &=&
\int_{-\infty}^{+\infty}\dtot t\;\dtot t\ppr\sum_{\vec{x},\vec{x}\ppr}\sum_{a,b=1,2}\sum_{p,q=\pm}
\Psi_{\vec{x}\ppr,q}^{+b}(t\ppr)\;\eta_{q}\;\hat{M}_{\vec{x}\ppr,q;\vec{x},p}^{ba}(t\ppr,t)\;\eta_{p}\;
\Psi_{\vec{x},p}^{a}(t)\;\;\;.
\eeq
However, the fields \(\Psi_{\vec{x}}^{a}(t_{-}=+T_{0}/2)\) and
\(\Psi_{\vec{x}}^{a}(t_{+}=+T_{0}/2)\) have to approach exactly the same values at the time $+T_{0}/2$
for continuity reasons. This must hold in both kinds of expressions with coherent state fields in (\ref{34})
(compare (\ref{30},\ref{32}) for contour time ordering and relations (\ref{31},\ref{33}) with prevailing
order for the anomalous parts).

The Gaussian ensemble averages of \(Z[\mcal{J},V_{I}]\), \(Z[\mcal{J},V_{II}]\) (\ref{25},\ref{26})
with second moments (\ref{5},\ref{6}) result in relations
\(\ovv{Z_{I}[\mcal{J}]}\), \(\ovv{Z_{II}[\mcal{J}]}\)
(\ref{35},\ref{36}) where the 'Nambu'-doubling has also been included for the
symmetry breaking source terms with
\(J_{\psi;\vec{x}}^{a}(t_{p})\), \(J_{\psi;\vec{x}}^{+a}(t_{p})\) and also
\(\hat{J}_{\psi\psi;\vec{x}}^{ba}(t_{q}\ppr,t_{p})\)
\beq \lb{35}
\ovv{Z_{I}[\mcal{J}]}&=&\int\dtot[\psi_{\vec{x}}(t_{p})]\;\exp\bigg\{-\frac{i}{\hbar}\int_{C}
\dtot t_{p}\sum_{\vec{x}}
\psi_{\vec{x}}^{*}(t_{p})\;\;\hat{h}_{p}(t_{p})\;\;\psi_{\vec{x}}(t_{p})\bigg\} \\ \no &\times&
\exp\bigg\{-\frac{\im}{2\hbar}\int_{C}\dtot t_{p}\sum_{\vec{x}}\Big[
J_{\psi;\vec{x}}^{+a}(t_{p})\;\Psi_{\vec{x}}^{a}(t_{p})+
\Psi_{\vec{x}}^{+a}(t_{p})\;J_{\psi;\vec{x}}^{a}(t_{p})\Big]\bigg\}
\\ \no &\times& \exp\bigg\{-\frac{\im}{2\hbar}\int_{C}\dtot t_{p}\sum_{\vec{x}}
\Psi_{\vec{x}}^{+b}(t_{p})\;\;\hat{J}_{\psi\psi;\vec{x}}^{ba}(t_{p})\;\;\Psi_{\vec{x}}^{a}(t_{p})\bigg\}
\\ \no &\times&
\exp\bigg\{-\frac{\im}{2\hbar}\int_{C}\dtot t_{p}\;\dtot t_{q}\ppr\sum_{\vec{x},\vec{x}\ppr}
\Psi_{\vec{x}\ppr}^{+b}(t_{q}\ppr)\;\;\mcal{J}_{\vec{x}\ppr,\vec{x}}^{ba}(t_{q}\ppr,t_{p})\;\;
\Psi_{\vec{x}}^{a}(t_{p})\bigg\}
\\ \no &\times& \exp\bigg\{-\frac{\im}{\hbar}
\int_{C}\dtot t_{p}\sum_{\vec{x}}V_{0}\;\big(\psi_{\vec{x}}^{*}(t_{p})\big)^{2}\;\;
\big(\psi_{\vec{x}}(t_{p})\big)^{2}\bigg\} \\ \no &\times&
\exp\bigg\{-\frac{R_{I}^{2}\Omega^{2}}{2\hbar^{2}\mcal{N}_{x}}\sum_{\vec{x}}
\bigg(\int_{C}\dtot t_{p}\;\;\psi_{\vec{x}}^{*}(t_{p})\;\psi_{\vec{x}}(t_{p})\bigg)\;\;
\bigg(\int_{C}\dtot t_{q}\ppr\;\;\psi_{\vec{x}}^{*}(t_{q}\ppr)\;\psi_{\vec{x}}(t_{q}\ppr)\bigg)\bigg\}
\eeq
\beq \lb{36}
\ovv{Z_{II}[\mcal{J}]}&=&
\int\dtot[\psi_{\vec{x}}(t_{p})]\;\exp\bigg\{-\frac{i}{\hbar}\int_{C}
\dtot t_{p}\sum_{\vec{x}}
\psi_{\vec{x}}^{*}(t_{p})\;\;\hat{h}_{p}(t_{p})\;\;\psi_{\vec{x}}(t_{p})\bigg\} \\ \no &\times&
\exp\bigg\{-\frac{\im}{2\hbar}\int_{C}\dtot t_{p}\sum_{\vec{x}}\Big[
J_{\psi;\vec{x}}^{+a}(t_{p})\;\Psi_{\vec{x}}^{a}(t_{p})+
\Psi_{\vec{x}}^{+a}(t_{p})\;J_{\psi;\vec{x}}^{a}(t_{p})\Big]\bigg\}
\\ \no &\times& \exp\bigg\{-\frac{\im}{2\hbar}\int_{C}\dtot t_{p}\sum_{\vec{x}}
\Psi_{\vec{x}}^{+b}(t_{p})\;\;\hat{J}_{\psi\psi;\vec{x}}^{ba}(t_{p})\;\;\Psi_{\vec{x}}^{a}(t_{p})\bigg\} \\ \no &\times&
\exp\bigg\{-\frac{\im}{2\hbar}\int_{C}\dtot t_{p}\;\dtot t_{q}\ppr\sum_{\vec{x},\vec{x}\ppr}
\Psi_{\vec{x}\ppr}^{+b}(t_{q}\ppr)\;\;
\mcal{J}_{\vec{x}\ppr,\vec{x}}^{ba}(t_{q}\ppr,t_{p})\;\;\Psi_{\vec{x}}^{a}(t_{p})\bigg\}
\\ \no &\times&
\exp\bigg\{-\frac{\im}{\hbar}\int_{C}\dtot t_{p}
\sum_{\vec{x}}V_{0}\;\big(\psi_{\vec{x}}^{*}(t_{p})\big)^{2}\;\;
\big(\psi_{\vec{x}}(t_{p})\big)^{2}\bigg\} \\ \no &\times&
\exp\bigg\{-\frac{R_{II}^{2}}{2\hbar^{2}}\int_{-\infty}^{\infty}\dtot t\sum_{\vec{x}}\sum_{p,q=\pm}
\Big(\psi_{\vec{x}}^{*}(t_{p})\;\eta_{p}\;\psi_{\vec{x}}(t_{p})\Big)\;\;
\Big(\psi_{\vec{x}}^{*}(t_{q})\;\eta_{q}\;\psi_{\vec{x}}(t_{q})\Big)\bigg\}\;\;.
\eeq
The 'Nambu'-doubling \(J_{\psi;\vec{x}}^{a(=1/2)}(t_{p})\) (\ref{37},\ref{38}) of the source term
\(j_{\psi;\vec{x}}(t_{p})\) is obtained in a similar manner as
that of the coherent state field \(\psi_{\vec{x}}(t_{p})\) (\ref{30},\ref{31}).
The 'Nambu'-doubled source field for \(j_{\psi;\vec{x}}(t_{p})\) is also defined by its
capital letter \(J_{\psi;\vec{x}}^{a}(t_{p})\) with the additional index (\(a=1,2\))
and can also be listed in the two manners of ordering (\ref{30},\ref{31})
as the doubled coherent state field \(\Psi_{\vec{x}}^{a}(t_{p})\)
\beq \lb{37}
 &  & \mbox{'contour time ordering' :}\\ \no
J_{\psi;\vec{x}}^{a(=1/2)}(t_{p(=\pm)})&=&\left(
\bea{c}
j_{\psi;\vec{x}}(t_{p})\;(a=1) \\ j_{\psi;\vec{x}}^{*}(t_{p})\;(a=2)
\eea\right)=\left(
\bea{c}
j_{\psi;\vec{x}}(t_{+}) \;\;(a=1)\\ j_{\psi;\vec{x}}^{*}(t_{+})\;\;(a=2) \\
j_{\psi;\vec{x}}(t_{-})\;\;(a=1) \\
j_{\psi;\vec{x}}^{*}(t_{-})\;\;(a=2)
\eea\right)
\eeq
\beq \lb{38}  &  & \mbox{'ordering for anomalous terms' :}\\ \no
J_{\psi;\vec{x}}^{a(=1/2)}(t_{p(=\pm)})&=&\left(
\bea{c}
j_{\psi;\vec{x}}(t_{p})\;(a=1) \\ j_{\psi;\vec{x}}^{*}(t_{p})\;(a=2)
\eea\right)=\left(
\bea{c}
j_{\psi;\vec{x}}(t_{+})\;\;(a=1) \\ j_{\psi;\vec{x}}(t_{-})\;\;(a=1) \\
j_{\psi;\vec{x}}^{*}(t_{+})\;\;(a=2) \\
j_{\psi;\vec{x}}^{*}(t_{-})\;\;(a=2)
\eea\right)_{\mbox{.}}
\eeq
The 'Nambu'-doubling of the source term \(j_{\psi\psi;\vec{x}}(t_{p})\)
for the pair condensate terms yields a matrix \(\hat{J}_{\psi\psi;\vec{x}}^{ab}(t_{p})\)
with local contour time dependence
which can also be ordered in the two analogous kinds as the $U(1)$ source term \(J_{\psi;\vec{x}}^{a}(t_{p})\)
(\ref{37},\ref{38}) or the field \(\Psi_{\vec{x}}^{a}(t_{p})\) (\ref{30},\ref{31})
\beq \lb{39}
&  &\mbox{'contour time ordering' :}  \\ \no
\hat{J}_{\psi\psi;\vec{x}}^{ab}(t_{p})&=&\left(
\bea{cccc}
0 & j_{\psi\psi;\vec{x}}(t_{+}) & 0 & 0 \\
j_{\psi\psi;\vec{x}}^{*}(t_{+}) & 0 & 0 & 0 \\
0 & 0 & 0 &  j_{\psi\psi;\vec{x}}(t_{-}) \\
0 & 0 &  j_{\psi\psi;\vec{x}}^{*}(t_{-}) & 0
\eea\right)
\eeq
\beq \lb{40}
 &  & \mbox{'ordering for anomalous terms' :}\\ \no
\hat{J}_{\psi\psi;\vec{x}}^{ab}(t_{p})&=&\left(
\bea{cccc}
0 & 0 & j_{\psi\psi;\vec{x}}(t_{+}) & 0 \\
0 & 0 & 0 & j_{\psi\psi;\vec{x}}(t_{-}) \\
j_{\psi\psi;\vec{x}}^{*}(t_{+}) & 0 & 0 &  0 \\
0 & j_{\psi\psi;\vec{x}}^{*}(t_{-}) &  0 & 0
\eea\right)_{\mbox{.}}
\eeq
The source terms $J_{\psi;\vec{x}}^{a}(t_{\pm})$ and $\hat{J}_{\psi\psi;\vec{x}}^{ab}(t_{\pm})$
have to be set to equivalent values concerning the two branches of
the contour time $t_{\pm}$ in the final relations for the observables or the saddle point equation (\ref{15}).
In the remainder the equivalent notations (\ref{41}-\ref{44}) for \(J_{\psi;\vec{x}}^{a}(t_{p})\) and
\(\hat{J}_{\psi\psi;\vec{x}}^{ab}(t_{p})\) will also temporarily occur as the equivalent notations for
the coherent state fields and 'Nambu'-doubled matrices (\ref{34})
\beq \lb{41}
 &  &\mbox{'contour time ordering' :} \\ \no
J_{\psi;\vec{x}}^{a(=1/2)}(t_{p(=\pm)})&=&J_{\psi;\vec{x},p(=\pm)}^{a(=1/2)}(t)=\left(
\bea{c}
j_{\psi;\vec{x},p}(t)\;(a=1) \\ j_{\psi;\vec{x},p}^{*}(t)\;(a=2)
\eea\right)=\left(
\bea{c}
j_{\psi;\vec{x},+}(t) \;\;(a=1)\\ j_{\psi;\vec{x},+}^{*}(t)\;\;(a=2) \\ j_{\psi;\vec{x},-}(t)\;\;(a=1) \\
j_{\psi;\vec{x},-}^{*}(t)\;\;(a=2)
\eea\right)
\eeq
\beq \lb{42} &  & \mbox{'ordering for anomalous terms' :} \\ \no
J_{\psi;\vec{x}}^{a(=1/2)}(t_{p(=\pm)})&=&J_{\psi;\vec{x},p(=\pm)}^{a(=1/2)}(t)=\left(
\bea{c}
j_{\psi;\vec{x},p}(t)\;(a=1) \\ j_{\psi;\vec{x},p}^{*}(t)\;(a=2)
\eea\right)=\left(
\bea{c}
j_{\psi;\vec{x},+}(t)\;\;(a=1) \\ j_{\psi;\vec{x},-}(t)\;\;(a=1) \\ j_{\psi;\vec{x},+}^{*}(t)\;\;(a=2) \\
j_{\psi;\vec{x},-}^{*}(t)\;\;(a=2)
\eea\right)
\eeq
\beq \lb{43}
 &  &\mbox{'contour time ordering' :} \\ \no
\hat{J}_{\psi\psi;\vec{x}}^{ab}(t_{p})&=&\hat{J}_{\psi\psi;\vec{x},p}^{ab}(t)=\left(
\bea{cccc}
0 & j_{\psi\psi;\vec{x},+}(t) & 0 & 0 \\
j_{\psi\psi;\vec{x},+}^{*}(t) & 0 & 0 & 0 \\
0 & 0 & 0 &  j_{\psi\psi;\vec{x},-}(t) \\
0 & 0 &  j_{\psi\psi;\vec{x},-}^{*}(t) & 0
\eea\right)
\eeq
\beq \lb{44}
 &  & \mbox{'ordering for anomalous terms' :}\\ \no
\hat{J}_{\psi\psi;\vec{x}}^{ab}(t_{p})&=&\hat{J}_{\psi\psi;\vec{x},p}^{ab}(t)=\left(
\bea{cccc}
0 & 0 & j_{\psi\psi;\vec{x},+}(t) & 0 \\
0 & 0 & 0 & j_{\psi\psi;\vec{x},-}(t) \\
j_{\psi\psi;\vec{x},+}^{*}(t) & 0 & 0 &  0 \\
0 & j_{\psi\psi;\vec{x},-}^{*}(t) &  0 & 0
\eea\right)_{\mbox{.}}
\eeq
One has to apply the matrix form \(\hat{J}_{\psi\psi;\vec{x}\ppr,\vec{x}}^{ba}(t_{q}\ppr,t_{p})\) or its
corresponding notation symbol \(\hat{J}_{\psi\psi;\vec{x}\ppr,q;\vec{x},p}^{ba}(t\ppr,t)\) (\ref{45})
in the case of two time contour integrations (\ref{46}) with
the bilinear fields \(\Psi_{\vec{x}\ppr}^{+,b}(t_{q}\ppr)\ldots\Psi_{\vec{x}}^{a}(t_{p})\) as with the
matrix \(\hat{M}_{\vec{x}\ppr,\vec{x}}^{ba}(t_{q}\ppr,t_{p}):=\hat{M}_{\vec{x}\ppr,q;\vec{x},p}^{ba}(t\ppr,t)\)
in (\ref{34})
\beq \lb{45}
\hat{J}_{\psi\psi;\vec{x}\ppr,\vec{x}}^{ba}(t_{q}\ppr,t_{p})&=&
\hat{J}_{\psi\psi;\vec{x}\ppr,q;\vec{x},p}^{ba}(t\ppr,t)
= \delta_{p,q}\;\eta_{p}\;\delta(t_{p}-t_{q}\ppr)\;\delta_{\vec{x},\vec{x}\ppr}\;\;
\hat{J}_{\psi\psi;\vec{x}}^{ab}(t_{p}) \\ \no &=& \delta_{p,q}\;\eta_{p}\;\delta(t_{p}-t_{q}\ppr)\;
\delta_{\vec{x},\vec{x}\ppr}\;\;\hat{J}_{\psi\psi;\vec{x},p}^{ab}(t)
\eeq
\beq \lb{46}
\lefteqn{\int_{C}\dtot t_{p}\;\dtot t_{q}\ppr\sum_{\vec{x},\vec{x}\ppr}\sum_{a,b=1,2}
\Psi_{\vec{x}\ppr}^{+b}(t_{q}\ppr)\;\;\hat{J}_{\psi\psi;\vec{x}\ppr,\vec{x}}^{ba}(t_{q}\ppr,t_{p})\;\;
\Psi_{\vec{x}}^{a}(t_{p}) = } \\ \no &=&
\int_{-\infty}^{+\infty}\dtot t\;\dtot t\ppr\sum_{\vec{x},\vec{x}\ppr}\sum_{a,b=1,2}\sum_{p,q=\pm}
\Psi_{\vec{x}\ppr,q}^{+b}(t\ppr)\;\eta_{q}\;\hat{J}_{\psi\psi;\vec{x}\ppr,q;\vec{x},p}^{ba}(t\ppr,t)\;\eta_{p}\;
\Psi_{\vec{x},p}^{a}(t) = \\ \no &=& \int_{C}\dtot t_{p}\sum_{\vec{x}}\sum_{a,b=1,2}
\Psi_{\vec{x}}^{+b}(t_{p})\;\;\hat{J}_{\psi\psi;\vec{x}}^{ba}(t_{p})\;\;
\Psi_{\vec{x}}^{a}(t_{p})\;\;\;.
\eeq
In this section we have achieved the ensemble averages of the disorder models I, II with the coherent
state path integrals \(\ovv{Z_{I}[\raisebox{-1pt}{\mbox{$\mcal{J}$}}]}\),
\(\ovv{Z_{II}[\raisebox{-1pt}{\mbox{$\mcal{J}$}}]}\) (\ref{35},\ref{36}).
We have described the various forms and equivalent notations concerning the coherent state fields and matrices
(\ref{30}-\ref{34}) on the Keldysh time contour $t_{p=\pm}$. The 'Nambu'-doubling of source
fields and matrices has also been incorporated for the creation of a coherent BE-wavefunction and
pair condensate terms (\ref{37}-\ref{46}).

\section{Hubbard-Stratonovich transformations for the repulsive and ensemble-averaged interactions
in model I and II}\lb{s3}

\subsection{Hubbard-Stratonovich transformation for repulsive interactions in model I and II}\lb{s31}

The repulsive interaction with parameter $V_{0}>0$ is a common part of the two disorder models I and II.
Its Hubbard-Stratonovich transformation (HST) to a density matrix
\(\hat{r}_{\vec{x}}^{ab}(t_{p})\) (\ref{47},\ref{48}) is accomplished by a dyadic product
of the fields in the repulsive interaction term where we already insert
the 'Nambu'-doubled form of the dyadic products with
\(\hat{r}_{\vec{x}}^{ab}(t_{p})=\Psi_{\vec{x}}^{a}(t_{p})\otimes\Psi_{\vec{x}}^{+b}(t_{p})\)
\beq \lb{47}
\lefteqn{\int_{C}\dtot t_{p}\sum_{\vec{x}}\big(\psi_{\vec{x}}^{*}(t_{p})\big)^{2}\;
\big(\psi_{\vec{x}}(t_{p})\big)^{2}=\frac{1}{4}
\int_{C}\dtot t_{p}\sum_{\vec{x}}\Big(\psi_{\vec{x}}^{*}(t_{p})\;\psi_{\vec{x}}(t_{p})+
\psi_{\vec{x}}(t_{p})\;\psi_{\vec{x}}^{*}(t_{p})\Big)^{2} } \\ \no &=&\frac{1}{4}\int_{C}\dtot t_{p}\sum_{\vec{x}}
\underbrace{\Psi_{\vec{x}}^{a}(t_{p})\otimes\Psi_{\vec{x}}^{+b}(t_{p})}_{\hat{r}_{\vec{x}}^{ab}(t_{p})}\;\;
\underbrace{\Psi_{\vec{x}}^{b}(t_{p})\otimes\Psi_{\vec{x}}^{+a}(t_{p})}_{\hat{r}_{\vec{x}}^{ba}(t_{p})}  =
\frac{1}{4}\int_{C}\dtot t_{p}\sum_{\vec{x}}\trab\Big[\hat{r}_{\vec{x}}^{ab}(t_{p})\;
\hat{r}_{\vec{x}}^{ba}(t_{p})\Big]\;\;\;.
\eeq
We introduce the self-energy matrix \(\hat{\sigma}_{\vec{x}}^{ab}(t_{p})\)
(\ref{49}) for the repulsive interaction term with $V_{0}>0$
in the models I, II with static and dynamic disorder
according to the symmetries of the resulting density matrix \(\hat{r}_{\vec{x}}^{ab}(t_{p})\)
(\ref{48}) in the trace '\(\mbox{tr}_{a,b}\)' over 'Nambu'-indices (\ref{47})
\beq \lb{48}
\hat{r}_{\vec{x}}^{ab}(t_{p})&=&\left(
\bea{cc}
\psi_{\vec{x}}(t_{p})\;\psi_{\vec{x}}^{*}(t_{p}) & \psi_{\vec{x}}(t_{p})\;\psi_{\vec{x}}(t_{p}) \\
\psi_{\vec{x}}^{*}(t_{p})\;\psi_{\vec{x}}^{*}(t_{p}) & \psi_{\vec{x}}^{*}(t_{p})\;\psi_{\vec{x}}(t_{p})
\eea\right) \\  \lb{49}
\hat{\sigma}_{\vec{x}}^{ab}(t_{p})&=&\left(
\bea{cc}
b_{\vec{x}}(t_{p}) & c_{\vec{x}}(t_{p}) \\
c_{\vec{x}}^{*}(t_{p}) & b_{\vec{x}}(t_{p})
\eea\right) \hspace*{1.0cm} b_{\vec{x}}(t_{p})\in{\sf R}\hspace*{0.5cm}c_{\vec{x}}(t_{p})\in{\sf C}\;\;\;.
\eeq
This self-energy matrix \(\hat{\sigma}_{\vec{x}}^{ab}(t_{p})\) (\ref{49})
has a local dependence in the spatial coordinates and also regarding the contour time.
Therefore, the HST of the repulsive interaction or of its density matrix form
with the trace '\(\mbox{tr}_{a,b}\)' over the 'Nambu'-indices $a,b=1,2$ yields
the relation (\ref{50}) with hermitian action in a Gaussian factor of the self-energy
and the hermitian coupling between density matrix \(\hat{r}_{\vec{x}}^{ab}(t_{p})\)
and \(\hat{\sigma}_{\vec{x}}^{ab}(t_{p})\)
\beq \lb{50}
\lefteqn{\exp\bigg\{-\frac{\im}{\hbar}\int_{C}\dtot t_{p}\sum_{\vec{x}}V_{0}\;
\big(\psi_{\vec{x}}^{*}(t_{p})\big)^{2}\;\;
\big(\psi_{\vec{x}}(t_{p})\big)^{2}\bigg\} =
\exp\bigg\{-\frac{\im}{4\hbar}\;V_{0}\int_{C}\dtot t_{p}\sum_{\vec{x}}\trab\Big[
\hat{r}_{\vec{x}}^{ab}(t_{p})\;\hat{r}_{\vec{x}}^{ba}(t_{p})\Big]\bigg\}  = }   \\ \no &=&
\int\dtot[\hat{\sigma}_{\vec{x}}^{ab}(t_{p})]\;\exp\bigg\{\frac{\im}{4\hbar}\frac{1}{V_{0}}
\int_{C}\dtot t_{p}\sum_{\vec{x}}
\trab\Big[\hat{\sigma}_{\vec{x}}^{ab}(t_{p})\;\hat{\sigma}_{\vec{x}}^{ba}(t_{p})\Big]\bigg\} \;
\exp\bigg\{ \frac{\im}{2\hbar}\int_{C}\dtot t_{p}\sum_{\vec{x}}
\trab\Big[\hat{\sigma}_{\vec{x}}^{ab}(t_{p})\;\hat{r}_{\vec{x}}^{ba}(t_{p})\Big]\bigg\}_{\mbox{.}}
\eeq
The real field \(b_{\vec{x}}(t_{p})=\sigma_{\vec{x}}^{11}(t_{p})=\sigma_{\vec{x}}^{22}(t_{p})\) in (\ref{49})
describes the density term of the self-energy for the repulsive interaction whereas the complex field
\(c_{\vec{x}}(t_{p})=\sigma_{\vec{x}}^{12}(t_{p})\) and its complex conjugate
\(c_{\vec{x}}^{*}(t_{p})=\sigma_{\vec{x}}^{21}(t_{p})\) in (\ref{49}) determine the
anomalous terms of the interaction with \(V_{0}>0\).

\subsection{Hubbard-Stratonovich transformation for the disorder term in model I and derivation of
the mean field equations with the disorder-self-energy}\lb{s32}

The HST of the quartic, non-hermitian 'interaction term' for the disorder in model I is
obtained by the dyadic product of fields in a similar manner as in section \ref{s31}.
The 'Nambu'-doubling with \(\Psi_{\vec{x}}^{a}(t_{p})\otimes\Psi_{\vec{x}}^{+b}(t_{q}\ppr)\)
leads to the disorder-density matrix
\(\hat{R}_{\vec{x}}^{ab}(t_{p},t_{q}\ppr)\) with inclusion of the anomalous terms
\beq \lb{51}
\lefteqn{\int_{C}\dtot t_{p}\;\dtot t_{q}\ppr\sum_{\vec{x}}\psi_{\vec{x}}^{*}(t_{p})\;\psi_{\vec{x}}(t_{p})\;\;
\psi_{\vec{x}}^{*}(t_{q}\ppr)\;\psi_{\vec{x}}(t_{q}\ppr) = \frac{1}{4}\int_{C}\dtot t_{p}\;\dtot t_{q}\ppr
\sum_{\vec{x}}\Big(\Psi_{\vec{x}}^{+a}(t_{p})\;\Psi_{\vec{x}}^{a}(t_{p})\Big)\;\;
\Big(\Psi_{\vec{x}}^{+b}(t_{q}\ppr)\;\Psi_{\vec{x}}^{b}(t_{q}\ppr)\Big) =} \\ \no &=&\frac{1}{4}
\int_{C}\dtot t_{p}\;\dtot t_{q}\ppr\sum_{\vec{x}}
\underbrace{\Psi_{\vec{x}}^{a}(t_{p})\otimes
\Psi_{\vec{x}}^{+b}(t_{q}\ppr)}_{\hat{R}_{\vec{x}}^{ab}(t_{p},t_{q}\ppr)}\;\;
\underbrace{\Psi_{\vec{x}}^{b}(t_{q}\ppr)\otimes
\Psi_{\vec{x}}^{+a}(t_{p})}_{\hat{R}_{\vec{x}}^{ba}(t_{q}\ppr,t_{p})}
= \frac{1}{4}\int_{C}\dtot t_{p}\;\dtot t_{q}\ppr\sum_{\vec{x}}
\trab\Big[\hat{R}_{\vec{x}}^{ab}(t_{p},t_{q}\ppr)\;\;\hat{R}_{\vec{x}}^{ba}(t_{q}\ppr,t_{p})\Big]
\\ \no &=&\frac{1}{4}\int_{-\infty}^{+\infty}\dtot t\;\dtot t\ppr\sum_{\vec{x}}
\TRAB\Big[\hat{R}_{\vec{x};pq}^{ab}(t,t\ppr)\;\eta_{q}\;
\hat{R}_{\vec{x};qp}^{ba}(t\ppr,t)\;\eta_{p}\Big]\;\;\;.
\eeq
We explicitly list the spatially local density matrix
\(\hat{R}_{\vec{x}}^{ab}(t_{p},t_{q}\ppr)\) in (\ref{52},\ref{51})
with 'Nambu'-indices (\(a,b=1,2\))
and {\it nonlocal} time contour dependence with $t_{p}$, $t_{q}\ppr$ (\(p,q=\pm\))
according to the contour ordering
of fields as for \(\Psi_{\vec{x}}^{a}(t_{p})=\Psi_{\vec{x},p}^{a}(t)\) (\ref{54},\ref{30},\ref{32}).
Note that two different notations for the disorder-density matrix (\ref{52},\ref{53})
can be used as in the cases of \(\Psi\raisebox{1pt}{\mbox{$_{\vec{x}}^{a}$}}(t_{p})=
\Psi\raisebox{2pt}{\mbox{$_{\vec{x},p}^{a}$}}(t)\) (\ref{30},\ref{32}),
\(J\raisebox{1pt}{\mbox{$_{\psi;\vec{x}}^{a}$}}(t_{p})=
J\raisebox{1pt}{\mbox{$_{\psi;\vec{x},p}^{a}$}}(t)\) (\ref{37},\ref{41}) or
\(\hat{J}_{\psi\psi;\vec{x}\ppr,\vec{x}}^{ba}(t_{q}\ppr,t_{p})=
\hat{J}_{\psi\psi;\vec{x}\ppr,q;\vec{x},p}^{ba}(t\ppr,t)\)
(\ref{45},\ref{46},\ref{39},\ref{43}).
Therefore, we have also added in relation (\ref{51}) the last line with the
trace '\(\mbox{Tr}_{p,q;a,b}\)' and the metric $\eta_{p}$, $\eta_{q}$
for the disorder-density matrix in notation (\ref{53}).
This clarifies the symmetry relations between the matrix elements of
\(\hat{R}_{\vec{x}}^{ab}(t_{p},t_{q}\ppr)=\hat{R}_{\vec{x};pq}^{ab}(t,t\ppr)\) (\ref{52},\ref{53}).
In the listings (\ref{52},\ref{53}) the contour time ordering prevails for the disorder-density matrices
\(\hat{R}_{\vec{x}}^{ab}(t_{p},t_{q}\ppr)=\hat{R}_{\vec{x};pq}^{ab}(t,t\ppr)\)  which follow by the dyadic
products \(\Psi_{\vec{x}}^{a}(t_{p})\otimes\Psi_{\vec{x}}^{+b}(t_{q}\ppr)\),
\(\Psi_{\vec{x},p}^{a}(t)\otimes\Psi_{\vec{x},q}^{+b}(t\ppr)\) of fields (\ref{54}) also applied with
respect to the contour time order
\be \lb{52}
\hat{R}_{\vec{x}}^{ab}(t_{p},t_{q}\ppr)=\left(
\bea{cccc}
\psi_{\vec{x}}(t_{+})\;\psi_{\vec{x}}^{*}(t_{+}\ppr) &
\psi_{\vec{x}}(t_{+})\;\psi_{\vec{x}}(t_{+}\ppr) &
\psi_{\vec{x}}(t_{+})\;\psi_{\vec{x}}^{*}(t_{-}\ppr) &
\psi_{\vec{x}}(t_{+})\;\psi_{\vec{x}}(t_{-}\ppr) \\
\psi_{\vec{x}}^{*}(t_{+})\;\psi_{\vec{x}}^{*}(t_{+}\ppr) &
\psi_{\vec{x}}^{*}(t_{+})\;\psi_{\vec{x}}(t_{+}\ppr) &
\psi_{\vec{x}}^{*}(t_{+})\;\psi_{\vec{x}}^{*}(t_{-}\ppr) &
\psi_{\vec{x}}^{*}(t_{+})\;\psi_{\vec{x}}(t_{-}\ppr) \\
\psi_{\vec{x}}(t_{-})\;\psi_{\vec{x}}^{*}(t_{+}\ppr) &
\psi_{\vec{x}}(t_{-})\;\psi_{\vec{x}}(t_{+}\ppr) &
\psi_{\vec{x}}(t_{-})\;\psi_{\vec{x}}^{*}(t_{-}\ppr) &
\psi_{\vec{x}}(t_{-})\;\psi_{\vec{x}}(t_{-}\ppr) \\
\psi_{\vec{x}}^{*}(t_{-})\;\psi_{\vec{x}}^{*}(t_{+}\ppr) &
\psi_{\vec{x}}^{*}(t_{-})\;\psi_{\vec{x}}(t_{+}\ppr) &
\psi_{\vec{x}}^{*}(t_{-})\;\psi_{\vec{x}}^{*}(t_{-}\ppr) &
\psi_{\vec{x}}^{*}(t_{-})\;\psi_{\vec{x}}(t_{-}\ppr)
\eea\right)
\ee
\be   \lb{53}
\hat{R}_{\vec{x};pq}^{ab}(t,t\ppr)=\left(
\bea{cccc}
\psi_{\vec{x},+}(t)\;\psi_{\vec{x},+}^{*}(t\ppr) &
\psi_{\vec{x},+}(t)\;\psi_{\vec{x},+}(t\ppr) &
\psi_{\vec{x},+}(t)\;\psi_{\vec{x},-}^{*}(t\ppr) &
\psi_{\vec{x},+}(t)\;\psi_{\vec{x},-}(t\ppr) \\
\psi_{\vec{x},+}^{*}(t)\;\psi_{\vec{x},+}^{*}(t\ppr) &
\psi_{\vec{x},+}^{*}(t)\;\psi_{\vec{x},+}(t\ppr) &
\psi_{\vec{x},+}^{*}(t)\;\psi_{\vec{x},-}^{*}(t\ppr) &
\psi_{\vec{x},+}^{*}(t)\;\psi_{\vec{x},-}(t\ppr) \\
\psi_{\vec{x},-}(t)\;\psi_{\vec{x},+}^{*}(t\ppr) &
\psi_{\vec{x},-}(t)\;\psi_{\vec{x},+}(t\ppr) &
\psi_{\vec{x},-}(t)\;\psi_{\vec{x},-}^{*}(t\ppr) &
\psi_{\vec{x},-}(t)\;\psi_{\vec{x},-}(t\ppr) \\
\psi_{\vec{x},-}^{*}(t)\;\psi_{\vec{x},+}^{*}(t\ppr) &
\psi_{\vec{x},-}^{*}(t)\;\psi_{\vec{x},+}(t\ppr) &
\psi_{\vec{x},-}^{*}(t)\;\psi_{\vec{x},-}^{*}(t\ppr) &
\psi_{\vec{x},-}^{*}(t)\;\psi_{\vec{x},-}(t\ppr)
\eea\right)
\ee
\be   \lb{54}
\Psi_{\vec{x}}^{a(=1/2)}(t_{p(=\pm)})=\left(
\bea{c}
\psi_{\vec{x}}(t_{p})\;(a=1) \\
\psi_{\vec{x}}^{*}(t_{p})\;(a=2)
\eea\right)=\left(
\bea{c}
\psi_{\vec{x}}(t_{+})\;(a=1) \\
\psi_{\vec{x}}^{*}(t_{+})\;(a=2) \\
\psi_{\vec{x}}(t_{-})\;(a=1) \\
\psi_{\vec{x}}^{*}(t_{-})\;(a=2)
\eea\right)
\left(\bea{l}
\mbox{'contour time} \\ \mbox{ordering'}
\eea\right)_{\mbox{.}}
\ee
The corresponding 'Nambu'-doubled disorder-self-energy
\(\hat{\Sigma}_{I;\vec{x}}^{ab}(t_{p},t_{q}\ppr)= \hat{\Sigma}_{I;\vec{x};pq}^{ab}(t,t\ppr)\) (\ref{55},\ref{56})
has to fulfill the equivalent symmetry relations between its matrix elements as the disorder-density matrix
\(\hat{R}_{\vec{x}}^{ab}(t_{p},t_{q}\ppr)\), \(\hat{R}_{\vec{x};pq}^{ab}(t,t\ppr)\) (\ref{52},\ref{53}).
We also consider in (\ref{55},\ref{56}) the two equivalent notations as for the disorder-density matrix
(\ref{52},\ref{53}). The disorder-self-energy (\ref{55},\ref{56}) has a nonlocal dependence with respect to
the contour times $t_{p}$, $t_{q}\ppr$ and contains density related terms labeled with the capital letter
'$\hat{B}$' and pair condensate terms labeled with the capital letter '$\hat{C}$'.
The basic matrices of the disorder self-energy \(\hat{\Sigma}_{I;\vec{x}}^{ab}(t_{p},t_{q}\ppr)\) are
\(\hat{B}_{\vec{x}}(t_{+},t_{+}\ppr)\), \(\hat{B}_{\vec{x}}(t_{-},t_{-}\ppr)\) and
\(\hat{B}_{\vec{x}}(t_{+},t_{-}\ppr)\) for density related terms.
The basic matrices for the anomalous parts are given by \(\hat{C}_{\vec{x}}(t_{+},t_{+}\ppr)\),
\(\hat{C}_{\vec{x}}(t_{-},t_{-}\ppr)\) and \(\hat{C}_{\vec{x}}(t_{+},t_{-}\ppr)\).
Taking into account the symmetries of \(\hat{R}_{\vec{x}}^{ab}(t_{p},t_{q}\ppr)\),
one has to place the basic matrices \(\hat{B}_{\vec{x}}(t_{+},t_{+}\ppr)\),
\(\hat{B}_{\vec{x}}(t_{-},t_{-}\ppr)\) and \(\hat{B}_{\vec{x}}(t_{+},t_{-}\ppr)\) for densities and
the basic matrices \(\hat{C}_{\vec{x}}(t_{+},t_{+}\ppr)\), \(\hat{C}_{\vec{x}}(t_{-},t_{-}\ppr)\),
\(\hat{C}_{\vec{x}}(t_{+},t_{-}\ppr)\) for the pair condensates in the disorder-self-energy
as in relation (\ref{55},\ref{56}).
Furthermore, one has to require the symmetry restrictions (\ref{57}) for these basic matrices
of \(\hat{\Sigma}_{I;\vec{x}}^{ab}(t_{p},t_{q}\ppr)\), \(\hat{\Sigma}_{I;\vec{x};pq}^{ab}(t,t\ppr)\).
The two equivalent notations concerning the nonlocal contour time dependence are also tabulated
in relation (\ref{57}) for the symmetries of the '$\hat{B}$' and '$\hat{C}$' matrices.
These two notations clarify the symmetry relations with complex conjugation,
transposition and hermitian conjugation (see following examples for complex conjugation
\(\big(\hat{B}_{\vec{x}}(t_{p},t_{q}\ppr)\big)^{*}=\hat{B}_{\vec{x};pq}^{*}(t,t\ppr)\), transposition
  \(\big(\hat{B}_{\vec{x}}(t_{p},t_{q}\ppr)\big)^{T}=\hat{B}_{\vec{x};pq}^{T}(t,t\ppr)=
  \hat{B}_{\vec{x};pq}(t\ppr,t)=  \hat{B}_{\vec{x}}(t_{p}\ppr,t_{q})\) and hermitian conjugation
  \(\big(\hat{B}_{\vec{x}}(t_{p},t_{q}\ppr)\big)^{+}=\hat{B}_{\vec{x};pq}^{*}(t\ppr,t)=
  \hat{B}_{\vec{x}}^{*}(t_{p}\ppr,t_{q})\) within the two equivalent notations in (\ref{57}))
\beq \lb{55}
\hat{\Sigma}_{I;\vec{x}}^{ab}(t_{p},t_{q}\ppr)&=&\left(
\bea{cccc}
\hat{B}_{\vec{x}}(t_{+},t_{+}\ppr) & \hat{C}_{\vec{x}}(t_{+},t_{+}\ppr) & \hat{B}_{\vec{x}}(t_{+},t_{-}\ppr) & \hat{C}_{\vec{x}}(t_{+},t_{-}\ppr) \\
\hat{C}_{\vec{x}}^{+}(t_{+},t_{+}\ppr)&\hat{B}_{\vec{x}}^{T}(t_{+},t_{+}\ppr) & \hat{C}_{\vec{x}}^{*}(t_{+},t_{-}\ppr)& \hat{B}_{\vec{x}}^{*}(t_{+},t_{-}\ppr) \\
\hat{B}_{\vec{x}}^{+}(t_{+},t_{-}\ppr) & \hat{C}_{\vec{x}}^{T}(t_{+},t_{-}\ppr) & \hat{B}_{\vec{x}}(t_{-},t_{-}\ppr) & \hat{C}_{\vec{x}}(t_{-},t_{-}\ppr) \\
\hat{C}_{\vec{x}}^{+}(t_{+},t_{-}\ppr)& \hat{B}_{\vec{x}}^{T}(t_{+},t_{-}\ppr) & \hat{C}_{\vec{x}}^{+}(t_{-},t_{-}\ppr)& \hat{B}_{\vec{x}}^{T}(t_{-},t_{-}\ppr)
\eea\right)   \\  \lb{56}
\hat{\Sigma}_{I;\vec{x};pq}^{ab}(t,t\ppr)&=&\left(
\bea{cccc}
\hat{B}_{\vec{x};++}(t,t\ppr) & \hat{C}_{\vec{x};++}(t,t\ppr) & \hat{B}_{\vec{x};+-}(t,t\ppr) &
\hat{C}_{\vec{x};+-}(t,t\ppr) \\
\hat{C}_{\vec{x};++}^{+}(t,t\ppr)& \hat{B}_{\vec{x};++}^{T}(t,t\ppr) & \hat{C}_{\vec{x};+-}^{*}(t,t\ppr)&
\hat{B}_{\vec{x};+-}^{*}(t,t\ppr)
\\ \hat{B}_{\vec{x};+-}^{+}(t,t\ppr) & \hat{C}_{\vec{x};+-}^{T}(t,t\ppr) & \hat{B}_{\vec{x};--}(t,t\ppr) &
\hat{C}_{\vec{x};--}(t,t\ppr) \\
\hat{C}_{\vec{x};+-}^{+}(t,t\ppr)& \hat{B}_{\vec{x};+-}^{T}(t,t\ppr) & \hat{C}_{\vec{x};--}^{+}(t,t\ppr)&
\hat{B}_{\vec{x};--}^{T}(t,t\ppr)
\eea\right)
\eeq
\be  \lb{57}
\bea{rclrclrclrcl}
\psi_{\vec{x}}(t_{+})\;\psi_{\vec{x}}^{*}(t_{+}\ppr) &\propto &
\hat{B}_{\vec{x}}(t_{+},t_{+}\ppr)&=&\hat{B}_{\vec{x};++}(t,t\ppr)&=&
\Big(\hat{B}_{\vec{x}}(t_{+},t_{+}\ppr)\Big)^{+}&=&\Big(\hat{B}_{\vec{x};++}(t,t\ppr)\Big)^{+} \\
\psi_{\vec{x}}(t_{-})\;\psi_{\vec{x}}^{*}(t_{-}\ppr) &\propto &
\hat{B}_{\vec{x}}(t_{-},t_{-}\ppr)&=&\hat{B}_{\vec{x};--}(t,t\ppr)&=&
\Big(\hat{B}_{\vec{x}}(t_{-},t_{-}\ppr)\Big)^{+}&=&\Big(\hat{B}_{\vec{x};--}(t,t\ppr)\Big)^{+}  \\
\psi_{\vec{x}}(t_{-})\;\psi_{\vec{x}}^{*}(t_{+}\ppr) &\propto &
\hat{B}_{\vec{x}}(t_{-},t_{+}\ppr)&=&\hat{B}_{\vec{x};-+}(t,t\ppr)&=&\Big(\hat{B}_{\vec{x}}(t_{+},t_{-}\ppr)\Big)^{+}
&=&\Big(\hat{B}_{\vec{x};+-}(t,t\ppr)\Big)^{+} \\
\psi_{\vec{x}}(t_{+})\;\psi_{\vec{x}}(t_{+}\ppr) &\propto &
\hat{C}_{\vec{x}}(t_{+},t_{+}\ppr)&=&\hat{C}_{\vec{x};++}(t,t\ppr)&=&
\Big(\hat{C}_{\vec{x}}(t_{+},t_{+}\ppr)\Big)^{T}&=&\Big(\hat{C}_{\vec{x};++}(t,t\ppr)\Big)^{T} \\
\psi_{\vec{x}}(t_{-})\;\psi_{\vec{x}}(t_{-}\ppr) &\propto &
\hat{C}_{\vec{x}}(t_{-},t_{-}\ppr)&=&\hat{C}_{\vec{x};--}(t,t\ppr)&=&
\Big(\hat{C}_{\vec{x}}(t_{-},t_{-}\ppr)\Big)^{T}&=&\Big(\hat{C}_{\vec{x};--}(t,t\ppr)\Big)^{T}  \\
\psi_{\vec{x}}(t_{+})\;\psi_{\vec{x}}(t_{-}\ppr) &\propto &
\hat{C}_{\vec{x}}(t_{+},t_{-}\ppr)&=&\hat{C}_{\vec{x};+-}(t,t\ppr)&=&
\Big(\hat{C}_{\vec{x}}(t_{-},t_{+}\ppr)\Big)^{T}&=&\Big(\hat{C}_{\vec{x};-+}(t,t\ppr)\Big)^{T}.
\eea
\ee
We use relations (\ref{51}) to (\ref{57}) for the HST of the ensemble averaged
disorder term I with nonlocal time dependence and with the
symmetries following from the 'Nambu'-doubling. A 'non-hermitian' Gaussian factor of the
disorder-self-energy \(\hat{\Sigma}_{I;\vec{x}}^{ab}(t_{p},t_{q}\ppr)\) and its coupling to the disorder-density matrix
\(\hat{R}_{\vec{x}}^{ab}(t_{p},t_{q}\ppr)\) result in place of the quartic,
non-hermitian interaction of fields derived from the ensemble average with $V_{I}(\vec{x})$
\beq \lb{58}
\lefteqn{\exp\bigg\{-\frac{R_{I}^{2}\Omega^{2}}{2\hbar^{2}\mcal{N}_{x}}\int_{C}\dtot t_{p}\;\dtot t_{q}\ppr\sum_{\vec{x}}
\Big(\psi_{\vec{x}}^{*}(t_{p})\;\psi_{\vec{x}}(t_{p})\Big)\;
\Big(\psi_{\vec{x}}^{*}(t_{q}\ppr)\;\psi_{\vec{x}}(t_{q}\ppr)\Big)\bigg\}=} \\ \no &=&
\exp\bigg\{-\frac{1}{8}\frac{R_{I}^{2}\Omega^{2}}{\hbar^{2}\mcal{N}_{x}}\int_{C}\dtot t_{p}\;\dtot t_{q}\ppr\sum_{\vec{x}}
\trab\Big[\hat{R}_{\vec{x}}^{ab}(t_{p},t_{q}\ppr)\;\;\hat{R}_{\vec{x}}^{ba}(t_{q}\ppr,t_{p})\Big]\bigg\} = \\ \no &=&
\int\dtot[\hat{\Sigma}_{I;\vec{x}}^{ab}(t_{p},t_{q}\ppr)]\;\;
\exp\bigg\{-\frac{1}{8\hbar^{2}}\int_{C}\dtot t_{p}\;\dtot t_{q}\ppr\sum_{\vec{x}}
\trab\Big[\hat{\Sigma}_{I;\vec{x}}^{ab}(t_{p},t_{q}\ppr)\;
\hat{\Sigma}_{I;\vec{x}}^{ba}(t_{q}\ppr,t_{p})\Big]\bigg\}\times
\\ \no &\times&\exp\bigg\{-\frac{\im}{4\hbar^{2}}\frac{R_{I}\Omega}{\sqrt{\mcal{N}_{x}}}\int_{C}\dtot t_{p}\;
\dtot t_{q}\ppr\sum_{\vec{x}}
\trab\Big[\hat{\Sigma}_{I;\vec{x}}^{ab}(t_{p},t_{q}\ppr)\;
\hat{R}_{\vec{x}}^{ba}(t_{q}\ppr,t_{p})\Big]\bigg\}_{\mbox{.}}
\eeq
Substituting the terms of the HST's (\ref{50},\ref{58}) and the 'Nambu'-doubled symmetry breaking source terms into
\(\ovv{Z_{I}[\mcal{J}]}\) (\ref{35}), we acquire the ensemble averaged coherent state path integral of model I
with the self-energy \(\hat{\sigma}_{\vec{x}}^{ab}(t_{p})\) of the repulsive interaction and the disorder-self-energy
\(\hat{\Sigma}_{I;\vec{x}}^{ab}(t_{p},t_{q}\ppr)\). Moreover, we perform the 'Nambu'-doubling (\ref{60}-\ref{63})
on the one-particle terms in \(\ovv{Z_{I}[\mcal{J}]}\) so that relation (\ref{59})
for \(\ovv{Z_{I}[\mcal{J}]}\) only consists of 'Nambu'-doubled parts with
\(\hat{\Sigma}_{I;\vec{x}}^{ab}(t_{p},t_{q}\ppr)\), \(\hat{\sigma}_{\vec{x}}^{ab}(t_{p})\),
\(\Psi_{\vec{x}}^{a}(t_{p})\) and \(J_{\psi;\vec{x}}^{a}(t_{p})\),
\(\hat{J}_{\psi\psi;\vec{x}}^{ab}(t_{p})\) as well as \(\hat{\mcal{H}}_{\vec{x}\ppr,\vec{x}}^{ba}(t_{q}\ppr,t_{p})\)
\beq \no
\lefteqn{\hspace*{-1.0cm}\ovv{Z_{I}[\mcal{J}]}=\int\dtot[\hat{\Sigma}_{I;\vec{x}}^{ab}(t_{p},t_{q}\ppr)]\;
\dtot[\hat{\sigma}_{\vec{x}}^{ab}(t_{p})]\;
\exp\bigg\{-\frac{1}{8\hbar^{2}}\int_{C}\dtot t_{p}\;\dtot t_{q}\ppr\sum_{\vec{x}}\trab\Big[
\hat{\Sigma}_{I;\vec{x}}^{ab}(t_{p},t_{q}\ppr)\;\hat{\Sigma}_{I;\vec{x}}^{ba}(t_{q}\ppr,t_{p})\Big]\bigg\} }
\\ \lb{59} &\times& \exp\bigg\{\frac{\im}{4\hbar}\frac{1}{V_{0}}\int_{C}\dtot t_{p}\sum_{\vec{x}}\trab\Big[
\hat{\sigma}_{\vec{x}}^{ab}(t_{p})\;\hat{\sigma}_{\vec{x}}^{ba}(t_{p})\Big]\bigg\} \\ \no &\times&
\int\dtot[\psi_{\vec{x}}(t_{p})]\;
\exp\bigg\{-\frac{\im}{2\hbar}\int_{C}\dtot t_{p}\;\dtot t_{q}\ppr
\sum_{\vec{x},\vec{x}\ppr}
\Psi_{\vec{x}\ppr}^{+b}(t_{q}\ppr)\;\mcal{N}_{x}\;
\bigg[\hat{\mcal{H}}_{\vec{x}\ppr,\vec{x}}^{ba}(t_{q}\ppr,t_{p})+
\frac{\mcal{J}_{\vec{x}\ppr,\vec{x}}^{ba}(t_{q}\ppr,t_{p})}{\mcal{N}_{x}}+ \\ \no &+&
 \hat{J}_{\psi\psi;\vec{x}\ppr,\vec{x}}^{ba}(t_{q}\ppr,t_{p})- \delta(t_{p}-t_{q}\ppr)\;\delta_{p,q}\;\eta_{p}\;\delta_{\vec{x},\vec{x}\ppr}\;
\hat{\sigma}_{\vec{x}}^{ba}(t_{p})+
\frac{1}{2}\frac{R_{I}\Omega}{\hbar\sqrt{\mcal{N}_{x}}}\;\delta_{\vec{x},\vec{x}\ppr} \;
\hat{\Sigma}_{I;\vec{x}}^{ba}(t_{q}\ppr,t_{p})\bigg]\;
\Psi_{\vec{x}}^{a}(t_{p})\bigg\} \\ \no &\times&
\exp\bigg\{-\frac{\im}{2\hbar}\int_{C}\dtot t_{p}\sum_{\vec{x}}\Big[
J_{\psi;\vec{x}}^{+a}(t_{p})\;\Psi_{\vec{x}}^{a}(t_{p})+\Psi_{\vec{x}}^{+a}(t_{p})\;
J_{\psi;\vec{x}}^{a}(t_{p})\Big]\bigg\}_{\mbox{.}}
\eeq
The 'Nambu'-doubled one-particle Hamiltonian \(\hat{\mcal{H}}_{\vec{x}\ppr,\vec{x}}^{ba}(t_{q}\ppr,t_{p})\)
has to take the form as in (\ref{60}) for a chosen contour time ordering
with the part $\hat{h}_{p}(t_{p})$ (\ref{61}) and its transpose $\hat{h}_{p}^{T}(t_{p})$ (\ref{62})
(compare with the two notations of contour time ordering in (\ref{30},\ref{32},\ref{34},\ref{39},\ref{43},
\ref{45},\ref{46}))
\beq \lb{60}
\hat{\mcal{H}}_{\vec{x}\ppr,\vec{x}}^{ba}(t_{q}\ppr,t_{p})&=&\delta(t_{p}-t_{q}\ppr)\;
\delta_{p,q}\;\eta_{p}\;\delta_{\vec{x},\vec{x}\ppr}
\left(
\bea{cccc}
\hat{h}_{+}(t_{+}) &&& \\
&\hat{h}_{+}^{T}(t_{+}) && \\
&& \hat{h}_{-}(t_{-}) & \\
&&& \hat{h}_{-}^{T}(t_{-})
\eea\right) \\ \lb{61}
\hat{h}_{p}(t_{p})&=&-\im\hbar\frac{\pp}{\pp t_{p}}-\im\;\ve_{p}-\frac{\hbar^{2}}{2m}\Delta+u(\vec{x})-\mu_{0}
\\ \lb{62} \hat{h}_{p}^{T}(t_{p})&=&+\im\hbar\frac{\pp}{\pp t_{p}}-\im\;\ve_{p}-
\frac{\hbar^{2}}{2m}\Delta+u(\vec{x})-\mu_{0} \\ \lb{63}
\bigg(-\im\hbar\frac{\pp}{\pp t_{p}}\bigg)^{T}&=&\im\hbar\frac{\pp}{\pp t_{p}}_{\mbox{ .}}
\eeq
The contour energy operator \(\hat{E}_{p}=\im\hbar\;\pp/\pp t_{p}\) (\ref{63}) is antisymmetric with
respect to transposition whereas the other terms of \(\hat{h}_{p}(t_{p})\) are symmetric under transposition.
It has been mentioned in the introduction that the self-energy \(\hat{\sigma}_{\vec{x}}^{ab}(t_{p})\)
of the repulsive interaction can be considered as a subalgebra (with being the direct product of $M$ times \(sp(2)\)),
concerning the 'larger' disorder-self-energy \(\hat{\Sigma}_{I;\vec{x}}^{ab}(t_{p},t_{q}\ppr)\).
This disorder-self-energy can itself be regarded as an element of the symplectic Lie-Algebra $sp(4M)$,
with respect to the number of independent parameters which is given by \(4M\cdot(4M+1)/2\)
\cite{Corn,Luc},\cite{Bm2,Bm7}; (The parameter \(M\in \mbox{\sf N}>0\)
denotes the number of discrete time intervals or
steps during time development between \(-T_{0}/2<t_{p,j}<+T_{0}/2\) for times of
a single branch of the contour (\(p=\) fixed $\pm$ value, \(0<j<M-1\), \(t_{p,j}=\Delta t\cdot j\)).
This important observation allows to shift the disorder-self-energy
\(\hat{\Sigma}_{I;\vec{x}}^{ab}(t_{p},t_{q}\ppr)\) by the self-energy
\(\hat{\sigma}_{\vec{x}}^{ab}(t_{p})\) of the repulsive interaction (\ref{64}).
We can also transfer the source term \(\hat{J}_{\psi\psi;\vec{x}}^{ab}(t_{p})\) for the pair condensates
as a subset of the coset part
\(sp(4M)\backslash u(2M)\) of the symplectic Lie-Algebra $sp(4M)$ to the disorder-self-energy (\ref{65},\ref{66}).
Therefore, \(\hat{\Sigma}_{I;\vec{x}}^{ab}(t_{p},t_{q}\ppr)\), coupled to the bilinear fields
\(\Psi_{\vec{x}\ppr}^{+,b}(t_{q}\ppr)\ldots\Psi_{\vec{x}}^{a}(t_{p})\) in \(\ovv{Z_{I}[\mcal{J}]}\) (\ref{59}),
can absorb the self-energy of the repulsive interaction and the source matrix for the anomalous terms
\beq \lb{64}
\hat{\Sigma}_{I;\vec{x}}^{ab}(t_{p},t_{q}\ppr)&\to&
\hat{\Sigma}_{I;\vec{x}}^{ab}(t_{p},t_{q}\ppr)+2\frac{\hbar}{R_{I}}\frac{\sqrt{\mcal{N}_{x}}}{\Omega}\;
\delta_{p,q}\;\eta_{p}\;\delta(t_{p}-t_{q}\ppr)\;\hat{\sigma}_{\vec{x}}^{ab}(t_{p})  \\ \lb{65}
\hat{\Sigma}_{I;\vec{x}}^{ab}(t_{p},t_{q}\ppr)\;\delta_{\vec{x},\vec{x}\ppr}&\to&
\hat{\Sigma}_{I;\vec{x}}^{ab}(t_{p},t_{q}\ppr)\;\delta_{\vec{x},\vec{x}\ppr}
-2\frac{\hbar}{R_{I}}\frac{\sqrt{\mcal{N}_{x}}}{\Omega}\;
\hat{J}_{\psi\psi;\vec{x},\vec{x}\ppr}^{ab}(t_{p},t_{q}\ppr)  \\ \lb{66}
\hat{\Sigma}_{I;\vec{x}}^{ab}(t_{p},t_{q}\ppr)&\to&
\hat{\Sigma}_{I;\vec{x}}^{ab}(t_{p},t_{q}\ppr)
-2\frac{\hbar}{R_{I}}\frac{\sqrt{\mcal{N}_{x}}}{\Omega}\;\delta_{p,q}\;\eta_{p}\;\delta(t_{p}-t_{q}\ppr)\;
\hat{J}_{\psi\psi;\vec{x}}^{ab}(t_{p})\;\;\;.
\eeq
After these shifts of \(\hat{\Sigma}_{I;\vec{x}}^{ab}(t_{p},t_{q}\ppr)\) in \(\ovv{Z_{I}[\mcal{J}]}\) (\ref{59}),
we remove the 'Nambu'-doubled fields \(\Psi_{\vec{x}\ppr}^{+,b}(t_{q}\ppr)\ldots\Psi_{\vec{x}}^{a}(t_{p})\)
by integration. According to the doubling of the fields, we obtain the {\it square root} of the determinant with
the disorder-self-energy , the one-particle Hamiltonian \(\hat{\mcal{H}}_{\vec{x}\ppr,\vec{x}}^{ba}(t_{q}\ppr,t_{p})\)
and with the 'exterior' source term
\(\mcal{J}_{\vec{x}\ppr,\vec{x}}^{ba}(t_{q}\ppr,t_{p})\) for generating observables by differentiation. The
shifts of \(\hat{\Sigma}_{I;\vec{x}}^{ab}(t_{p},t_{q}\ppr)\) as in (\ref{64}-\ref{66})
have eliminated the self-energy \(\hat{\sigma}_{\vec{x}}^{ab}(t_{p})\)
of the repulsive interaction and the source matrix \(\hat{J}_{\psi\psi,\vec{x}}^{ab}(t_{p})\) from the determinant.
An additional Gaussian factor of \(\hat{\sigma}_{\vec{x}}^{ab}(t_{p})\) and Gaussian coupling terms with
\(\hat{\Sigma}_{I;\vec{x}}^{ab}(t_{p},t_{q}\ppr)\), \(\hat{J}_{\psi\psi;\vec{x}}^{ab}(t_{p})\) result
instead of the appearance in the functional determinant.
The Gaussian factors of the self-energy \(\hat{\sigma}_{\vec{x}}^{ab}(t_{p})\) disappear completely
from the coherent state path integral \(\ovv{Z_{I}[\mcal{J}]}\) (\ref{67}) after integration
as in (\ref{68}) so that the disorder-self-energy \(\hat{\Sigma}_{I;\vec{x}}^{ab}(t_{p},t_{q}\ppr)\)
remains as the only integration variable
\beq \lb{67}
\lefteqn{\ovv{Z_{I}[\mcal{J}]}=\int\dtot[\hat{\Sigma}_{I;\vec{x}}^{ab}(t_{p},t_{q}\ppr)]\;
\exp\bigg\{-\frac{1}{8\hbar^{2}}\int_{C}\dtot t_{p}\;\dtot t_{q}\ppr
\sum_{\vec{x}}\trab\Big[
\hat{\Sigma}_{I;\vec{x}}^{ab}(t_{p},t_{q}\ppr)\;\hat{\Sigma}_{I;\vec{x}}^{ba}(t_{q}\ppr,t_{p})\Big]\bigg\} }
\\ \no &\times & \exp\bigg\{\frac{1}{2}\frac{\sqrt{\mcal{N}_{x}}}{\hbar R_{I}\Omega}
\int_{C}\dtot t_{p}\sum_{\vec{x}}\trab\Big[\hat{\Sigma}_{I;\vec{x}}^{ab}(t_{p},t_{p})\;
\hat{J}_{\psi\psi;\vec{x}}^{ba}(t_{p})\Big]\bigg\}  \\ \no &\times&
\exp\bigg\{-\frac{1}{2}\frac{\mcal{N}_{x}}{R_{I}^{2}\;\Omega}\int_{-\infty}^{+\infty}\dtot t\;
\sum_{\vec{x}}\sum_{p=\pm}\trab\Big[\hat{J}_{\psi\psi;\vec{x}}^{ab}(t_{p})\;
\hat{J}_{\psi\psi;\vec{x}}^{ba}(t_{p})\Big]\bigg\}
\\ \no &\times&
\exp\bigg\{-\frac{1}{2}\int_{C}\frac{\dtot t_{p}}{\hbar}\eta_{p}\sum_{\vec{x}}\hbar\Omega\mcal{N}_{x} \\ \no &&
\trab\ln\bigg[\eta_{q}\bigg(
\hat{\mcal{H}}_{\vec{x}\ppr,\vec{x}}^{ba}(t_{q}\ppr,t_{p})+
\frac{\mcal{J}_{\vec{x}\ppr,\vec{x}}^{ba}(t_{q}\ppr,t_{p})}{\mcal{N}_{x}}+
\frac{1}{2}\frac{R_{I}\Omega}{\sqrt{\mcal{N}_{x}}\hbar}\;\delta_{\vec{x},\vec{x}\ppr}\;
\hat{\Sigma}_{I;\vec{x}}^{ba}(t_{q}\ppr,t_{p})\bigg)\eta_{p}\bigg]\bigg\} \\ \no &\times&
\exp\bigg\{\frac{\im}{2}\frac{\Omega^{2}}{\hbar}\int_{C}\dtot t_{p}\;\dtot t_{q}\ppr
\sum_{\vec{x},\vec{x}\ppr}\mcal{N}_{x}\sum_{a,b=1,2}
J_{\psi;\vec{x}\ppr}^{+b}(t_{q}\ppr)\;
\bigg[\eta_{q}\bigg(\hat{\mcal{H}}_{\vec{x}\ppr,\vec{x}}^{ba}(t_{q}\ppr,t_{p})+
\frac{\mcal{J}_{\vec{x}\ppr,\vec{x}}^{ba}(t_{q}\ppr,t_{p})}{\mcal{N}_{x}}+   \\ \no &+&
\frac{1}{2}\frac{R_{I}\Omega}{\sqrt{\mcal{N}_{x}}\hbar}\;\delta_{\vec{x},\vec{x}\ppr}\;
\hat{\Sigma}_{I;\vec{x}}^{ba}(t_{q}\ppr,t_{p})\bigg)\eta_{p}\bigg]^{-1,ba}_{\vec{x}\ppr,\vec{x}}
\!\!\!\Big(t_{q}\ppr,t_{p}\Big)\;\;\;J_{\psi,\vec{x}}^{a}(t_{p})\bigg\} \\ \no &\times&
\int\dtot[\hat{\sigma}_{\vec{x}}^{ab}(t_{p})]\;
\exp\bigg\{-\frac{1}{2}\int_{-\infty}^{+\infty}\dtot t
\sum_{\vec{x}}\sum_{p=\pm}\bigg(\frac{\mcal{N}_{x}}{R_{I}^{2}\Omega}-
\frac{\im\;\eta_{p}}{2\hbar V_{0}}\bigg)\trab\Big[
\hat{\sigma}_{\vec{x}}^{ab}(t_{p})\;\hat{\sigma}_{\vec{x}}^{ba}(t_{p})\Big]\bigg\} \\ \no &\times&
\exp\bigg\{-\frac{1}{2}\frac{\sqrt{\mcal{N}_{x}}}{\hbar R_{I}\Omega}\int_{C}\dtot t_{p}\sum_{\vec{x}}
\trab\Big[\hat{\sigma}_{\vec{x}}^{ab}(t_{p})\;\Big(\hat{\Sigma}_{I;\vec{x}}^{ba}(t_{p},t_{p})-
2\;\eta_{p}\;\frac{\hbar\sqrt{\mcal{N}_{x}}}{R_{I}}\;\hat{J}_{\psi\psi;\vec{x}}^{ba}(t_{p})\Big)
\Big]\bigg\}
\eeq
\beq \lb{68}
\lefteqn{\int\dtot[\hat{\sigma}_{\vec{x}}^{ab}(t_{p})]\;
\exp\bigg\{-\frac{1}{2}\int_{-\infty}^{+\infty}\dtot t
\sum_{\vec{x}}\sum_{p=\pm}\bigg(\frac{\mcal{N}_{x}}{R_{I}^{2}\Omega}-
\frac{\im\;\eta_{p}}{2\hbar V_{0}}\bigg)\trab\Big[
\hat{\sigma}_{\vec{x}}^{ab}(t_{p})\;\hat{\sigma}_{\vec{x}}^{ba}(t_{p})\Big]\bigg\}  } \\ \no &\times&
\exp\bigg\{-\frac{1}{2}\frac{\sqrt{\mcal{N}_{x}}}{\hbar R_{I}\Omega}\int_{C}\dtot t_{p}\sum_{\vec{x}}
\trab\Big[\hat{\sigma}_{\vec{x}}^{ab}(t_{p})\;\Big(\hat{\Sigma}_{I;\vec{x}}^{ba}(t_{p},t_{p})-
2\;\eta_{p}\;\frac{\hbar\sqrt{\mcal{N}_{x}}}{R_{I}}\;\hat{J}_{\psi\psi;\vec{x}}^{ba}(t_{p})\Big)
\Big]\bigg\} = \\ \no &=&
\exp\bigg\{\frac{1}{8}\frac{1}{\hbar^{2}\Omega}
\int_{-\infty}^{+\infty}\dtot t \sum_{\vec{x}}\sum_{p=\pm}
\frac{1}{\Big(1-\im\;\eta_{p}\;\frac{R_{I}^{2}\Omega}{\mcal{N}_{x}2\hbar V_{0}}\Big)}
\trab\Big[\hat{\Sigma}_{I;\vec{x}}^{ab}(t_{p},t_{p})\;
\hat{\Sigma}_{I;\vec{x}}^{ba}(t_{p},t_{p})\Big]\bigg\}   \\ \no &\times &
\exp\bigg\{-\frac{1}{2}\frac{\sqrt{\mcal{N}_{x}}}{\hbar R_{I}\Omega}
\int_{C}\dtot t_{p} \sum_{\vec{x}}\frac{1}{\Big(1-\im\;\eta_{p}\;\frac{R_{I}^{2}\Omega}{\mcal{N}_{x}2\hbar V_{0}}\Big)}
\trab\Big[\hat{\Sigma}_{I;\vec{x}}^{ab}(t_{p},t_{p})\;
\hat{J}_{\psi\psi;\vec{x}}^{ba}(t_{p})\Big]\bigg\}   \\ \no &\times &
\exp\bigg\{\frac{1}{2}\frac{\mcal{N}_{x}}{R_{I}^{2}\;\Omega}
\int_{-\infty}^{+\infty}\dtot t \sum_{\vec{x}}\sum_{p=\pm}\frac{1}{\Big(1-\im\;\eta_{p}\;
\frac{R_{I}^{2}\Omega}{\mcal{N}_{x}2\hbar V_{0}}\Big)}
\trab\Big[\hat{J}_{\psi\psi;\vec{x}}^{ab}(t_{p})\;
\hat{J}_{\psi\psi;\vec{x}}^{ba}(t_{p})\Big]\bigg\}_{\mbox{.}}
\eeq
Finally, we obtain the ensemble averaged generating function \(\ovv{Z_{I}[\mcal{J}]}\) (\ref{69})
for the disorder model I at zero temperature. It only contains as single integration variables the
disorder-self-energy matrix elements \(\hat{\Sigma}_{I;\vec{x}}^{ab}(t_{p},t_{q}\ppr)\). The generating function
\(\ovv{Z_{I}[\mcal{J}]}\) (\ref{69}) consists of Gaussian factors with
\(\hat{\Sigma}_{I;\vec{x}}^{ab}(t_{p},t_{q}\ppr)\), one functional determinant and the bilinear source term
with \(J_{\psi;\vec{x}\ppr}^{+b}(t_{q}\ppr)\ldots J_{\psi;\vec{x}}^{a}(t_{p})\) for the coherent
BE-condensate wavefunction. The final expression for \(\ovv{Z_{I}[\mcal{J}]}\) is listed in Eq. (\ref{69})
with the matrix \(\hat{M}_{I;\vec{x}\ppr,\vec{x}}^{ba}(t_{q}\ppr,t_{p})\) (\ref{71}) as an important ingredient
apart from the Gaussian factors with the complex parameter $\mu_{p}^{(I)}$ (\ref{70})
\beq \no
\lefteqn{\hspace*{-2.26cm}
\ovv{Z_{I}[\mcal{J}]}=\int\dtot[\hat{\Sigma}_{I;\vec{x}}^{ab}(t_{p},t_{q}\ppr)]\;
\exp\bigg\{-\frac{1}{8\hbar^{2}}\int_{C}\dtot t_{p}\;\dtot t_{q}\ppr \sum_{\vec{x}} {\ts
\bigg(1-\frac{\delta_{p,q}\;\delta(t_{p}-t_{q}\ppr)}{\Omega}\;\mu_{p}^{(I)}\bigg)}\;
\trab\Big[\hat{\Sigma}_{I;\vec{x}}^{ab}(t_{p},t_{q}\ppr)\;
\hat{\Sigma}_{I;\vec{x}}^{ba}(t_{q}\ppr,t_{p})\Big]\bigg\}  }
\\ \lb{69} &\times & \exp\bigg\{\frac{1}{2}\frac{\sqrt{N}_{x}}{\hbar R_{I}\Omega}\int_{C}\dtot t_{p}\sum_{\vec{x}}
\big(1-\mu_{p}^{(I)}\big)\;\trab\Big[\hat{\Sigma}_{I;\vec{x}}^{ab}(t_{p},t_{p})\;
\hat{J}_{\psi\psi;\vec{x}}^{ba}(t_{p})\Big]\bigg\}   \\ \no &\times&
\exp\bigg\{-\frac{1}{2}\frac{\mcal{N}_{x}}{R_{I}^{2}\;\Omega}\sum_{p=\pm}
\int_{-\infty}^{+\infty}\dtot t\sum_{\vec{x}}
\big(1-\mu_{p}^{(I)}\big)\;\trab\Big[\hat{J}_{\psi\psi;\vec{x}}^{ab}(t_{p})\;
\hat{J}_{\psi\psi;\vec{x}}^{ba}(t_{p})\Big]\bigg\}  \\ \no &\times &
\exp\bigg\{-\frac{1}{2}\int_{C}\frac{\dtot t_{p}}{\hbar}\eta_{p}\sum_{\vec{x}}\hbar\Omega\mcal{N}_{x}
\trab\ln\Big[\hat{M}_{I;\vec{x}\ppr,\vec{x}}^{ba}(t_{q}\ppr,t_{p})\Big]\bigg\}   \\ \no &\times &
\exp\bigg\{\frac{\im}{2}\frac{\Omega^{2}}{\hbar}\int_{C}\dtot t_{p}\;\dtot t_{q}\ppr\sum_{\vec{x},\vec{x}\ppr}
\mcal{N}_{x}\sum_{a,b=1,2}J_{\psi;\vec{x}\ppr}^{+b}(t_{q}\ppr)\;\;
\hat{M}^{-1;ba}_{I;\vec{x}\ppr,\vec{x}}(t_{q}\ppr,t_{p})\;\;J_{\psi;\vec{x}}^{a}(t_{p})\bigg\}
\eeq
\beq \lb{70}
\mu_{p}^{(I)}&=&\frac{1}{\Big(1-\im\;\eta_{p}\;\frac{R_{I}^{2}\Omega}{\mcal{N}_{x}2\hbar V_{0}}\Big)}=
\frac{1}{\Big(1-\frac{\im}{2}\;\eta_{p}\;\Big(\frac{R_{I}}{\hbar}\Big)^{2}\;
\Big(\frac{V_{0}}{(\hbar\Omega/\mcal{N}_{x})}\Big)^{-1}\Big)}  \\ \no &=&
\frac{1+\frac{\im}{2}\;\eta_{p}\;\Big(\frac{R_{I}}{\hbar}\Big)^{2}\;
\Big(\frac{V_{0}}{(\hbar\Omega/\mcal{N}_{x})}\Big)^{-1}}
{1+\frac{1}{4}\;\Big[\Big(\frac{R_{I}}{\hbar}\Big)^{2}\;
\Big(\frac{V_{0}}{(\hbar\Omega/\mcal{N}_{x})}\Big)^{-1}\Big]^{2}} \hspace*{1.0cm}
\Big(0<\Re(\mu_{p}^{(I)})<1\Big)
\eeq
\beq \lb{71}
\hat{M}_{I;\vec{x}\ppr,\vec{x}}^{ba}(t_{q}\ppr,t_{p}) &=&\eta_{q}\bigg(
\hat{\mcal{H}}_{\vec{x}\ppr,\vec{x}}^{ba}(t_{q}\ppr,t_{p})+
\frac{\mcal{J}_{\vec{x}\ppr,\vec{x}}^{ba}(t_{q}\ppr,t_{p})}{\mcal{N}_{x}}+
\frac{1}{2}\frac{R_{I}\Omega}{\sqrt{\mcal{N}_{x}}\hbar}\;\delta_{\vec{x},\vec{x}\ppr}\;
\hat{\Sigma}_{I;\vec{x}}^{ba}(t_{q}\ppr,t_{p})\bigg)\eta_{p}
 \\ \no &=&\hspace*{-0.28cm}\delta_{p,q}\;\eta_{p}\;
\delta(t_{p}-t_{q}\ppr)\;\delta_{\vec{x},\vec{x}\ppr}\left(
\bea{cccc}
\hat{h}_{+}(t_{+}) & 0 & 0 & 0 \\
0 & \hat{h}_{+}^{T}(t_{+}) & 0 & 0 \\
0 & 0 & \hat{h}_{-}(t_{-}) & 0 \\
0 & 0 & 0 & \hat{h}_{-}^{T}(t_{-})
\eea\right)  + \\ \no &+&\eta_{q}\bigg(
\frac{\mcal{J}_{\vec{x}\ppr,\vec{x}}^{ba}(t_{q}\ppr,t_{p})}{\mcal{N}_{x}}+
\frac{1}{2}\frac{R_{I}\Omega}{\sqrt{\mcal{N}_{x}}\hbar}\;\delta_{\vec{x}\ppr,\vec{x}}\;
\hat{\Sigma}_{I;\vec{x}}^{ba}(t_{q}\ppr,t_{p})\bigg)\eta_{p}\;\;\;.
\eeq
In order to derive the saddle point equation, we have to perform the first order variation of the actions in
\(\ovv{Z_{I}[\mcal{J}]}\) (\ref{69}) with respect to the disorder-self-energy
\(\hat{\Sigma}\raisebox{1.5pt}{\mbox{$_{I;\vec{x}}^{ab}$}}(t_{p},\raisebox{1.5pt}{\mbox{$t_{q}\ppr$}})\).
In principle one can continue the first order variations of the
actions in \(\ovv{Z_{I}[\raisebox{-1pt}{\mbox{$\mcal{J}$}}]}\) (\ref{69})
to second or even higher order variations as a kind of
functional Taylor expansion with the disorder-self-energy. We restrict in model I only to solutions following
from the first order variations \(\delta\hat{\Sigma}_{I;\vec{x}}^{ab}(t_{p},t_{q}\ppr)\) and have to scale all
parameters and self-energy matrix fields to dimensionless values.
This scaling to dimensionless quantities is given in relations (\ref{72}) to (\ref{83})
\footnote{A tilde '\(\wtilde{\ph{V}}\)' over the self-energy, the one particle Hamilton operator or other
parameters refers to the corresponding dimensionless, scaled quantity.}
\beq \lb{72}
\mbox{dim}[\hat{\mcal{H}}_{\vec{x}\ppr,\vec{x}}^{ba}(t_{q}\ppr,t_{p})]&=&\frac{[\mbox{energy}]}{[\mbox{time}]}
\\ \lb{73}
\hat{\mcal{H}}_{\vec{x}\ppr,\vec{x}}^{ba}(t_{q}\ppr,t_{p}) &\to &
\wtilde{\mcal{H}}_{\vec{x}_{j}\ppr,\vec{x}_{i}}^{ba}(t_{q,l}\ppr,t_{p,k})=\frac{
\hat{\mcal{H}}_{\vec{x}_{j}\ppr,\vec{x}_{i}}^{ba}(t_{q,l}\ppr,t_{p},k)}{\hbar\Omega^{2}}
\eeq
\beq \lb{74}
\hat{\Sigma}_{I;\vec{x}}^{ba}(t_{q}\ppr,t_{p}) &\to&
\wtilde{\Sigma}_{I;\vec{x}_{i}}^{ba}(t_{q,l}\ppr,t_{p,k})=\frac{1}{\sqrt{\mcal{N}_{x}}}\;
\frac{\hat{\Sigma}_{I;\vec{x}_{i}}^{ba}(t_{q,l}\ppr,t_{p,k})}{\hbar\Omega}   \\ \lb{75}
R_{I}&\to&\wtilde{R}_{I}=\frac{R_{I}}{\hbar}_{\mbox{.}}
\eeq
Moreover, we have to consider a kind of typical 'level spacing' '$e$' (\ref{76},\ref{77}) which follows from
the fundamental discreteness of spatial and time-like variables and fields in \(\ovv{Z_{I}[\mcal{J}]}\) (\ref{69}).
A dimensionless parameter \(\wtilde{V}_{0}\) (\ref{76}) replaces the repulsive interaction strength $V_{0}$
after scaling with a kind of 'mean level spacing' '$e$' (\ref{77}) known from random matrix theories.
Additionally, we introduce the dimensionless quantity $\xi_{I}$ (\ref{78}) as the quotient of the second moment
related disorder parameter \(\wtilde{R}_{I}^{2}\) to the parameter $\wtilde{V}_{0}$ of the repulsive interaction
so that the complex parameter $\mu_{p}^{(I)}$ (\ref{70}) in \(\ovv{Z[\mcal{J}]}\) (\ref{69}) is determined
by relations (\ref{78},\ref{79})
\beq \lb{76}
V_{0}&\to&\wtilde{V_{0}}= \frac{V_{0}}{e} \\  \lb{77}
e&=&\frac{\hbar\Omega}{\mcal{N}_{x}}  \\ \lb{78}
\xi_{I} &=&\bigg(\frac{R_{I}}{\hbar}\bigg)^{2}\frac{1}{(V_{0}/e)}=\wtilde{R_{I}}^{2}/\wtilde{V_{0}} \\ \lb{79}
\mu_{p}^{(I)}&=&\frac{1}{1-\frac{\im}{2}\;\eta_{p}\;\xi_{I}}  \\ \lb{80}
J_{\psi;\vec{x}}^{a}(t_{p})&\to&\wtilde{J}_{\psi;\vec{x}_{i}}^{a}(t_{p,k})=\frac{1}{\sqrt{\mcal{N}_{x}}}\;
\frac{J_{\psi;\vec{x}_{i}}^{a}(t_{p,k})}{\hbar\Omega}  \\ \lb{81}
\hat{J}_{\psi\psi;\vec{x}\ppr,\vec{x}}^{ba}(t_{q}\ppr,t_{p}) &\to &
\wtilde{J}_{\psi\psi;\vec{x}_{j}\ppr,\vec{x}_{i}}^{ba}(t_{q,l}\ppr,t_{p,k})=\frac{
\hat{J}_{\psi\psi;\vec{x}_{j}\ppr,\vec{x}_{i}}^{ba}(t_{q,l}\ppr,t_{p},k)}{\hbar\Omega^{2}} =
\\ \no &&=\delta_{p,q}\;\eta_{p}\;\delta(t_{p,k},t_{q,l}\ppr)\;
\delta_{\vec{x}_{i},\vec{x}_{j}\ppr}\;\wtilde{J}_{\psi\psi;\vec{x}_{i}}^{ba}(t_{p,k})
\\ \lb{82} \hat{J}_{\psi\psi;\vec{x}}^{ba}(t_{p}) &\to &
\wtilde{J}_{\psi\psi;\vec{x}_{i}}^{ba}(t_{p,k})=
\frac{\hat{J}_{\psi\psi;\vec{x}_{i}}^{ba}(t_{p,k})}{\hbar\Omega} \\
\lb{83}
\frac{\mcal{J}_{\vec{x}\ppr,\vec{x}}^{ba}(t_{q}\ppr,t_{p})}{\mcal{N}_{x}}
&\to &
\wtilde{\mcal{J}}_{\vec{x}_{j}\ppr,\vec{x}_{i}}^{ba}(t_{q,l}\ppr,t_{p,k})=
\frac{\mcal{J}_{\vec{x}_{j}\ppr,\vec{x}_{i}}^{ba}(t_{q,l}\ppr,t_{p,k})}{\mcal{N}_{x}\hbar\Omega^{2}}_{\mbox{.}}
\eeq
A scaling of the source terms (\ref{80}-\ref{83}) has also to
be included for the derivation of the saddle equation. We list the
ensemble averaged generating function
\(\ovv{Z_{I}[\mcal{J}]}\) (\ref{69}-\ref{71}) in terms of the
scaled disorder-self-energy
\(\wtilde{\Sigma}_{I;\vec{x}_{i}}^{\raisebox{-3pt}{\mbox{$\scr
ba$}}} (t\raisebox{0.5pt}{\mbox{$_{q,l}$}}\ppr,t_{p,k})\) (\ref{74})
and corresponding scaled operators, fields and parameters
(\ref{72}-\ref{83}) in discrete space-time coordinates in relation
(\ref{84}). The generating function
\(\ovv{Z_{I}[\raisebox{-1.5pt}{\mbox{$\wtilde{\mcal{J}}$}}]}\)
(\ref{84}) only consists of discrete sums with space-time points
\(\vec{x}_{i}\), \(\vec{x}_{j}\) and \(t_{p,k}\), \(t_{q,l}\ppr\)
(\(p,q=\pm\;;\;t_{p,k}=k\cdot\Delta
t_{p}\;;\;t_{q,l}\ppr=l\cdot\Delta t_{q}\ppr\;;\;k,l=0,\ldots,M-1\))
\footnote{In the remainder the symbol
'\(\delta(t_{p,k},t_{q,l}\ppr)\)' as in
\(\ovv{Z_{I}[\wtilde{\mcal{J}}]}\) (\ref{84},\ref{81}) denotes
the Kronecker-delta for the discrete times \(t_{p,k}\) and
\(t_{q,l}\ppr\).}
\beq \lb{84}
\lefteqn{\ovv{Z_{I}[\wtilde{\mcal{J}}]}=
\int\dtot[\wtilde{\Sigma}_{I;\vec{x}_{i}}^{ab}(t_{p,k},t_{q,l}\ppr)]\;
\exp\bigg\{-\frac{1}{2}\frac{1}{\wtilde{R}_{I}^{2}}\sum_{p=\pm}\sum_{t_{p,k}}\sum_{\vec{x}_{i}}
(1-\mu_{p}^{(I)})\;\trab\Big[\wtilde{J}_{\psi\psi;\vec{x}_{i}}^{ab}(t_{p,k})\;
\wtilde{J}_{\psi\psi;\vec{x}_{i}}^{ba}(t_{p,k})\Big]\bigg\} }
\\ \no  &\times&
\exp\bigg\{-\frac{1}{8}\sum_{p,q=\pm}\sum_{t_{p,k},t_{q,l}\ppr}\sum_{\vec{x}_{i}}
\Big(1-\delta_{p,q}\;\delta(t_{p,k},t_{q,l}\ppr)\;\mu_{p}^{(I)}\Big)\;
\trab\Big[\eta_{p}\;\wtilde{\Sigma}_{I;\vec{x}_{i}}^{ab}(t_{p,k},t_{q,l}\ppr)\;\eta_{q}\;
\wtilde{\Sigma}_{I;\vec{x}_{i}}^{ba}(t_{q,l}\ppr,t_{p,k})\Big]\bigg\}
\times \eeq \beq  \no &\times &
\exp\bigg\{\frac{1}{2}\frac{1}{\wtilde{R}_{I}}\sum_{p=\pm}\sum_{t_{p,k}}\sum_{\vec{x}_{i}}
(1-\mu_{p}^{(I)})\;\eta_{p}\;\trab\Big[\wtilde{\Sigma}_{I;\vec{x}_{i}}^{ab}(t_{p,k},t_{p,k})\;
\wtilde{J}_{\psi\psi;\vec{x}_{i}}^{ba}(t_{p,k})\Big]\bigg\} \\
\no &\times &
\exp\bigg\{-\frac{1}{2}\sum_{p=\pm}\sum_{t_{p,k}}\frac{\Delta
t_{p,k}}{\hbar}\sum_{\vec{x}_{i}}
\hbar\Omega\;\;\trab\ln\Big[\wtilde{M}_{I;\vec{x}_{j}\ppr,\vec{x}_{i}}^{ba}(t_{q,l}\ppr,t_{p,k})\Big]\bigg\}
\\ \no &\times &
\exp\bigg\{\frac{\im}{2}\sum_{p,q=\pm}\;\;\sum_{t_{p,k},t_{q,l}\pr}\;\;\sum_{\vec{x}_{i},\vec{x}_{j}\ppr}
\wtilde{J}_{\psi;\vec{x}_{j}\ppr}^{+b}(t_{q,l}\ppr)\;\eta_{q}\;
\wtilde{M}_{I;\vec{x}_{j}\ppr,\vec{x}_{i}}^{-1;ba}(t_{q,l}\ppr,t_{p,k})\;\eta_{p}\;
\wtilde{J}_{\psi;\vec{x}_{i}}^{a}(t_{p,k}) \bigg\}
\eeq
\be \lb{85}
\wtilde{M}_{I;\vec{x}_{j}\ppr,\vec{x}_{i}}^{ba}(t_{q,l}\ppr,t_{p,k})=\eta_{q}\bigg(
\wtilde{\mcal{H}}_{\vec{x}_{j}\ppr,\vec{x}_{i}}^{ba}(t_{q,l}\ppr,t_{p,k})+
\wtilde{\mcal{J}}_{\vec{x}_{j}\ppr,\vec{x}_{i}}^{ba}(t_{q,l}\ppr,t_{p,k})+\frac{1}{2}\;\wtilde{R}_{I}\;
\wtilde{\Sigma}_{I;\vec{x}_{i}}^{ba}(t_{q,l}\ppr,t_{p,k})\;\delta_{\vec{x}_{i},\vec{x}_{j}\ppr}
\bigg)\eta_{p}
\ee
\be \lb{86}
\int_{C}\dtot t_{p}\underbrace{\int_{L^{d}}\frac{d^{d}x}{L^{d}}}_{\sum_{\vec{x}}}\ldots
\mbox{fields}(\vec{x},t_{p})\ldots\to
\sum_{p=\pm}\hspace*{0.28cm}\sum_{\stackrel{k=0,\ldots,M-1}{t_{p,k}}}\hspace*{0.28cm}\sum_{\vec{x}_{i}}\eta_{p}\ldots
\wtilde{\mbox{fields}}(\vec{x}_{i},t_{p,k})\ldots\;\;\;.
\ee
The sum \(\sum_{t_{p,k},t_{q,l}\ppr}\) over discrete time contour variables in (\ref{84}) is defined
{\it without} the metric \(\eta_{p}\), \(\eta_{q}\) which has therefore to be included separately with
the sum \(\sum\raisebox{1.5pt}{\mbox{$_{p,q=\pm}$}}\) over the contour time branches.
However, the metric \(\eta_{p}\), \(\eta_{q}\)
has to appear in the appropriate terms of
\(\ovv{Z_{I}[\raisebox{-2pt}{\mbox{$\wtilde{\mcal{J}}$}}]}\) (\ref{84}) where a contour
integration is really performed with the negative sign for the propagation in the backward
direction of a time development.

The variation of \(\ovv{Z_{I}[\wtilde{\mcal{J}}]}\) (\ref{84}) with respect to
\(\delta\wtilde{\Sigma}_{I;\vec{x}_{i}}^{ab}(t_{p,k},t_{q,l}\ppr)\) has to be accomplished with great care,
implementing the particular discreteness of the contour time with its two branches and the sign
(\(\eta_{p}=p\), \(\eta_{\pm}=\pm\)) for forward and backward propagation
\beq \lb{87}
\lefteqn{\hspace*{-1.0cm}\Big(1-\delta_{p,q}\;\delta(t_{p,k},t_{q,l}\ppr)\;\mu_{p}^{(I)}\Big)\;
\wtilde{\Sigma}_{I;\vec{x}_{i}}^{ba}(t_{q,l}\ppr,t_{p,k}) =2\;
\frac{(1-\mu_{p}^{(I)})\;\eta_{p}}{\wtilde{R}_{I}}\;
\delta_{p,q}\;\delta(t_{p,k},t_{q,l}\ppr)\;\;
\wtilde{J}_{\psi\psi;\vec{x}_{i}}^{ba}(t_{p,k}) +} \\ \no &-&  \wtilde{R}_{I}\;\;
\wtilde{M}_{I;\vec{x}_{i},\vec{x}_{i}}^{-1;ba}(t_{q,l}\ppr,t_{p,k})
-\im\;\wtilde{R}_{I}
\sum_{p\ppr,q\ppr=\pm}\hspace*{0.19cm}\sum_{\tau_{p\ppr,k\ppr},\tau_{q\ppr,l\ppr}\ppr}\hspace*{0.19cm}
\sum_{\vec{y}_{i\ppr},\vec{y}_{j\ppr}\ppr}\hspace*{0.19cm}\sum_{c,d=1,2} \\ \no &&
\wtilde{J}_{\psi;\vec{y}_{j\ppr}\ppr}^{+d}(\tau_{q\ppr,l\ppr}\ppr)\;\eta_{q\ppr}\;
\wtilde{M}_{I;\vec{y}_{j\ppr}\ppr,\vec{x}_{i}}^{-1;da}(\tau_{q\ppr,l\ppr}\ppr,t_{p,k})\;\;\;
\wtilde{M}_{I;\vec{x}_{i},\vec{y}_{i\ppr}}^{-1;bc}(t_{q,l}\ppr,\tau_{p\ppr,k\ppr})\;\eta_{p\ppr}\;
\wtilde{J}_{\psi;\vec{y}_{i\ppr}}^{c}(\tau_{p\ppr,k\ppr})
\eeq
\beq \lb{88}
\wtilde{J}_{\psi;\vec{x}_{i}}^{a}(t_{p=+,k})&=&\wtilde{J}_{\psi;\vec{x}_{i}}^{a}(t_{p=-,k})\hspace*{1.0cm}
\wtilde{J}_{\psi\psi;\vec{x}_{i}}^{ab}(t_{p=+,k})=\wtilde{J}_{\psi\psi;\vec{x}_{i}}^{ab}(t_{p=-,k})\;\;\;.
\eeq
A solution of the saddle point equation (\ref{87},\ref{88}) directly follows from continued
fractions with the disorder-self-energy \(\wtilde{\Sigma}_{I;\vec{x}_{i}}^{ab}(t_{p,k},t_{q,l}\ppr)\)
as a mean field solution \cite{Groe,Viswa}.
The various steps towards a converging solution of the continued
fractions simplify in the case of time independent source terms and trap potential.
Assuming further spatial symmetries (isotropic or translational invariance),
one can reduce the expenditure for solving (\ref{87},\ref{88}) with continued fractions.
The continued fraction of (\ref{87},\ref{88}) leads to a solution under very general assumptions
regardless of spatial symmetries or time independence of source terms and the trap potential. However, one has
to fulfill the pole structure in the continued fractions, originally defined
by the imaginary increment \(-\im\;\ve_{p}\).

The iteration \(m\to m+1\) from \(\wtilde{\Sigma}_{I;\vec{x}_{i}}^{ab}(m;t_{p,k},t_{q,l}\ppr)\) to
\(\wtilde{\Sigma}_{I;\vec{x}_{i}}^{ab}(m+1;t_{p,k},t_{q,l}\ppr)\) proceeds according to relation (\ref{89})
via continued fraction. One starts from the noninteracting, 'free' Green function with vanishing disorder-self-energy
\(\wtilde{\Sigma}_{I;\vec{x}_{i}}^{ab}(m=0;t_{p,k},t_{q,l}\ppr)\equiv 0\) and obtains a disorder-self-energy
\(\wtilde{\Sigma}_{I;\vec{x}_{i}}^{ab}(m=1;t_{p,k},t_{q,l}\ppr)\) with non-vanishing non-diagonal
'\(+-\)' and '\(-+\)' parts (see the notations for contour time ordering)
\beq \lb{89}
\lefteqn{\Big(1-\delta_{p,q}\;\delta(t_{p,k},t_{q,l}\ppr)\;\mu_{p}^{(I)}\Big)\;
\wtilde{\Sigma}_{I;\vec{x}_{i}}^{ba}(m+1;t_{q,l}\ppr,t_{p,k}) = }   \\ \no &= &
2\;\frac{(1-\mu_{p}^{(I)})\;\eta_{p}}{\wtilde{R}_{I}}\;
\delta_{p,q}\;\delta(t_{p,k},t_{q,l}\ppr)\;\;
\wtilde{J}_{\psi\psi;\vec{x}_{i}}^{ba}(t_{p,k}) -  \wtilde{R}_{I}\;\;
\wtilde{M}_{I;\vec{x}_{i},\vec{x}_{i}}^{-1;ba}(m;t_{q,l}\ppr,t_{p,k})+
\\ \no &-&\im\;\wtilde{R}_{I}
\sum_{p\ppr,q\ppr=\pm}\hspace*{0.19cm}\sum_{\tau_{p\ppr,k\ppr},\tau_{q\ppr,l\ppr}\ppr}\hspace*{0.19cm}
\sum_{\vec{y}_{i\ppr},\vec{y}_{j\ppr}\ppr}\hspace*{0.19cm}\sum_{c,d=1,2}
\wtilde{J}_{\psi;\vec{y}_{j\ppr}\ppr}^{+d}(\tau_{q\ppr,l\ppr}\ppr)\;\eta_{q\ppr}\;
\wtilde{M}_{I;\vec{y}_{j\ppr}\ppr,\vec{x}_{i}}^{-1;da}(m;\tau_{q\ppr,l\ppr}\ppr,t_{p,k})\;\times \\ \no &\times&
\wtilde{M}_{I;\vec{x}_{i},\vec{y}_{i\ppr}}^{-1;bc}(m;t_{q,l}\ppr,\tau_{p\ppr,k\ppr})\;\eta_{p\ppr}\;
\wtilde{J}_{\psi;\vec{y}_{i\ppr}}^{c}(\tau_{p\ppr,k\ppr})
\eeq
\beq \lb{90}
\wtilde{M}_{I;\vec{x}_{i},\vec{x}_{j}\ppr}^{ab}(m;t_{p,k},t_{q,l}\ppr)&=&\eta_{p}\bigg(
\wtilde{\mcal{H}}_{\vec{x}_{i},\vec{x}_{j}\ppr}^{ab}(t_{p,k},t_{q,l}\ppr)+
\frac{1}{2}\;\wtilde{R}_{I}\;
\wtilde{\Sigma}_{I;\vec{x}_{i}}^{ab}(m;t_{p,k},t_{q,l}\ppr)\;\delta_{\vec{x}_{i},\vec{x}_{j}\ppr}
\bigg)\eta_{q}  \\ \lb{91}
\wtilde{J}_{\psi;\vec{x}_{i}}^{a}(t_{p=+,k})&=&\wtilde{J}_{\psi;\vec{x}_{i}}^{a}(t_{p=-,k})\hspace*{1.0cm}
\wtilde{J}_{\psi\psi;\vec{x}_{i}}^{ab}(t_{p=+,k})=\wtilde{J}_{\psi\psi;\vec{x}_{i}}^{ab}(t_{p=-,k})\;\;\;.
\eeq
The nondiagonal parts \(\wtilde{\Sigma}_{I;\vec{x}_{i}}^{ab}(m+1;t_{p=\pm,k},t_{q=\mp,l}\ppr)\)
reappear at every iteration step due to the source field
\(\wtilde{J}_{\psi;\vec{y}_{i\ppr}}^{c}(\tau_{p\ppr,k\ppr})\)
for the creation of a coherent BE-wavefunction. The anomalous parts
\(\langle\psi_{\vec{x}}(t_{p})\;\psi_{\vec{x}}(t_{p})\rangle\) follow from the source matrix
\(\wtilde{J}_{\psi\psi;\vec{x}_{i}}^{ab}(t_{p,k})\neq 0\) (for \(a\neq b\); \(a,b=1,2\))
which has a diagonal contour time dependence, but nondiagonal terms (\(a\neq b\)) for the creation
of pair condensates.

In the case of a time independent trap potential, the solution
\(\wtilde{\Sigma}_{I;\vec{x}_{i}}^{ab}(t_{p,k},t_{q,l}\ppr)\) simplifies to
\(\wtilde{\Sigma}_{I;\vec{x}_{i};pq}^{ab}(t_{k}-t_{l}\ppr)\)
whose Fourier transform therefore takes the form
\(\wtilde{\Sigma}_{I;\vec{\xi};pq}^{ab}(\omega)\) with a
dimensionless frequency $\omega$ and dimensionless spatial vector
$\vec{\xi}$. In this case the disorder-self-energy and one-particle Hamiltonian reduce to
\(\wtilde{\Sigma}_{I;\vec{\xi};pq}^{ab}(\omega)\)
(\ref{95}) and \(\wtilde{\mcal{H}}_{\vec{\xi},p;\vec{\xi}\ppr,q}^{ab}(\omega)\)
(\ref{93},\ref{94}) in the matrix
\(\wtilde{M}_{I;\vec{\xi},p;\vec{\xi}\ppr,q}^{ab}(\omega)\)
(\ref{92})
\beq \lb{92}\hspace*{-0.91cm}
\wtilde{M}_{I;\vec{\xi},p;\vec{\xi}\ppr,q}^{ab}(\omega)&=&
\eta_{p}\bigg(\wtilde{\mcal{H}}_{\vec{\xi},p;\vec{\xi}\ppr,q}^{ab}(\omega)
+\frac{1}{2}\;\wtilde{R}_{I}\;\;
\wtilde{\Sigma}_{I;\vec{\xi};pq}^{ab}(\omega)\;\;\delta_{\vec{\xi},\vec{\xi}\ppr}\bigg)\eta_{q}
\\ \lb{93}
\wtilde{\mcal{H}}_{\vec{\xi},p;\vec{\xi}\ppr,q}^{ab}(\omega)&=&\delta_{p,q}\;\eta_{p}\;\delta_{a,b}\;
\delta_{\vec{\xi},\vec{\xi}\ppr}\;\;\left[-\omega\;\hat{1}_{4\times4}+
\left(\bea{cccc}
\hat{\wtilde{\mbox{\sf h}}}_{+}(\vec{\xi}) & 0 & 0 & 0  \\
0 & \hat{\wtilde{\mbox{\sf h}}}_{+}^{T}(\vec{\xi}) & 0 & 0 \\
0 & 0 & \hat{\wtilde{\mbox{\sf h}}}_{-}(\vec{\xi}) &  0 \\
0 & 0 & 0 & \hat{\wtilde{\mbox{\sf h}}}_{-}^{T}(\vec{\xi})
\eea\right)\right]_{pq}^{ab}
\\ \lb{94} \hat{\wtilde{\mbox{\sf h}}}_{p}(\vec{\xi}) &=&
-\im\;\wtilde{\ve}_{p}-\pp_{\vec{\xi}}\cdot\pp_{\vec{\xi}}+\wtilde{u}(\vec{\xi})-\wtilde{\mu}_{0}\;\;;
\hspace*{0.55cm}\hat{\wtilde{\mbox{\sf h}}}_{p}^{T}(\vec{\xi})=\hat{\wtilde{\mbox{\sf h}}}_{p}(\vec{\xi})  \\ \lb{95}
\wtilde{\Sigma}_{I;\vec{\xi};pq}^{ab}(\omega)&=&\left( \bea{cccc} \wtilde{B}_{\vec{\xi};++}(\omega) &
\wtilde{C}_{\vec{\xi};++}(\omega) &
\wtilde{B}_{\vec{\xi};+-}(\omega) & \wtilde{C}_{\vec{\xi};+-}(\omega) \\
\wtilde{C}_{\vec{\xi};++}^{*}(\omega)& \wtilde{B}_{\vec{\xi};++}(\omega) & \wtilde{C}_{\vec{\xi};+-}^{*}(\omega)&
\wtilde{B}_{\vec{\xi};+-}^{*}(\omega)
\\ \wtilde{B}_{\vec{\xi};+-}^{*}(\omega) & \wtilde{C}_{\vec{\xi};+-}(\omega) &
\wtilde{B}_{\vec{\xi};--}(\omega) & \wtilde{C}_{\vec{\xi};--}(\omega) \\
\wtilde{C}_{\vec{\xi};+-}^{*}(\omega)& \wtilde{B}_{\vec{\xi};+-}(\omega) & \wtilde{C}_{\vec{\xi};--}^{*}(\omega)&
\wtilde{B}_{\vec{\xi};--}(\omega) \eea\right)_{\mbox{.}} \eeq The Green function
\(\wtilde{M}_{I;\vec{\xi},p;\vec{\xi}\ppr,q}^{-1;ab}(\omega)\) is obtained from solving the generalized eigenvalue
problem (\ref{96}) with the eigenfunction \(\Psi_{\vec{\xi},p}^{R,a}(\omega_{N})\) and eigenvalue $\omega_{N}$
\beq\lb{96} \lefteqn{\hspace*{-0.55cm}\sum_{q=\pm}\sum_{b=1,2} \left(\bea{cccc}
\hat{\wtilde{\mbox{\sf h}}}(\vec{\xi}) & 0 & 0 & 0  \\
0 & \hat{\wtilde{\mbox{\sf h}}}^{T}(\vec{\xi}) & 0 & 0 \\
0 & 0 & -\hat{\wtilde{\mbox{\sf h}}}(\vec{\xi}) &  0 \\
0 & 0 & 0 & -\hat{\wtilde{\mbox{\sf h}}}^{T}(\vec{\xi})
\eea\right)_{pq}^{ab}
\underbrace{\left(\bea{c}
\psi_{\vec{\xi},+}(\omega_{N}) \\ \psi_{\vec{\xi},+}^{*}(\omega_{N}) \\ \psi_{\vec{\xi},-}(\omega_{N}) \\
\psi_{\vec{\xi},-}^{*}(\omega_{N})
\eea\right)_{q}^{R,b}}_{\Psi_{\vec{\xi},q}^{R,b}(\omega_{N})} + \frac{\wtilde{R}_{I}}{2}\sum_{q=\pm}\sum_{b=1,2} }
\\ \no && \eta_{p}\;
\left( \bea{cccc} \wtilde{B}_{\vec{\xi};++}(\omega_{N}) & \wtilde{C}_{\vec{\xi};++}(\omega_{N})
& \wtilde{B}_{\vec{\xi};+-}(\omega_{N}) & \wtilde{C}_{\vec{\xi};+-}(\omega_{N}) \\
\wtilde{C}_{\vec{\xi};++}^{*}(\omega_{N})& \wtilde{B}_{\vec{\xi};++}(\omega_{N}) &
\wtilde{C}_{\vec{\xi};+-}^{*}(\omega_{N})& \wtilde{B}_{\vec{\xi};+-}^{*}(\omega_{N})
\\ \wtilde{B}_{\vec{\xi};+-}^{*}(\omega_{N}) & \wtilde{C}_{\vec{\xi};+-}(\omega_{N})
& \wtilde{B}_{\vec{\xi};--}(\omega_{N}) & \wtilde{C}_{\vec{\xi};--}(\omega_{N}) \\
\wtilde{C}_{\vec{\xi};+-}^{*}(\omega_{N})& \wtilde{B}_{\vec{\xi};+-}(\omega_{N}) &
\wtilde{C}_{\vec{\xi};--}^{*}(\omega_{N})& \wtilde{B}_{\vec{\xi};--}(\omega_{N}) \eea\right)_{pq}^{ab} \eta_{q}
\underbrace{\left(\bea{c}
\psi_{\vec{\xi},+}(\omega_{N}) \\ \psi_{\vec{\xi},+}^{*}(\omega_{N}) \\ \psi_{\vec{\xi},-}(\omega_{N}) \\
\psi_{\vec{\xi},-}^{*}(\omega_{N})
\eea\right)_{q}^{R,b}}_{\Psi_{\vec{\xi},q}^{R,b}(\omega_{N})} \hspace*{-0.37cm}=  \\ \no &=&
 \omega_{N}\;\;\eta_{p}\;\;
\underbrace{\left(\bea{c}
\psi_{\vec{\xi},+}(\omega_{N}) \\ \psi_{\vec{\xi},+}^{*}(\omega_{N}) \\ \psi_{\vec{\xi},-}(\omega_{N}) \\
\psi_{\vec{\xi},-}^{*}(\omega_{N})
\eea\right)_{p}^{R,a}}_{\Psi_{\vec{\xi},p}^{R,a}(\omega_{N})}
\eeq
\beq \lb{97}
 \hat{\wtilde{\mbox{\sf h}}}(\vec{\xi}) &=&
-\pp_{\vec{\xi}}\cdot\pp_{\vec{\xi}}+\wtilde{u}(\vec{\xi})-\wtilde{\mu}_{0} \;\;;
\hspace*{0.55cm}\hat{\wtilde{\mbox{\sf h}}}_{p}^{T}(\vec{\xi})=\hat{\wtilde{\mbox{\sf h}}}_{p}(\vec{\xi})
\\ \lb{98}\delta_{\omega_{N},\omega_{N\ppr}\ppr} &=&
\int\dtot[\vec{\xi}]\sum_{p=\pm}\sum_{a=1,2}\;\;\Psi_{\vec{\xi},p}^{L,a}(\omega_{N\ppr}\ppr)\;\;\eta_{p}\;\;
\Psi_{\vec{\xi},p}^{R,a}(\omega_{N})\;\;\;.
\eeq
Note that the disorder-self-energy \(\wtilde{\Sigma}_{I;\vec{\xi};pq}^{ab}(\omega)\) in (\ref{96})
also depends on the frequency $\omega$ and therefore has to coincide with the resulting eigenvalue
$\omega_{N}$ on the right hand-side of (\ref{96}).  Since the generalized eigenvalue problem (\ref{96})
becomes non-hermitian due to the iterations for the continued fraction, we have to introduce left and
right eigenvectors \(\Psi_{\vec{\xi},p}^{L,a}(\omega_{N\ppr}\ppr)\),
\(\Psi_{\vec{\xi},p}^{R,a}(\omega_{N})\) for the ortho-normalization (\ref{98}).
Assuming completeness of the eigenfunctions \(\Psi_{\vec{\xi},p}^{(R/L),a}(\omega_{N})\) (\ref{99}),
one can construct the non-equilibrium Green function \(\wtilde{M}_{I;\vec{\xi},p;\vec{\xi}\ppr,q}^{-1;ab}(\omega)\)
(\ref{100}) from the eigenfunctions \(\Psi_{\vec{\xi},p}^{(R/L),a}(\omega_{N})\) and
eigenvalues \(\omega_{N}\) of (\ref{96})
\beq \lb{99}
\sum_{\{\omega_{N}\}}\eta_{p}\;\;\Psi_{\vec{\xi},p}^{R,a}(\omega_{N})\;\;\;\Psi_{\vec{\xi}\ppr,q}^{L,b}(\omega_{N})
&=&\delta_{\vec{\xi},\vec{\xi}\ppr}\;\;\delta_{p,q}\;\;\delta_{a,b}
\eeq
\beq  \lb{100}
\wtilde{M}_{I;\vec{\xi},p;\vec{\xi}\ppr,q}^{-1;ab}(\omega)&=&\sum_{\{\omega_{N}\}}
\frac{\Psi_{\vec{\xi},p}^{R,a}(\omega_{N})\;\otimes\;
\Psi_{\vec{\xi}\ppr,q}^{L,b}(\omega_{N})}{-\omega-\im\;\wtilde{\ve} +\omega_{N}}\hspace*{1.0cm}(\wtilde{\ve}>0)\;.
\eeq
The generalized eigenvalue problem (\ref{96}) has to be considered at every iteration step \(m\to m+1\)
of the continued fraction, but reduces to the solution
of the radial part of (\ref{96}) for a rotational symmetry. The non-diagonal parts of the disorder-self-energy
\(\wtilde{\Sigma}_{I;\vec{\xi};pq}^{ab}(\omega)\)
are reminiscent of the Bogoliubov-de Gennes equations for superconductivity where the nondiagonal, anomalous parts
are an important ingredient of BCS theory.
The nondiagonal 'Nambu' parts (\(a\neq b\), \(a,b=1,2\)) in (\ref{96}) correspond to such pair condensates as
\(\langle\psi_{\vec{\xi},p}(\omega)\;\;\psi_{\vec{\xi},p}(\omega)\rangle\) (in the bosonic case),
and the nondiagonal contour time parts '+$-$' and '$-$+' are related to quasiparticles created by
defects and disorder (for a classification of various types of disorder see \cite{Groe}).
Since both anomalous parts (pair condensates and anomalous disorder effects)
are taken into account in (\ref{96}), we have described the exact mean field theory with
the \(4\times 4\) disorder-self-energy \(\wtilde{\Sigma}_{I;\vec{\xi};pq}^{ab}(\omega)\).

\subsection{Hubbard-Stratonovich transformation for the disorder term in model II and derivation of
the mean field equations with the disorder-self-energy of a local time dependence}\lb{s33}

The various steps of the derivation for the saddle point equation (\ref{87},\ref{88}) in section \ref{s32}
can be conveyed to model II with dynamic disorder.
Apart from the common repulsive interaction with strength $V_{0}$ in both models,
the 'non-hermitian' ensemble averaged quartic interaction depends only on a single time variable,
but includes the averaging effect of \(V_{II}(\vec{x},t)\) in the two time contour indices (\(p,q=\pm\)) (\ref{36}).
The resulting 'Nambu'-doubled density matrix  \(\hat{R}_{\vec{x};pq}^{ab}(t)\) (\ref{101},\ref{102}),
following from the dyadic products of 'Nambu'-doubled Bose fields, therefore has only a single time variable, but
two contour indices (\(p,q=\pm\)) for forward and backward propagation apart
from the 'Nambu' indices (\(a,b=1,2\))
\beq \lb{101}
\lefteqn{\int_{-\infty}^{\infty}\dtot t\sum_{\vec{x}}\sum_{p,q=\pm}
\Big(\psi_{\vec{x}}^{*}(t_{p})\;\eta_{p}\;\psi_{\vec{x}}(t_{p})\Big)\;\;
\Big(\psi_{\vec{x}}^{*}(t_{q})\;\eta_{q}\;\psi_{\vec{x}}(t_{q})\Big)=} \\ \no &=&\frac{1}{4}
\int_{-\infty}^{\infty}\dtot t\sum_{\vec{x}}\sum_{p,q=\pm}
\bigg(\psi_{\vec{x}}^{*}(t_{p})\;\eta_{p}\;\psi_{\vec{x}}(t_{p})+
\psi_{\vec{x}}(t_{p})\;\eta_{p}\;\psi_{\vec{x}}^{*}(t_{p})\bigg) \times \\ \no &\times&
\bigg(\psi_{\vec{x}}^{*}(t_{q})\;\eta_{q}\;\psi_{\vec{x}}(t_{q})+
\psi_{\vec{x}}(t_{q})\;\eta_{q}\;\psi_{\vec{x}}^{*}(t_{q})\bigg)
\\ \no &=&\frac{1}{4}\int_{-\infty}^{\infty}\dtot t\sum_{\vec{x}}
\sum_{p,q=\pm}\bigg(\Psi_{\vec{x}}^{+a}(t_{p})\;\eta_{p}\;\Psi_{\vec{x}}^{a}(t_{p})\bigg)
\bigg(\Psi_{\vec{x}}^{+b}(t_{q})\;\eta_{q}\;\Psi_{\vec{x}}^{b}(t_{q})\bigg) \\ \no &=&
\frac{1}{4}\int_{-\infty}^{\infty}\dtot t \sum_{\vec{x}}\sum_{p,q=\pm}
\eta_{p}\;\underbrace{\Psi_{\vec{x}}^{a}(t_{p})\otimes
\Psi_{\vec{x}}^{+b}(t_{q})}_{\hat{R}_{\vec{x};pq}^{ab}(t)}\;\eta_{q}\;
\underbrace{\Psi_{\vec{x}}^{b}(t_{q})\otimes
\Psi_{\vec{x}}^{+a}(t_{p})}_{\hat{R}_{\vec{x};qp}^{ba}(t)}  \\ \no &=&
\frac{1}{4}\int_{-\infty}^{\infty}\dtot t\sum_{\vec{x}}\sum_{p,q=\pm}\trab\Big[\eta_{p}\;
\hat{R}_{\vec{x};pq}^{ab}(t)\;\eta_{q}\;
\hat{R}_{\vec{x};qp}^{ba}(t)\Big]
\eeq
\be \lb{102}
\hat{R}_{\vec{x};pq}^{ab}(t)=\left(
\bea{cccc}
\psi_{\vec{x},+}(t)\;\psi_{\vec{x},+}^{*}(t) &
\psi_{\vec{x},+}(t)\;\psi_{\vec{x},+}(t) &
\psi_{\vec{x},+}(t)\;\psi_{\vec{x},-}^{*}(t) &
\psi_{\vec{x},+}(t)\;\psi_{\vec{x},-}(t) \\
\psi_{\vec{x},+}^{*}(t)\;\psi_{\vec{x},+}^{*}(t) &
\psi_{\vec{x},+}^{*}(t)\;\psi_{\vec{x},+}(t) &
\psi_{\vec{x},+}^{*}(t)\;\psi_{\vec{x},-}^{*}(t) &
\psi_{\vec{x},+}^{*}(t)\;\psi_{\vec{x},-}(t) \\
\psi_{\vec{x},-}(t)\;\psi_{\vec{x},+}^{*}(t) &
\psi_{\vec{x},-}(t)\;\psi_{\vec{x},+}(t) &
\psi_{\vec{x},-}(t)\;\psi_{\vec{x},-}^{*}(t) &
\psi_{\vec{x},-}(t)\;\psi_{\vec{x},-}(t) \\
\psi_{\vec{x},-}^{*}(t)\;\psi_{\vec{x},+}^{*}(t) &
\psi_{\vec{x},-}^{*}(t)\;\psi_{\vec{x},+}(t) &
\psi_{\vec{x},-}^{*}(t)\;\psi_{\vec{x},-}^{*}(t) &
\psi_{\vec{x},-}^{*}(t)\;\psi_{\vec{x},-}(t)
\eea\right)_{\mbox{.}}
\ee
Consequently, the HST for the disorder term in model II can be performed as in relation (\ref{103})
where the doubled disorder-self-energy \(\hat{\Sigma}_{II;\vec{x};pq}^{ab}(t)\;\in sp(4)\) (\ref{104})
has to fulfill the equivalent symmetry relations (\ref{105},\ref{106}) as the density matrix
\(\hat{R}_{\vec{x};pq}^{ab}(t)\) (\ref{102})
\beq \lb{103}
\lefteqn{\exp\bigg\{-\frac{R_{II}^{2}}{2\hbar^{2}}\int_{-\infty}^{\infty}\dtot t\sum_{\vec{x}}\sum_{p,q=\pm}
\psi_{\vec{x}}^{*}(t_{p})\;\eta_{p}\;\psi_{\vec{x}}(t_{p})\;\;\psi_{\vec{x}}^{*}(t_{q})\;\eta_{q}\;\psi_{\vec{x}}(t_{q})
\bigg\}=} \\ \no &=&\exp\bigg\{-\frac{1}{8}\frac{R_{II}^{2}}{\hbar^{2}}\int_{-\infty}^{\infty}
\dtot t\sum_{\vec{x}}\sum_{p,q=\pm}
\trab\Big[\eta_{p}\;\hat{R}_{\vec{x};pq}^{ab}(t)\;\eta_{q}\;\hat{R}_{\vec{x};qp}^{ba}(t)\Big]\bigg\} = \\ \no &=&
\int\dtot[\hat{\Sigma}_{II;\vec{x};pq}^{ab}(t)]\;\;\exp\bigg\{-\frac{1}{8\;R_{II}^{2}}\int_{-\infty}^{\infty}
\dtot t\sum_{\vec{x}}\sum_{p,q=\pm}
\trab\Big[\eta_{p}\;\hat{\Sigma}_{II;\vec{x};pq}^{ab}(t)\;\eta_{q}\;
\hat{\Sigma}_{II;\vec{x};qp}^{ba}(t)\Big]\bigg\} \\ \no &\times&
\exp\bigg\{\frac{\im}{4\hbar}\int_{-\infty}^{\infty}\dtot t\sum_{\vec{x}}\sum_{p,q=\pm}\trab\Big[\eta_{p}\;
\hat{\Sigma}_{II;\vec{x};pq}^{ab}(t)\;\eta_{q}\;\hat{R}_{\vec{x};qp}^{ba}(t)\Big]\bigg\}
\eeq
\beq \lb{104}
\hat{\Sigma}_{II;\vec{x};pq}^{ab}(t)&=&\left(
\bea{cccc}
B_{\vec{x};++}(t) & C_{\vec{x};++}(t) & B_{\vec{x};+-}(t) & C_{\vec{x};+-}(t) \\
C_{\vec{x};++}^{*}(t)& B_{\vec{x};++}(t) & C_{\vec{x};+-}^{*}(t)& B_{\vec{x};+-}^{*}(t)
\\ B_{\vec{x};+-}^{*}(t) & C_{\vec{x};+-}(t) & B_{\vec{x};--}(t) & C_{\vec{x};--}(t) \\
C_{\vec{x};+-}^{*}(t)& B_{\vec{x};+-}(t) & C_{\vec{x};--}^{*}(t)&
B_{\vec{x};--}(t)
\eea\right)
\eeq
\be \lb{105}
\Big(B_{\vec{x};++}(t)\;,\;B_{\vec{x};--}(t)\in\;\mbox{\sf R}\Big) \hspace*{0.64cm}
\Big(B_{\vec{x};+-}(t)\;,\;B_{\vec{x};-+}(t)\in\;\mbox{\sf C}\Big);\;\;
\Big(B_{\vec{x};-+}(t)=B_{\vec{x};+-}^{*}(t)\Big)
\ee
\be \lb{106}
\Big(C_{\vec{x};++}(t)\;,\;C_{\vec{x};--}(t)\;,\;C_{\vec{x};+-}(t)\in\;\mbox{\sf C}\Big)\hspace*{0.64cm}
\Big(C_{\vec{x};-+}(t)=C_{\vec{x};+-}(t)\Big)\;\;\;.
\ee
Inserting the two HST's (\ref{103}) and (\ref{50}) into \(\ovv{Z_{II}[\mcal{J}]}\) (\ref{36}),
we obtain the ensemble averaged coherent state path integral \(\ovv{Z_{II}[\mcal{J}]}\) (\ref{107})
for dynamic disorder and also extend the 'Nambu'-doubling to the source terms and one-particle part
\(\hat{\mcal{H}}_{\vec{x}\ppr,\vec{x}}^{ba}(t_{q}\ppr,t_{p})\) (\ref{60}-\ref{63})
\beq \no
\lefteqn{\hspace*{-1.0cm}\ovv{Z_{II}[\mcal{J}]}=\int\dtot[\hat{\Sigma}_{II;\vec{x};pq}^{ab}(t)]\;
\dtot[\hat{\sigma}_{\vec{x}}^{ab}(t_{p})]\;\;
\exp\bigg\{\frac{\im}{4\hbar}\frac{1}{V_{0}}\int_{C}\dtot t_{p}\sum_{\vec{x}}\trab\Big[
\hat{\sigma}_{\vec{x}}^{ab}(t_{p})\;\hat{\sigma}_{\vec{x}}^{ba}(t_{p})\Big]\bigg\}  } \\ \lb{107} &\times&
\exp\bigg\{-\frac{1}{8}\frac{1}{R_{II}^{2}}\sum_{p,q=\pm}\int_{-\infty}^{\infty}\dtot t\sum_{\vec{x}}
\trab\Big[\eta_{p}\;\hat{\Sigma}_{II;\vec{x};pq}^{ab}(t)\;\eta_{q}\;
\hat{\Sigma}_{II;\vec{x};qp}^{ba}(t)\Big]\bigg\}   \\ \no &\times&
\int\dtot[\psi_{\vec{x}}(t_{p})]\;
\exp\bigg\{-\frac{\im}{2\hbar}\int_{C}\dtot t_{p}\;\dtot t_{q}\ppr
\sum_{\vec{x},\vec{x}\ppr} \Psi_{\vec{x}\ppr}^{+b}(t_{q}\ppr)\;\mcal{N}_{x}
\bigg[\hat{\mcal{H}}_{\vec{x}\ppr,\vec{x}}^{ba}(t_{q}\ppr,t_{p})+
\frac{\mcal{J}_{\vec{x}\ppr,\vec{x}}^{ba}(t_{q}\ppr,t_{p})}{\mcal{N}_{x}}+  \\ \no &+&
\delta(t-t\ppr)\;\delta_{\vec{x},\vec{x}\ppr}\Big(\delta_{p,q}\;\eta_{p}\;\big(
\hat{J}_{\psi\psi;\vec{x}}^{ba}(t_{p})-\hat{\sigma}_{\vec{x}}^{ba}(t_{p})\big)+\frac{1}{2}\;
\hat{\Sigma}_{II;\vec{x};qp}^{ba}(t)\Big)\bigg]\;
\Psi_{\vec{x}}^{a}(t_{p})\bigg\} \\ \no &\times&
\exp\bigg\{-\frac{\im}{2\hbar}\int_{C}\dtot t_{p}\sum_{\vec{x}}\Big[
J_{\psi;\vec{x}}^{+a}(t_{p})\;\Psi_{\vec{x}}^{a}(t_{p})+\Psi_{\vec{x}}^{+a}(t_{p})\;
J_{\psi;\vec{x}}^{a}(t_{p})\Big]\bigg\}_{\mbox{.}}
\eeq
In analogy to (\ref{59}-\ref{71}) in model I, the disorder-self-energy \(\hat{\Sigma}_{II;\vec{x};pq}^{ab}(t)\)
with local time dependence can be shifted by the self-energy \(\hat{\sigma}_{\vec{x}}^{ab}(t_{p})\)
of the repulsive interaction and by the source term \(\hat{J}_{\psi\psi;\vec{x}}^{ab}(t_{p})\)
for the pair condensates
\beq \lb{108}
\hat{\Sigma}_{II;\vec{x};pq}^{ab}(t) &\to& \hat{\Sigma}_{II;\vec{x};pq}^{ab}(t)+2\;
\delta_{p,q}\;\eta_{p}\;\hat{\sigma}_{\vec{x}}^{ab}(t_{p}) \\  \lb{109}
\hat{\Sigma}_{II;\vec{x};pq}^{ab}(t) &\to& \hat{\Sigma}_{II;\vec{x};pq}^{ab}(t)-2\;
\delta_{p,q}\;\eta_{p}\;\hat{J}_{\psi\psi;\vec{x}}^{ab}(t_{p})\;\;\;,
\eeq
so that the matrix \(\hat{M}_{II;\vec{x}\ppr,\vec{x}}^{ba}(t_{q}\ppr,t_{p})\) (\ref{111})
coupled to the bilinear Bose fields \(\Psi_{\vec{x}\ppr}^{+b}(t_{q}\ppr)\ldots\Psi_{\vec{x}}^{a}(t_{p})\) in
\(\ovv{Z_{II}[\mcal{J}]}\) (\ref{110})
only contains the disorder-self-energy \(\hat{\Sigma}_{II;\vec{x};pq}^{ab}(t)\) (\ref{104})
\beq \lb{110}
\lefteqn{\ovv{Z_{II}[\mcal{J}]}=\int\dtot[\hat{\Sigma}_{II;\vec{x};pq}^{ab}(t)]\;\;
\exp\bigg\{-\frac{1}{8\;R_{II}^{2}}\sum_{p,q=\pm}\int_{-\infty}^{\infty}\hspace*{-0.28cm}\dtot t \sum_{\vec{x}}
\trab\Big[\eta_{p}\;\hat{\Sigma}_{II;\vec{x};pq}^{ab}(t)\;\eta_{q}\;
\hat{\Sigma}_{II;\vec{x};qp}^{ba}(t)\Big]\bigg\} } \\ \no &\times&
\hspace*{-0.28cm}\exp\bigg\{\frac{1}{2\;R_{II}^{2}}\sum_{p=\pm}\int_{-\infty}^{+\infty}\hspace*{-0.28cm}
\dtot t\sum_{\vec{x}}\bigg(\eta_{p}\;
\trab\Big[\hat{\Sigma}_{II;\vec{x};pp}^{ab}(t)\;\;\hat{J}_{\psi\psi;\vec{x}}^{ba}(t_{p})\Big]-
\trab\Big[\hat{J}_{\psi\psi;\vec{x}}^{ab}(t_{p})\;\;\hat{J}_{\psi\psi;\vec{x}}^{ba}(t_{p})\Big]
\bigg)\bigg\}  \\ \no &\times&\hspace*{-0.28cm}
\exp\bigg\{-\frac{1}{2}\int_{C}\hspace*{-0.1cm}
\frac{\dtot t_{p}}{\hbar}\eta_{p}\sum_{\vec{x}}\hbar\Omega\mcal{N}_{x}
\trab\ln\bigg[\eta_{q}\bigg(\hat{\mcal{H}}_{\vec{x}\ppr,\vec{x}}^{ba}(t_{q}\ppr,t_{p})+
\frac{\mcal{J}_{\vec{x}\ppr,\vec{x}}^{ba}(t_{q}\ppr,t_{p})}{\mcal{N}_{x}}+
\frac{1}{2}\delta(t-t\ppr)\;\delta_{\vec{x},\vec{x}\ppr}\;
\hat{\Sigma}_{II;\vec{x};qp}^{ba}(t)\bigg)\eta_{p}\bigg]\bigg\} \\ \no &\times&\hspace*{-0.28cm}
\exp\bigg\{\frac{\im}{2}\frac{\Omega^{2}}{\hbar}\int_{C}\dtot t_{p}\;\dtot t_{q}\ppr
\sum_{\vec{x},\vec{x}\ppr}\mcal{N}_{x} \sum_{a,b=1,2}
J_{\psi;\vec{x}\ppr}^{+b}(t_{q}\ppr)\\ \no &&
\bigg[\eta_{q}\bigg(\hat{\mcal{H}}_{\vec{x}\ppr,\vec{x}}^{ba}(t_{q}\ppr,t_{p})+
\frac{\mcal{J}_{\vec{x}\ppr,\vec{x}}^{ba}(t_{q}\ppr,t_{p})}{\mcal{N}_{x}}+
\frac{1}{2}\delta(t-t\ppr)\;\delta_{\vec{x},\vec{x}\ppr}\;\hat{\Sigma}_{II;\vec{x};qp}^{ba}(t)
\bigg)\eta_{p}\bigg]^{-1;ba}_{\vec{x}\ppr,\vec{x}}\!\!\!\!\!\!(t_{q}\ppr,t_{p})\;\;
J_{\psi;\vec{x}}^{a}(t_{p})\bigg\}
\\ \no &\times&\hspace*{-0.28cm}
\int\dtot[\hat{\sigma}_{\vec{x}}^{ab}(t_{p})]\;
\exp\bigg\{-\frac{1}{2}\sum_{p=\pm}\int_{-\infty}^{+\infty}\dtot t\sum_{\vec{x}}
\bigg(\frac{1}{R_{II}^{2}}-\frac{\im}{2}\eta_{p}\;\frac{1}{\hbar V_{0}}\bigg)
\trab\Big[\hat{\sigma}_{\vec{x}}^{ab}(t_{p})\;
\hat{\sigma}_{\vec{x}}^{ba}(t_{p})\Big]\bigg\} \\ \no &\times&\hspace*{-0.28cm}
\exp\bigg\{-\frac{1}{2\;R_{II}^{2}}\sum_{p=\pm}\int_{-\infty}^{+\infty}\dtot t\sum_{\vec{x}}
\trab\Big[\hat{\sigma}_{\vec{x}}^{ab}(t_{p})\;\Big(\eta_{p}\;\hat{\Sigma}_{II;\vec{x};pp}^{ba}(t)
-2\;\hat{J}_{\psi\psi;\vec{x}}^{ba}(t_{p})\Big)\Big]\bigg\}
\eeq
\beq \lb{111}
\lefteqn{\hspace*{-0.37cm}\hat{M}_{II;\vec{x}\ppr,\vec{x}}^{ba}(t_{q}\ppr,t_{p}) = } \\ \no &=&
\eta_{q}\bigg(\hat{\mcal{H}}_{\vec{x}\ppr,\vec{x}}^{ba}(t_{q}\ppr,t_{p})+
\frac{\mcal{J}_{\vec{x}\ppr,\vec{x}}^{ba}(t_{q}\ppr,t_{p})}{\mcal{N}_{x}}+
\frac{1}{2}\;\delta(t-t\ppr)\;\delta_{\vec{x},\vec{x}\ppr}\;
\hat{\Sigma}_{II;\vec{x};qp}^{ba}(t)\bigg)\eta_{p} =\eta_{q}
\frac{\mcal{J}_{\vec{x}\ppr,\vec{x}}^{ba}(t_{q}\ppr,t_{p})}{\mcal{N}_{x}}\eta_{p}+
\\ \no &+&\hspace*{-0.28cm}
\delta(t-t\ppr)\;\delta_{\vec{x},\vec{x}\ppr}\left[\delta_{p,q}\;\eta_{p}\;\left(
\bea{cccc}
\hat{h}_{+}(t_{+}) & 0 & 0 & 0 \\
0 & \hat{h}_{+}^{T}(t_{+}) & 0 & 0 \\
0 & 0 & \hat{h}_{-}(t_{-}) & 0 \\
0 & 0 & 0 & \hat{h}_{-}^{T}(t_{-}) \eea\right)+
\frac{1}{2}\; \eta_{q}\;\hat{\Sigma}_{II;\vec{x};qp}^{ba}(t)\;\eta_{p}\right]_{\mbox{.}}
\eeq
The remaining Gaussian factors with the quadratic self-energy \(\hat{\sigma}_{\vec{x}}^{ab}(t_{p})\) of the repulsive
interaction and its coupling to \(\hat{\Sigma}_{II;\vec{x};pq}^{ab}(t)\), \(\hat{J}_{\psi\psi;\vec{x}}^{ab}(t_{p})\) can
be integrated out completely as in section \ref{s32}
\beq \lb{112}
\lefteqn{\int\dtot[\hat{\sigma}_{\vec{x}}^{ab}(t_{p})]\;\; \exp\bigg\{-\frac{1}{2}\sum_{p=\pm}
\int_{-\infty}^{+\infty}\dtot t\sum_{\vec{x}}
\bigg(\frac{1}{R_{II}^{2}}-\frac{\im}{2}\eta_{p}\;\frac{1}{\hbar V_{0}}\bigg)
\trab\Big[\hat{\sigma}_{\vec{x}}^{ab}(t_{p})\; \hat{\sigma}_{\vec{x}}^{ba}(t_{p})\Big]\bigg\} } \\ \no &\times&
\exp\bigg\{-\frac{1}{2\;R_{II}^{2}}\sum_{p=\pm}\int_{-\infty}^{+\infty}\dtot t\sum_{\vec{x}}
\trab\Big[\hat{\sigma}_{\vec{x}}^{ab}(t_{p})\;\Big(\eta_{p}\;\hat{\Sigma}_{II;\vec{x};pp}^{ba}(t)
-2\;\hat{J}_{\psi\psi;\vec{x}}^{ba}(t_{p})\Big)\Big]\bigg\} = \\ \no &=&
\exp\bigg\{\frac{1}{8}\frac{1}{R_{II}^{2}}\sum_{p=\pm}\int_{-\infty}^{+\infty}\dtot t\sum_{\vec{x}}
\frac{1}{\Big(1-\im\eta_{p}\frac{R_{II}^{2}}{2\hbar\;V_{0}}\Big)}\;
\trab\Big[\hat{\Sigma}_{II;\vec{x};pp}^{ab}(t)\;\;\hat{\Sigma}_{II;\vec{x};pp}^{ba}(t)\Big]\bigg\}  \\ \no &\times&
\exp\bigg\{-\frac{1}{2}\frac{1}{R_{II}^{2}}\sum_{p=\pm}\int_{-\infty}^{+\infty}\dtot t\sum_{\vec{x}}
\eta_{p}\;\frac{1}{\Big(1-\im\eta_{p}\frac{R_{II}^{2}}{2\hbar\;V_{0}}\Big)}\;
\trab\Big[\hat{\Sigma}_{II;\vec{x};pp}^{ab}(t)\;\;\hat{J}_{\psi\psi;\vec{x}}^{ba}(t_{p})\Big]\bigg\}  \\ \no &\times&
\exp\bigg\{\frac{1}{2}\frac{1}{R_{II}^{2}}\sum_{p=\pm}\int_{-\infty}^{+\infty}\dtot t\sum_{\vec{x}}
\frac{1}{\Big(1-\im\eta_{p}\frac{R_{II}^{2}}{2\hbar\;V_{0}}\Big)}\;
\trab\Big[\hat{J}_{\psi\psi;\vec{x}}^{ab}(t_{p})\;\;\hat{J}_{\psi\psi;\vec{x}}^{ba}(t_{p})\Big]\bigg\}\;\;\;.
\eeq
Substituting (\ref{112}) into (\ref{110}), we finally achieve the ensemble averaged generating function
\(\ovv{Z_{II}[\mcal{J}]}\) (\ref{113}) for zero temperature with the disorder-self-energy
\(\hat{\Sigma}_{II;\vec{x};pq}^{ab}(t)\) (\ref{104}) as the only field variable and the matrix
\(\hat{M}_{II;\vec{x}\ppr,\vec{x}}^{ba}(t_{q}\ppr,t_{p})\) (\ref{111})
\beq \lb{113}
\lefteqn{\ovv{Z_{II}[\mcal{J}]}=\int\dtot[\hat{\Sigma}_{II;\vec{x};pq}^{ab}(t)] \;\;\;\times} \\ \no
&\times& \exp\bigg\{-\frac{1}{8}\frac{1}{R_{II}^{2}}\sum_{p,q=\pm}\int_{-\infty}^{+\infty}\dtot t
\sum_{\vec{x}}\Big(1-\delta_{p,q}\;\mu_{p}^{(II)}\Big)
\trab\Big[\eta_{p}\;\hat{\Sigma}_{II;\vec{x};pq}^{ab}(t)\;\eta_{q}\;
\hat{\Sigma}_{II;\vec{x};qp}^{ba}(t)\Big]\bigg\}
\\ \no &\times& \exp\bigg\{\frac{1}{2}\frac{1}{R_{II}^{2}}\sum_{p=\pm}
\int_{-\infty}^{+\infty}\dtot t\sum_{\vec{x}}\eta_{p}\;
\Big(1-\mu_{p}^{(II)}\Big)\trab\Big[\hat{\Sigma}_{II;\vec{x};pp}^{ab}(t)\;\;
\hat{J}_{\psi\psi;\vec{x}}^{ba}(t_{p})\Big]\bigg\}
\\ \no &\times& \exp\bigg\{-\frac{1}{2}\frac{1}{R_{II}^{2}}\sum_{p=\pm}\int_{-\infty}^{+\infty}\dtot t\sum_{\vec{x}}
\Big(1-\mu_{p}^{(II)}\Big)\trab\Big[\hat{J}_{\psi\psi;\vec{x}}^{ab}(t_{p})\;\;
\hat{J}_{\psi\psi;\vec{x}}^{ba}(t_{p})\Big]\bigg\} \\ \no &\times& \exp\bigg\{-\frac{1}{2}\int_{C}\frac{\dtot
t_{p}}{\hbar}\eta_{p}\sum_{\vec{x}}\hbar\Omega\mcal{N}_{x}
\trab\ln\Big[\hat{M}_{II;\vec{x}\ppr,\vec{x}}^{ba}(t_{q}\ppr,t_{p})\Big]\bigg\} \\ \no &\times&
\exp\bigg\{\frac{\im}{2}\frac{\Omega^{2}}{\hbar}\int_{C}\dtot t_{p}\;\dtot t_{q}\ppr\sum_{\vec{x},\vec{x}\ppr}
\mcal{N}_{x}\sum_{a,b=1,2}J_{\psi;\vec{x}\ppr}^{+b}(t_{q}\ppr)\;\;
\hat{M}_{II;\vec{x}\ppr,\vec{x}}^{-1;ba}(t_{q}\ppr,t_{p})\;\;J_{\psi;\vec{x}}^{a}(t_{p})\bigg\} \eeq \beq \lb{114}
\mu_{p}^{(II)}&=&\frac{1}{\Big(1-\frac{\im}{2}\eta_{p}\;\frac{R_{II}^{2}}{\hbar V_{0}}\Big)}=
\frac{1+\frac{\im}{2}\eta_{p}\; \Big(\frac{R_{II}^{2}}{\hbar V_{0}}\Big)}{1+\frac{1}{4}\Big(\frac{R_{II}^{2}}{\hbar
V_{0}}\Big)^{2}} \hspace*{0.37cm}\Big(0<\Re(\mu_{p}^{(II)})<1\Big)\;\;\;.
\eeq
The scaling of the disorder-self-energy
and the other energy parameters to dimensionless quantities is listed in relations (\ref{115}-\ref{124})
\beq \lb{115}
\hat{\mcal{H}}_{\vec{x}\ppr,\vec{x}}^{ba}(t_{q}\ppr,t_{p}) &\to&
\wtilde{\mcal{H}}_{\vec{x}_{j}\ppr,\vec{x}_{i}}^{ba}(t_{q,l}\ppr,t_{p,k})=
\frac{\hat{\mcal{H}}_{\vec{x}_{j}\ppr,\vec{x}_{i}}^{ba}(t_{q,l}\ppr,t_{p,k})}{\hbar\Omega^{2}\mcal{N}_{x}} \\ \lb{116}
\delta(t-t\ppr)\;\;\hat{\Sigma}_{II;\vec{x};qp}^{ba}(t) &\to&\delta(t_{l}\ppr,t_{k})\;\;
\wtilde{\Sigma}_{II;\vec{x}_{i};qp}^{ba}(t_{k})=
\frac{\hat{\Sigma}_{II;\vec{x}_{i};qp}^{ba}(t_{k})}{\hbar\Omega^{2}\mcal{N}_{x}}\;\;\delta(t_{k}-t_{l}\ppr) \\ \lb{117}
R_{II}^{2} &\to& \wtilde{R}_{II}^{2}=\frac{R_{II}^{2}}{\hbar^{2}\Omega\mcal{N}_{x}} \\ \lb{118}
V_{0}&\to&\wtilde{V}_{0}=\frac{V_{0}}{\hbar\Omega\mcal{N}_{x}} \\  \lb{119} \xi_{II}&=&\frac{R_{II}^{2}}{\hbar
V_{0}}=\frac{\wtilde{R}_{II}^{2}}{\wtilde{V}_{0}} \\  \lb{120}
\mu_{p}^{(II)}&=&\frac{1}{1-\frac{\im}{2}\;\eta_{p}\;\xi_{II}}  \\  \lb{121} J_{\psi;\vec{x}}^{a}(t_{p})
&\to&\wtilde{J}_{\psi;\vec{x}_{i}}^{a}(t_{p,k})= \frac{J_{\psi;\vec{x}_{i}}^{a}(t_{p,k})}{\hbar\Omega\mcal{N}_{x}} \\
\lb{122} \hat{J}_{\psi\psi;\vec{x}\ppr,\vec{x}}^{ba}(t_{q}\ppr,t_{p})&\to&
\wtilde{J}_{\psi\psi;\vec{x}_{j}\ppr,\vec{x}_{i}}^{ba}(t_{q,l}\ppr,t_{p,k})=
\frac{\hat{J}_{\psi\psi;\vec{x}_{j}\ppr,\vec{x}_{i}}^{ba}(t_{q,l}\ppr,t_{p,k})}{\hbar\Omega^{2}\mcal{N}_{x}} =
\\ \no &=&\delta_{p,q}\;\eta_{p}\;\delta(t_{p,k},t_{q,l}\ppr)\;
\delta_{\vec{x}_{i},\vec{x}_{j}\ppr}\;\;
\wtilde{J}_{\psi\psi;\vec{x}_{i}}^{ba}(t_{p,k})    \\   \lb{123}
\hat{J}_{\psi\psi;\vec{x}}^{ba}(t_{p}) &\to&
\wtilde{J}_{\psi\psi;\vec{x}_{i}}^{ba}(t_{p,k})=
\frac{\hat{J}_{\psi\psi;\vec{x}_{i}}^{ba}(t_{p,k})}{\hbar\Omega\mcal{N}_{x}} \\  \lb{124}
\frac{\mcal{J}_{\vec{x}\ppr,\vec{x}}^{ba}(t_{q}\ppr,t_{p})}{\mcal{N}_{x}}&\to&
\wtilde{\mcal{J}}_{\vec{x}_{j}\ppr,\vec{x}_{i}}^{ba}(t_{q,l}\ppr,t_{p,k})=
\frac{\mcal{J}_{\vec{x}_{j},\vec{x}_{i}}^{ba}(t_{q,l}\ppr,t_{p,k})}{\hbar\Omega^{2}\mcal{N}_{x}^{2}}_{\mbox{.}}
\eeq
The coherent state path integral \(\ovv{Z_{II}[\mcal{J}]}\) (\ref{113})
is transformed to relation (\ref{125})
with discrete spatial and time-like variables of the dimensionless parameters and fields defined in (\ref{115}-\ref{124})
\beq \no
\lefteqn{\hspace*{-1.37cm}\ovv{Z_{II}[\mcal{J}]}=\int\hspace*{-0.1cm}
\dtot[\wtilde{\Sigma}_{II;\vec{x}_{i};pq}^{ab}(t_{k})]\;\;
\exp\bigg\{-\frac{1}{2}\frac{1}{\wtilde{R}_{II}^{2}}\sum_{p=\pm}\sum_{t_{k}}\sum_{\vec{x}_{i}}
\Big(1-\mu_{p}^{(II)}\Big)\hspace*{-0.19cm}\;\trab\Big[\wtilde{J}_{\psi\psi;\vec{x}_{i},p}^{ab}(t_{k})\;\;
\wtilde{J}_{\psi\psi;\vec{x}_{i},p}^{ba}(t_{k})\Big]\bigg\}  } \\ \no &\times&
\exp\bigg\{-\frac{1}{8}\frac{1}{\wtilde{R}_{II}^{2}}\sum_{p,q=\pm}\sum_{t_{k}}\sum_{\vec{x}_{i}}
\Big(1-\delta_{p,q}\;\mu_{p}^{(II)}\Big)\;\trab
\Big[\eta_{p}\;\wtilde{\Sigma}_{II;\vec{x}_{i};pq}^{ab}(t_{k})\;\eta_{q}\;
\wtilde{\Sigma}_{II;\vec{x}_{i};qp}^{ba}(t_{k})\Big]\bigg\}
\\ \lb{125} &\times& \exp\bigg\{\frac{1}{2}\frac{1}{\wtilde{R}_{II}^{2}}\sum_{p=\pm}\sum_{t_{k}}\sum_{\vec{x}_{i}}
\eta_{p}\;\Big(1-\mu_{p}^{(II)}\Big)\trab\Big[\wtilde{\Sigma}_{II;\vec{x}_{i};pp}^{ab}(t_{k})\;\;
\wtilde{J}_{\psi\psi;\vec{x}_{i},p}^{ba}(t_{k})\Big]\bigg\}
 \\ \no &\times &
\exp\bigg\{-\frac{1}{2}\sum_{p=\pm}\sum_{t_{k}}\frac{\Delta t_{k}}{\hbar}\sum_{\vec{x}_{i}}\hbar\Omega
\trab\ln\Big[\wtilde{M}_{II;\vec{x}_{j}\ppr,\vec{x}_{i}}^{ba}(t_{q,l}\ppr,t_{p,k})\Big]\bigg\} \\ \no &\times&
\exp\bigg\{\frac{\im}{2}\sum_{p,q=\pm}\sum_{t_{k},t_{l}\ppr}\sum_{\vec{x}_{i},\vec{x}_{j}\ppr}
\wtilde{J}_{\psi;\vec{x}_{j}\ppr,q}^{+b}(t_{l}\ppr)\;\eta_{q}\;
\wtilde{M}_{II;\vec{x}_{j}\ppr,\vec{x}_{i}}^{-1;ba}(t_{q,l}\ppr,t_{p,k})\;\eta_{p}\;
\wtilde{J}_{\psi;\vec{x}_{i},p}^{a}(t_{k}) \bigg\} \eeq \be \lb{126}\hspace*{-0.46cm}
\wtilde{M}_{II;\vec{x}_{j}\ppr,\vec{x}_{i}}^{ba}(t_{q,l}\ppr,t_{p,k})=\eta_{q}\bigg(
\wtilde{\mcal{H}}_{\vec{x}_{j}\ppr,\vec{x}_{i}}^{ba}(t_{q,l}\ppr,t_{p,k})+
\wtilde{\mcal{J}}_{\vec{x}_{j}\ppr,\vec{x}_{i}}^{ba}(t_{q,l}\ppr,t_{p,k})+
\frac{1}{2}\;\delta(t_{k},t_{l}\ppr)\;\;\delta_{\vec{x}_{j}\ppr,\vec{x}_{i}}\;\;
\wtilde{\Sigma}_{II;\vec{x}_{i};qp}^{ba}(t_{k})\bigg)\eta_{p}\;. \ee A functional Taylor expansion can be performed on
the actions in (\ref{125},\ref{126}) with respect to \(\delta\wtilde{\Sigma}_{II;\vec{x}_{i};pq}^{ab}(t_{k})\) as in
section \ref{s32}. We restrict to the vanishing of the first order variation in (\ref{125}) and so derive a mean field
equation (\ref{127},\ref{128}) for dynamic disorder with a dependence on the disorder-self-energy as the only remaining
field variable \beq \lb{127}
\lefteqn{\hspace*{-1.0cm}\frac{1}{\wtilde{R}_{II}^{2}}\;\Big(1-\delta_{p,q}\;\mu_{p}^{(II)}\Big)\;\;
\wtilde{\Sigma}_{II;\vec{x}_{i};qp}^{ba}(t_{k})=
\frac{2}{\wtilde{R}_{II}^{2}}\;\delta_{p,q}\;\eta_{p}\;\Big(1-\mu_{p}^{(II)}\Big)\;\;
\wtilde{J}_{\psi\psi;\vec{x}_{i},p}^{ba}(t_{k}) + } \\ \no
&-&\wtilde{M}_{II;\vec{x}_{i},\vec{x}_{i}}^{-1;ba}(t_{q,k},t_{p,k})  - \im
\sum_{p\ppr,q\ppr=\pm}\;\;\sum_{\tau_{k\ppr},\tau_{l\ppr}\ppr}\;\;\sum_{\vec{y}_{i\ppr},\vec{y}_{j\ppr}\ppr}
\sum_{c,d=1,2} \\ \no && \wtilde{J}_{\psi;\vec{y}_{j\ppr}\ppr}^{+d}(\tau_{q\ppr,l\ppr}\ppr)\;\;\eta_{q\ppr}\;
\wtilde{M}_{II;\vec{y}_{j\ppr}\ppr,\vec{x}_{i}}^{-1;da}(\tau_{q\ppr,l\ppr}\ppr,t_{p,k}) \hspace*{1.0cm}
\wtilde{M}_{II;\vec{x}_{i},\vec{y}_{i\ppr}}^{-1;bc}(t_{q,k},\tau_{p\ppr,k\ppr})\;\eta_{p\ppr}\;\;
\wtilde{J}_{\psi;\vec{y}_{i\ppr}}^{c}(\tau_{p\ppr,k\ppr}) \eeq \be  \lb{128}
\wtilde{J}_{\psi;\vec{x}_{i}}^{a}(t_{p=+,k})=\wtilde{J}_{\psi;\vec{x}_{i}}(t_{p=-,k})\hspace*{1.0cm}
\wtilde{J}_{\psi\psi;\vec{x}_{i}}^{ab}(t_{p=+,k})=\wtilde{J}_{\psi\psi;\vec{x}_{i}}^{ab}(t_{p=-,k})\;\;\;. \ee We also
have to consider the originally introduced imaginary increment \(-\im\;\ve_{p}\) (\ref{26}) for time reversal symmetry
breaking so that these convergence properties are consequently transferred to the disorder-self-energy in the continued
fractions. The iteration \(m\to m+1\) of \(\wtilde{\Sigma}_{II;\vec{x}_{i};pq}^{ab}(m;t_{k})\) in the continued
fractions follows in analogy to section \ref{s32} and equations (\ref{89}) to (\ref{91}) starting from the free Green
function with \(\wtilde{\Sigma}_{II;\vec{x}_{i};pq}^{ab}(m=0;t_{k})\equiv 0\) \beq \lb{129}
\lefteqn{\hspace*{-1.0cm}\frac{1}{\wtilde{R}_{II}^{2}}\;\Big(1-\delta_{p,q}\;\mu_{p}^{(II)}\Big)\;\;
\wtilde{\Sigma}_{II;\vec{x}_{i};qp}^{ba}(m+1;t_{k})=
\frac{2}{\wtilde{R}_{II}^{2}}\;\delta_{p,q}\;\eta_{p}\;\Big(1-\mu_{p}^{(II)}\Big)\;\;
\wtilde{J}_{\psi\psi;\vec{x}_{i},p}^{ba}(t_{k}) + } \\ \no
&-&\wtilde{M}_{II;\vec{x}_{i},\vec{x}_{i}}^{-1;ba}(m;t_{q,k},t_{p,k}) - \im
\sum_{p\ppr,q\ppr=\pm}\sum_{\tau_{k\ppr},\tau_{l\ppr}\ppr}\sum_{\vec{y}_{i\ppr},\vec{y}_{j\ppr}\ppr} \sum_{c,d=1,2}  \\
\no && \wtilde{J}_{\psi;\vec{y}_{j\ppr}\ppr}^{+d}(\tau_{q\ppr,l\ppr}\ppr)\;\;\eta_{q\ppr}\;
\wtilde{M}_{II;\vec{y}_{j\ppr}\ppr,\vec{x}_{i}}^{-1;da}(m;\tau_{q\ppr,l\ppr}\ppr,t_{p,k}) \hspace*{0.64cm}
\wtilde{M}_{II;\vec{x}_{i},\vec{y}_{i\ppr}}^{-1;bc}(m;t_{q,k},\tau_{p\ppr,k\ppr})\;\eta_{p\ppr}\;
\wtilde{J}_{\psi;\vec{y}_{i\ppr}}^{c}(\tau_{p\ppr,k\ppr}) \eeq \beq \lb{130}
\lefteqn{\wtilde{M}_{II;\vec{x}_{i},\vec{x}_{j}\ppr}^{ab}(m;t_{p,k},t_{q,l}\ppr)=} \\ \no &=&\eta_{p}\bigg(
\wtilde{\mcal{H}}_{\vec{x}_{i},\vec{x}_{j}\ppr}^{ab}(t_{p,k},t_{q,l}\ppr)+
\wtilde{\mcal{J}}_{\vec{x}_{i},\vec{x}_{j}\ppr}^{ab}(t_{p,k},t_{q,l}\ppr)+
\frac{1}{2}\;\delta(t_{k},t_{l}\ppr)\;\;\delta_{\vec{x}_{i},\vec{x}_{j}\ppr}\;\;
\wtilde{\Sigma}_{II;\vec{x}_{i};pq}^{ab}(m;t_{k})\bigg)\eta_{q} \eeq \beq  \lb{131}
\wtilde{J}_{\psi;\vec{x}_{i}}^{a}(t_{p=+,k})&=&\wtilde{J}_{\psi;\vec{x}_{i}}(t_{p=-,k})\hspace*{1.0cm}
\wtilde{J}_{\psi\psi;\vec{x}_{i}}^{ab}(t_{p=+,k})=\wtilde{J}_{\psi\psi;\vec{x}_{i}}^{ab}(t_{p=-,k})\;\;\;. \eeq The
nondiagonal parts \(\wtilde{\Sigma}_{II;\vec{x}_{i};p=\pm,q=\mp}^{ab}(m+1;t_{k})\) of the contour time also occur at
every iteration step due to the source field \(\wtilde{J}_{\psi;\vec{y}_{i\ppr}}^{c}(\tau_{p\ppr,k\ppr})\) as in the
disorder model I of section \ref{s32}, and the source matrix \(\wtilde{J}_{\psi\psi;\vec{x}_{i},p}^{a\neq b}(t_{k})\)
creates the anomalous terms. The solution \(\wtilde{\Sigma}_{II;\vec{x}_{i};pq}^{ab}(t_{k})\) of (\ref{129}-\ref{131})
reduces to a time independent function \(\wtilde{\Sigma}_{II;\vec{x}_{i};pq}^{ab}\) in the case of a time independent
trap potential \(\wtilde{u}(\vec{\xi})\) where \(\vec{\xi}\) is the dimensionless spatial vector. In order to obtain
the corresponding Green function \(\wtilde{M}_{II;\vec{x}_{i},\vec{x}_{j}\ppr}^{-1;ab}(m;t_{p,k},t_{q,l}\ppr)\)
(\ref{130},\ref{129}), one has to solve the eigenvalue problem (\ref{136}-\ref{137}) as in section \ref{s32} which
results from the following relations (\ref{132}-\ref{135}) for the matrix
\(\wtilde{M}_{II;\vec{\xi},p;\vec{\xi}\ppr,q}^{-1;ab}(\omega)\) after Fourier transformation to dimensionless frequency
$\omega$ \beq \lb{132} \wtilde{M}_{II;\vec{\xi},p;\vec{\xi}\ppr,q}^{ab}(\omega)&=&\eta_{p}\bigg(
\wtilde{\mcal{H}}_{\vec{\xi},p;\vec{\xi}\ppr,q}^{ab}(\omega) +\frac{1}{2}\;\;
\wtilde{\Sigma}_{II;\vec{\xi};pq}^{ab}\;\;\delta_{\vec{\xi},\vec{\xi}\ppr}\bigg)\eta_{q} \\ \lb{133}
\wtilde{\Sigma}_{II;\vec{\xi};pq}^{ab}&=&\left( \bea{cccc} \wtilde{B}_{\vec{\xi};++} & \wtilde{C}_{\vec{\xi};++} &
\wtilde{B}_{\vec{\xi};+-} &
\wtilde{C}_{\vec{\xi};+-} \\
\wtilde{C}_{\vec{\xi};++}^{*} & \wtilde{B}_{\vec{\xi};++} & \wtilde{C}_{\vec{\xi};+-}^{*} &
\wtilde{B}_{\vec{\xi};+-}^{*}
\\ \wtilde{B}_{\vec{\xi};+-}^{*} & \wtilde{C}_{\vec{\xi};+-} & \wtilde{B}_{\vec{\xi};--} &
\wtilde{C}_{\vec{\xi};--} \\ \wtilde{C}_{\vec{\xi};+-}^{*} & \wtilde{B}_{\vec{\xi};+-} &
\wtilde{C}_{\vec{\xi};--}^{*} & \wtilde{B}_{\vec{\xi};--} \eea\right) \\
\lb{134}
\wtilde{\mcal{H}}_{\vec{\xi},p;\vec{\xi}\ppr,q}^{ab}(\omega)&=&\delta_{p,q}\;\eta_{p}\;\delta_{a,b}\;
\delta_{\vec{\xi},\vec{\xi}\ppr}\left[-\omega\;\hat{1}_{4\times4}+ \left(\bea{cccc}
\hat{\wtilde{\mbox{\sf h}}}_{+}(\vec{\xi}) & 0 & 0 & 0  \\
0 & \hat{\wtilde{\mbox{\sf h}}}_{+}^{T}(\vec{\xi}) & 0 & 0 \\
0 & 0 & \hat{\wtilde{\mbox{\sf h}}}_{-}(\vec{\xi}) &  0 \\
0 & 0 & 0 & \hat{\wtilde{\mbox{\sf h}}}_{-}^{T}(\vec{\xi})
\eea\right)\right]_{pq}^{ab}   \\ \lb{135}
\hat{\wtilde{\mbox{\sf h}}}_{p}(\vec{\xi}) &=&
-\im\;\wtilde{\ve}_{p}-\pp_{\vec{\xi}}\cdot\pp_{\vec{\xi}}+\wtilde{u}(\vec{\xi})-\wtilde{\mu}_{0}
\hspace*{1.0cm}\hat{\wtilde{\mbox{\sf h}}}_{p}^{T}(\vec{\xi})=\hat{\wtilde{\mbox{\sf h}}}_{p}(\vec{\xi})\;\;\;.
\eeq
In comparison to the disorder model I, one has also to compute the right and left eigenfunctions
\(\Psi_{\vec{\xi},p}^{R/L,a}(\omega_{N})\) and eigenvalues \(\omega_{N}\), but without a
dependence on the eigenvalue of the disorder-self-energy  \(\wtilde{\Sigma}_{II;\vec{\xi};pq}^{ab}\)
for stationary states
\beq\lb{136}
\lefteqn{\sum_{q=\pm}\sum_{b=1,2}
\left(\bea{cccc}
\hat{\wtilde{\mbox{\sf h}}}(\vec{\xi}) & 0 & 0 & 0  \\
0 & \hat{\wtilde{\mbox{\sf h}}}^{T}(\vec{\xi}) & 0 & 0 \\
0 & 0 & -\hat{\wtilde{\mbox{\sf h}}}(\vec{\xi}) &  0 \\
0 & 0 & 0 & -\hat{\wtilde{\mbox{\sf h}}}^{T}(\vec{\xi})
\eea\right)_{pq}^{ab}
\underbrace{\left(\bea{c}
\psi_{\vec{\xi},+}(\omega_{N}) \\ \psi_{\vec{\xi},+}^{*}(\omega_{N}) \\ \psi_{\vec{\xi},-}(\omega_{N}) \\
\psi_{\vec{\xi},-}^{*}(\omega_{N}) \eea\right)_{q}^{R,b}}_{\Psi_{\vec{\xi},q}^{R,b}(\omega_{N})}+ } \\ \no &+&
\frac{1}{2}\sum_{q=\pm}\sum_{b=1,2}\eta_{p}\; \left( \bea{cccc} \wtilde{B}_{\vec{\xi};++} & \wtilde{C}_{\vec{\xi};++}
& \wtilde{B}_{\vec{\xi};+-} & \wtilde{C}_{\vec{\xi};+-} \\
\wtilde{C}_{\vec{\xi};++}^{*} & \wtilde{B}_{\vec{\xi};++} & \wtilde{C}_{\vec{\xi};+-}^{*} &
\wtilde{B}_{\vec{\xi};+-}^{*}
\\ \wtilde{B}_{\vec{\xi};+-}^{*} & \wtilde{C}_{\vec{\xi};+-}
& \wtilde{B}_{\vec{\xi};--} & \wtilde{C}_{\vec{\xi};--} \\
\wtilde{C}_{\vec{\xi};+-}^{*} & \wtilde{B}_{\vec{\xi};+-} & \wtilde{C}_{\vec{\xi};--}^{*} & \wtilde{B}_{\vec{\xi};--}
\eea\right)_{pq}^{ab}\;\eta_{q}\; \underbrace{\left(\bea{c}
\psi_{\vec{\xi},+}(\omega_{N}) \\ \psi_{\vec{\xi},+}^{*}(\omega_{N}) \\ \psi_{\vec{\xi},-}(\omega_{N}) \\
\psi_{\vec{\xi},-}^{*}(\omega_{N})
\eea\right)_{q}^{R,b}}_{\Psi_{\vec{\xi},q}^{R,b}(\omega_{N})}=  \\ \no &=& \omega_{N}\;\;\eta_{p}\;\;
\underbrace{\left(\bea{c}
\psi_{\vec{\xi},+}(\omega_{N}) \\ \psi_{\vec{\xi},+}^{*}(\omega_{N}) \\ \psi_{\vec{\xi},-}(\omega_{N}) \\
\psi_{\vec{\xi},-}^{*}(\omega_{N})
\eea\right)_{p}^{R,a}}_{\Psi_{\vec{\xi},p}^{R,a}(\omega_{N})}
\eeq
\beq \lb{137}
 \hat{\wtilde{\mbox{\sf h}}}(\vec{\xi}) &=&
-\pp_{\vec{\xi}}\cdot\pp_{\vec{\xi}}+\wtilde{u}(\vec{\xi})-\wtilde{\mu}_{0}
\hspace*{0.55cm}\hat{\wtilde{\mbox{\sf h}}}_{p}^{T}(\vec{\xi})=\hat{\wtilde{\mbox{\sf h}}}_{p}(\vec{\xi})
\\ \lb{138}\delta_{\omega_{N},\omega_{N\ppr}\ppr} &=&
\int\dtot[\vec{\xi}]\sum_{p=\pm}\sum_{a=1,2}\;\;\Psi_{\vec{\xi},p}^{L,a}(\omega_{N\ppr}\ppr)\;\;\eta_{p}\;\;
\Psi_{\vec{\xi},p}^{R,a}(\omega_{N})\;\;\;.
\eeq
We transfer the generalized eigenvalue problem (\ref{96}-\ref{100}) in section \ref{s32}
to the case with dynamic disorder
so that the Green function \(\wtilde{M}_{II;\vec{\xi},p;\vec{\xi}\ppr,q}^{-1;ab}(\omega)\)
(\ref{140}) is also determined
by relations (\ref{136}) to (\ref{139}) in terms of the orthonormalized eigenfunctions (\ref{138}) and eigenvalues
\beq \lb{139}
\sum_{\{\omega_{N}\}}\eta_{p}\;\;\Psi_{\vec{\xi},p}^{R,a}(\omega_{N})\;\;\;\Psi_{\vec{\xi}\ppr,q}^{L,b}(\omega_{N})
&=&\delta_{\vec{\xi},\vec{\xi}\ppr}\;\;\delta_{p,q}\;\;\delta_{a,b} \\  \lb{140}
\wtilde{M}_{II;\vec{\xi},p;\vec{\xi}\ppr,q}^{-1;ab}(\omega)&=&\sum_{\{\omega_{N}\}}
\frac{\Psi_{\vec{\xi},p}^{R,a}(\omega_{N})\;\otimes\;
\Psi_{\vec{\xi}\ppr,q}^{L,b}(\omega_{N})}{-\omega-\im\;\wtilde{\ve} +\omega_{N}}\hspace*{1.0cm}(\wtilde{\ve}>0)\;.
\eeq
We also assume the completeness of the orthonormalized eigenfunctions (\ref{139},\ref{138}) as in section \ref{s32}
for model I.  The eigenvalue problem (\ref{136}-\ref{140}) is reminiscent of the Bogoliubov-de Gennes equations and
the Nambu-Gorkov Green functions in the theory for superconductivity \cite{Zago}-\cite{Mosk},\cite{mbody1}-\cite{mbody3}.
It has to be solved at every iteration step of
the continued fractions, but is easier to solve as in the case of the static
disorder because the disorder-self-energy is time independent for a trap potential having only a spatial dependence.
The iteration procedure can be further simplified in the case of spatial symmetries.

\section{Determination of the observables from the derivative with the source term
$\boldsymbol{\mcal{J}_{\vec{x},\vec{x}\ppr}^{ab}(t_{p},t_{q}\ppr)}$} \lb{s4}

The $U(1)$ invariant density '\(\lim_{\delta t_{+}\to 0_{+}}\langle\Psi_{\vec{x}}^{+,b=1}(t_{+})\;\;
\Psi_{\vec{x}}^{a=1}(t_{+}+\delta t_{+})\rangle\)' of non-condensed atoms follows from differentiating
\(\ovv{Z_{I}[\mcal{J}]}\) (\ref{69}), \(\ovv{Z_{II}[\mcal{J}]}\) (\ref{113}) with respect to
\(\mcal{J}_{\vec{x},\vec{x}}^{b=1,a=1}(t_{+},t_{+}+\delta t_{+})\) and the appropriate normalization.
Since a field operator \(\hat{\psi}_{\vec{x}}(t_{p})\) and its hermitian conjugate \(\hat{\psi}_{\vec{x}}^{+}(t_{p})\)
must not act at the same space time point due to the infinite delta
function of the corresponding commutator at coincidence of time,
a limit process (\(\lim_{\delta t_{p}\to 0_{p}}\ldots\), \(p=\pm\))
has to be performed for the density terms at the same branch of the contour time.
The corresponding relations for static and dynamic disorder are tabulated in Eqs. (\ref{141}) to (\ref{145})
with the matrices
\(\hat{\mcal{O}}_{\vec{x},\vec{x}\ppr}^{cd}(t_{p},t_{q}\ppr)=
\hat{M}_{I;\vec{x},\vec{x}\ppr}^{cd}(t_{p},t_{q}\ppr)\)
(\ref{71}) and
\(\hat{\mcal{O}}_{\vec{x},\vec{x}\ppr}^{cd}(t_{p},t_{q}\ppr)=
\hat{M}_{II;\vec{x},\vec{x}\ppr}^{cd}(t_{p},t_{q}\ppr)\)
(\ref{111}). The one particle part
\(\hat{\mcal{H}}_{\vec{x},\vec{x}\ppr}^{cd}(t_{p},t_{q}\ppr)\) is defined
in Eqs. (\ref{60}-\ref{63}) ('\(c,d=1,2\)' are
'Nambu'-indices as '\(a,b=1,2\)')
\beq\lb{141}
\mbox{static disorder} &:&\mbox{see }\;\;\ovv{Z_{I}[\mcal{J}]}\;\;\;
\mbox{(\ref{69})}\;\;\mbox{ with }\;\;\hat{M}_{I;\vec{x},\vec{x}\ppr}^{cd}(t_{p},t_{q}\ppr)
\;\;\;\mbox{(\ref{71})}    \\ \lb{142}
\mbox{dynamic disorder} &:&\mbox{see }\;\;\ovv{Z_{II}[\mcal{J}]}\;\;\;
\mbox{(\ref{113})}\;\;\mbox{ with }\;\;\hat{M}_{II;\vec{x},\vec{x}\ppr}^{cd}(t_{p},t_{q}\ppr)
\;\;\;\mbox{(\ref{111})}    \\ \lb{143}
\hat{\mcal{O}}_{\vec{x},\vec{x}\ppr}^{cd}(t_{p},t_{q}\ppr)&=&
\hat{M}_{I;\vec{x},\vec{x}\ppr}^{cd}(t_{p},t_{q}\ppr)
\hspace*{0.64cm}\mbox{ or }\hspace*{0.64cm}
\hat{\mcal{O}}_{\vec{x},\vec{x}\ppr}^{cd}(t_{p},t_{q}\ppr)=
\hat{M}_{II;\vec{x},\vec{x}\ppr}^{cd}(t_{p},t_{q}\ppr)
\eeq
\beq \lb{144}
\lefteqn{\hspace*{-1.0cm}\lim_{\delta t_{+}\to 0_{+}}\langle\Psi_{\vec{x}}^{+,b=1}(t_{+})\;\;
\Psi_{\vec{x}}^{a=1}(t_{+}+\delta t_{+})\rangle =2\im\hbar\Omega^{2}\mcal{N}_{x}^{2}\;\;
\left(\frac{\pp \ovv{Z_{I,II}[\mcal{J}]}}{\pp
\mcal{J}_{\vec{x},\vec{x}}^{b=1,a=1}(t_{+},t_{+}+\delta t_{+})}\right)\bigg|_{
\mcal{J}\equiv 0,\{j_{\psi},j_{\psi\psi}\}}  }  \\ \no
&\stackrel{\delta t_{+}\to 0_{+}}{=}& -\im\hbar\Omega^{2}\mcal{N}_{x}\;\;
\hat{\mcal{O}}_{\vec{x},\vec{x}}^{-1;a=1,b=1}(t_{+}+\delta t_{+},t_{+}) +\Omega^{4}\;\mcal{N}_{x}^{2}
\int_{C}\dtot t_{p}^{\prime}\;\dtot t_{q}^{\prime\prime}\sum_{\vec{x}_{1},\vec{x}_{2}}\sum_{c,d=1,2} \\ \no &&
J_{\psi;\vec{x}_{2}}^{+d}(t_{q}^{\prime\prime})\;\;
\hat{\mcal{O}}_{\vec{x}_{2},\vec{x}}^{-1;d,b=1}(t_{q}^{\prime\prime},t_{+})\;\;\;\;
\hat{\mcal{O}}_{\vec{x},\vec{x}_{1}}^{-1;a=1,c}(t_{+}+\delta t_{+},t_{p}^{\prime})\;\;
J_{\psi;\vec{x}_{1}}^{c}(t_{p}^{\prime})
\eeq
\beq\lb{145}
\hat{M}_{I;\vec{x},\vec{x}\ppr}^{cd}(t_{p},t_{q}\ppr)&=&\eta_{p}\bigg(
\hat{\mcal{H}}_{\vec{x},\vec{x}\ppr}^{cd}(t_{p},t_{q}\ppr)+\frac{1}{2}\;
\frac{R_{I}\;\Omega}{\sqrt{\mcal{N}_{x}}\;\hbar}\;\delta_{\vec{x},\vec{x}\ppr}\;\;
\hat{\Sigma}_{I;\vec{x}}^{cd}(t_{p},t_{q}\ppr)\bigg)\eta_{q}  \\  \lb{146}
\hat{M}_{II;\vec{x},\vec{x}\ppr}^{cd}(t_{p},t_{q}\ppr)&=&\eta_{p}\bigg(
\hat{\mcal{H}}_{\vec{x},\vec{x}\ppr}^{cd}(t_{p},t_{q}\ppr)+\frac{1}{2}\;\delta(t-t\ppr)\;\;
\hat{\Sigma}_{II;\vec{x};pq}^{cd}(t)\bigg)\eta_{q}  \\   \lb{147}
J_{\psi;\vec{x}}^{c}(t_{+})&=&J_{\psi;\vec{x}}^{c}(t_{-})\hspace*{1.9cm}
\hat{J}_{\psi\psi;\vec{x}}^{cd}(t_{+})=\hat{J}_{\psi\psi;\vec{x}}^{cd}(t_{-})\;\;\;.
\eeq
We can disentangle the general relations (\ref{141}-\ref{147}) by Fourier transformation to energy momentum
space for a time independent trap potential. The non-equilibrium Green functions
\(\hat{\mcal{O}}_{\vec{x},\vec{x}\ppr}^{-1;cd}(t_{p},t_{q}\ppr)\) become more transparent in energy momentum
space \(\hat{\mcal{O}}_{\vec{k},p;\vec{k}\ppr,q}^{cd}(\omega)\) (\ref{148}-\ref{155}) and are analogous to the
'Nambu-Gorkov' Green function formalism for superconductivity \cite{Zago}-\cite{Mosk},\cite{mbody1}-\cite{mbody3}.
In the case of $d=3$ spatial
dimensions, we have to consider the zero momentum state in the summation over wave-vectors
\(\sum_{\vec{k}}\ldots\) explicitly (\ref{148}) \cite{Pita,Peth}. The mean field equations (\ref{87}-\ref{91}),
(\ref{127}-\ref{131}) are mainly applied in three spatial dimensions whereas the two dimensional case
should be treated preferably by spontaneous symmetry breaking for the derivation of a nonlinear sigma model
and requires different HST transformations than the ones described in this paper
(see section \ref{s5} and \cite{Bm7,Bm2})
\beq \lb{148}
\lefteqn{\sum_{\vec{k}}\ldots(\mbox{fields},\vec{k})\ldots=} \\ \no &\hspace*{-0.37cm}=&\hspace*{-0.37cm}\Bigg\{
\bea{rcl}
(d=3)&:& (\mbox{fields},\vec{k}\equiv\vec{0})+{\ds\int_{0}^{2\pi}\dtot\varphi
\int_{0}^{\pi}\dtot\theta\;\sin\theta\int_{0}^{\infty}\dtot k\bigg(\frac{k^{2}}{\big(\frac{2\pi}{L}\big)^{d}}\bigg)
\ldots(\mbox{fields},\vec{k}\neq \vec{0})   } \\
(d=1,2) &:& {\ds \int\bigg(\frac{\dtot^{d}k}{(\frac{2\pi}{L})^{d}}\bigg)}\ldots(\mbox{fields},\vec{k})\ldots \eea \eeq
\beq  \lb{149} \lefteqn{\hspace*{-1.0cm}\lim_{\delta t_{+}\to 0_{+}}\langle\Psi_{\vec{x}}^{+,b=1}(t_{+})\;\;
\Psi_{\vec{x}}^{a=1}(t_{+}+\delta t_{+})\rangle\stackrel{\delta t_{+}\to 0_{+}}{=}
\im\hbar\Omega\mcal{N}_{x}\sum_{\vec{k}} \int\frac{\dtot\omega}{(\frac{2\pi}{T_{0}})}\;\exp\{-\im\;\delta t_{+}\;\omega\}\;\;
\hat{\mcal{O}}_{\vec{k},p=+;\vec{k},q=+}^{-1;a=1,b=1}(\omega) + }
\\ \no &+& \Omega^{2}\;T_{0}^{2}\;\mcal{N}_{x}^{2}\sum_{\vec{k},\vec{k}\ppr}\sum_{\vec{k}_{1},\vec{k}_{2}}\sum_{c,d=1,2}
\int_{C}\frac{\dtot\omega_{p}^{\prime}}{(\frac{2\pi}{T_{0}})}\; \frac{\dtot\omega_{q}^{\prime\prime}}{(\frac{2\pi}{T_{0}})}\;
J_{\psi;\vec{k}_{2}}^{+d}(\omega_{q}^{\prime\prime})\;\; \hat{\mcal{O}}_{\vec{k}_{2},q;\vec{k}\ppr,+}^{-1;d,b=1}(\omega^{\prime\prime})
\;\exp\{-\im\;\omega^{\prime}\;\delta t_{+}\}\;\times \\ \no &\times&
\exp\{\im((\vec{k}-\vec{k}\ppr)\cdot\vec{x}-(\omega^{\prime}-\omega^{\prime\prime})\;t_{+})\}\;\;
\hat{\mcal{O}}_{\vec{k},+;\vec{k}_{1},p}^{-1;a=1,c}(\omega^{\prime})\;\; J_{\psi;\vec{k}_{1}}^{c}(\omega_{p}^{\prime}) \eeq \beq \lb{150}
\hat{\mcal{O}}_{\vec{k},p;\vec{k}\ppr,q}^{cd}(\omega) &=&- \hat{M}_{I;\vec{k},p;\vec{k}\ppr,q}^{cd}(\omega)  \;\;\;\mbox{ for }
\;\;\;\ovv{Z_{I}[\mcal{J}]}\;\;\;\mbox{(\ref{69})}  \\  \lb{151} \hat{\mcal{O}}_{\vec{k},p;\vec{k}\ppr,q}^{cd}(\omega) &=&-
\hat{M}_{II;\vec{k},p;\vec{k}\ppr,q}^{cd}(\omega)  \;\;\;\mbox{ for } \;\;\;\ovv{Z_{II}[\mcal{J}]}\;\;\;\mbox{(\ref{113})}  \\   \lb{152}
\hat{M}_{I;\vec{k},p;\vec{k}\ppr,q}^{cd}(\omega)&=&\eta_{p}\bigg( \hat{\mcal{H}}_{\vec{k},p;\vec{k}\ppr,q}^{cd}(\omega)+\frac{1}{2}\;
\frac{R_{I}}{\sqrt{\mcal{N}_{x}}\;\hbar}\;\hat{\Sigma}_{I;\vec{k}- \vec{k}\ppr;pq}^{cd}(\omega)\bigg)\eta_{q} \\  \lb{153}
\hat{M}_{II;\vec{k},p;\vec{k}\ppr,q}^{cd}(\omega)&=&\eta_{p}\bigg( \hat{\mcal{H}}_{\vec{k},p;\vec{k}\ppr,q}^{cd}(\omega)+\frac{1}{2}\;
\;\hat{\Sigma}_{II;\vec{k}-\vec{k}\ppr;pq}^{cd}\bigg)\eta_{q} \\  \lb{154}
\hat{\mcal{H}}_{\vec{k},p;\vec{k}\ppr,q}^{cd}(\omega)&=&-\delta_{p,q}\;\eta_{p}\;\delta_{c,d}\;
\delta_{\vec{k},\vec{k}\ppr}\;\;\hbar\omega\;\;\hat{1}_{4\times 4} + \\ \no &+& \delta_{p,q}\;\eta_{p}\;\delta_{c,d}\left( \bea{cccc}
\hat{\mbox{\sf h}}_{+}(\vec{k}-\vec{k}\ppr) & 0 & 0 & 0  \\
0 & \hat{\mbox{\sf h}}_{+}^{T}(\vec{k}-\vec{k}\ppr) & 0 & 0 \\
0 & 0 & \hat{\mbox{\sf h}}_{-}(\vec{k}-\vec{k}\ppr) & 0 \\
0 & 0 & 0 & \hat{\mbox{\sf h}}_{-}^{T}(\vec{k}-\vec{k}\ppr)
\eea\right)_{pq}^{cd}   \\ \lb{155}
\hat{\mbox{\sf h}}_{p}(\vec{k}-\vec{k}\ppr) &=&
\left(\frac{\hbar^{2}\;|\vec{k}|^{2}}{2m}-\im\;\ve_{p}-\mu_{0}\right)\;\;\delta_{\vec{k},\vec{k}\ppr}+
u(\vec{k}-\vec{k}\ppr)\;\;\;.
\eeq
If we further assume spatial independence of the source field
\(J_{\psi;\vec{k}_{1}}^{c}(\omega_{p}\ppr)\) and of the trap potential
\(\ovv{u}=u(\vec{k}-\vec{k}\ppr\equiv\vec{0})\) and also a constant creation rate with
\(J_{\psi;\vec{k}_{1}}^{c}(\omega_{p}\ppr)\) (\ref{156}), the $U(1)$ invariant density terms in (\ref{144},\ref{149})
reduce to integrals over energy momentum space and a contribution from \(\vec{k}\equiv\vec{0}\),
\(\omega\equiv0\) which is related to the 'negative' density of the coherent BE-wavefunction
\be\lb{156}
J_{\psi;\vec{k}}^{c}(\omega_{p})=\frac{2\pi}{T_{0}}\;\;
\delta(\omega)\;\;\delta_{\vec{k},\vec{0}}\;\;\frac{j}{2}\;;\hspace*{1.0cm}
j\;\in\;\mbox{\sf R}
\ee
\beq \lb{157}
\lefteqn{\lim_{\delta t_{+}\to 0_{+}}\langle\Psi_{\vec{x}}^{+,b=1}(t_{+})\;\;
\Psi_{\vec{x}}^{a=1}(t_{+}+\delta t_{+})\rangle
\stackrel{\mbox{\scriptsize static disorder}}{=}  \im\hbar\Omega\mcal{N}_{x}\sum_{\vec{k}}
\int\frac{\dtot\omega}{(\frac{2\pi}{T_{0}})} \; \exp\{-\im\;\delta t_{+}\;\omega\}\;\times } \\ \no &\times &
\bigg\{\delta_{p\ppr,q\ppr}\;\eta_{p\ppr}\;\delta_{c\ppr,d\ppr}\;\hat{1}_{4\times 4}
\bigg[\hbar\omega+\im\;\ve_{p\ppr}-\bigg(\frac{\hbar^{2}\;|\vec{k}|^{2}}{2m}-\mu_{0}+\ovv{u}\bigg)\bigg] -
\frac{R_{I}}{\sqrt{\mcal{N}_{x}}\;\hbar}\;\eta_{p\ppr}\;
\hat{\Sigma}_{\Delta\vec{k}=\vec{0};p\ppr q\ppr}^{c\ppr d\ppr}(\omega)\;\eta_{q\ppr}
\bigg\}_{++}^{-1;a=1,b=1}\hspace*{-1.0cm}(\vec{k},\omega) + \\ \no   &+&
\big(\Omega T_{0}\mcal{N}_{x}\big)^{2}\sum_{p,q=\pm}\sum_{c,d=1,2} \;\;\;\times \\ \no &\times&
\frac{j^{*}}{2}\;\eta_{q}
\bigg\{\delta_{p\ppr,q\ppr}\;\eta_{p\ppr}\;\delta_{c\ppr,d\ppr}\;\hat{1}_{4\times 4}
\Big(\im\;\ve_{p\ppr}+\mu_{0}-\ovv{u}\Big)-\eta_{p\ppr}\;\frac{R_{I}\;
\hat{\Sigma}_{\Delta\vec{k}=\vec{0};p\ppr q\ppr}^{c\ppr d\ppr}(\omega\equiv0)}{\sqrt{\mcal{N}_{x}}\;\hbar}\;
\eta_{q\ppr}\bigg\}_{q+}^{-1;d,b=1}\hspace*{-1.27cm}(\vec{k}\equiv\vec{0},\omega\equiv0)\;\;\;\times
\\ \no &\times&
\bigg\{\delta_{p\ppr,q\ppr}\;\eta_{p\ppr}\;\delta_{c\ppr,d\ppr}\;\hat{1}_{4\times 4}
\Big(\im\;\ve_{p\ppr}+\mu_{0}-\ovv{u}\Big)
-\eta_{p\ppr}\;
\frac{R_{I}\;\hat{\Sigma}_{\Delta\vec{k}
\equiv\vec{0};p\ppr q\ppr}^{c\ppr d\ppr}(\omega\equiv0)}{\sqrt{\mcal{N}_{x}}
\;\hbar}\;\eta_{q\ppr} \bigg\}_{+p}^{-1;a=1,c}\hspace*{-1.27cm}(\vec{k}\equiv\vec{0},\omega\equiv0)
\;\;\;\;\eta_{p}\;\frac{j}{2} \\ \no
\eeq
\beq
\no \\  \lb{158}
\lefteqn{\lim_{\delta t_{+}\to 0_{+}}\langle\Psi_{\vec{x}}^{+,b=1}(t_{+})\;\;
\Psi_{\vec{x}}^{a=1}(t_{+}+\delta t_{+})\rangle
\stackrel{\mbox{\scriptsize dynamic disorder}}{=} \im\hbar\Omega\mcal{N}_{x}\sum_{\vec{k}}
\int\frac{\dtot\omega}{(\frac{2\pi}{T_{0}})} \;\exp\{-\im\;\delta t_{+}\;\omega\}  \;\times } \\ \no&\times&
\bigg\{\delta_{p\ppr,q\ppr}\;\eta_{p\ppr}\;\delta_{c\ppr,d\ppr}\;\hat{1}_{4\times 4}
\bigg[\hbar\omega+\im\;\ve_{p\ppr}-\bigg(\frac{\hbar^{2}\;|\vec{k}|^{2}}{2m}-\mu_{0}+\ovv{u}\bigg)\bigg]
-\frac{1}{2}\;\eta_{p\ppr}\;
\hat{\Sigma}_{\Delta\vec{k}\equiv\vec{0};p\ppr q\ppr}^{c\ppr d\ppr}\;\eta_{q\ppr}
\bigg\}_{++}^{-1;a=1,b=1}\hspace*{-1.27cm}(\vec{k},\omega) + \\ \no &+&
\big(\Omega T_{0}\mcal{N}_{x}\big)^{2}\sum_{p,q=\pm}\sum_{c,d=1,2}\;\times \\ \no &\times&
\frac{j^{*}}{2}\;\eta_{q}
\bigg\{\delta_{p\ppr,q\ppr}\;\eta_{p\ppr}\;\delta_{c\ppr,d\ppr}\;\hat{1}_{4\times 4}
\Big(\im\;\ve_{p\ppr}+\mu_{0}-\ovv{u}\Big)
-\frac{1}{2}\;\eta_{p\ppr}\;
\hat{\Sigma}_{\Delta\vec{k}\equiv\vec{0};p\ppr q\ppr}^{c\ppr d\ppr}\;\eta_{q\ppr}
\bigg\}_{q+}^{-1;d,b=1}\hspace*{-1.27cm}(\vec{k}\equiv\vec{0},\omega\equiv0)\;\;\;\times    \\ \no &\times&
\bigg\{\delta_{p\ppr,q\ppr}\;\eta_{p\ppr}\;\delta_{c\ppr,d\ppr}\;\hat{1}_{4\times 4}
\Big(\im\;\ve_{p\ppr}+\mu_{0}-\ovv{u}\Big)
-\frac{1}{2}\;\eta_{p\ppr}\;
\hat{\Sigma}_{\Delta\vec{k}\equiv\vec{0};p\ppr q\ppr}^{c\ppr d\ppr}\;\eta_{q\ppr}
\bigg\}_{+p}^{-1;a=1,c}\hspace*{-1.27cm}(\vec{k}\equiv\vec{0},\omega\equiv0)\;\;\;\;\;\;
\eta_{p}\;\frac{j}{2}_{\mbox{.}}
\eeq
The coherent BE-wavefunction \(\psi_{BEC}(\vec{x},t_{+})\) is obtained by differentiating
\(\ovv{Z_{I,II}[\mcal{J}]}\) (\ref{69},\ref{113}) with respect to the $U(1)$ symmetry breaking source field
\(J_{\psi;\vec{x}}^{+,a=1}(t_{+})\). The general case for a
spatial and time dependence of the coherent BE-wavefunction is listed in the following Eq. (\ref{159})
where the Green function \(\hat{\mcal{O}}_{\vec{x},\vec{x}\ppr}^{-1;a=1,c}(t_{+},t_{p}\ppr)\) refers to
\(\hat{M}_{I;\vec{k},p;\vec{k}\ppr,q}^{cd}(\omega)\) (\ref{150},\ref{152}) and to
\(\hat{M}_{II;\vec{k},p;\vec{k}\ppr,q}^{cd}(\omega)\) (\ref{151},\ref{153})
\beq \lb{159}
\psi_{BEC}(\vec{x},t_{+})=\langle\Psi_{\vec{x}}^{a=1}(t_{+})\rangle &=& 2\im\hbar\Omega\mcal{N}_{x}
\left(\frac{\pp\ovv{Z_{I,II}[\mcal{J}]}}{\pp J_{\psi;\vec{x}}^{+,a=1}(t_{+})}\right)
\bigg|_{\mcal{J}\equiv0,\{j_{\psi},j_{\psi\psi}\}} =  \\ \no &=&
\mcal{N}_{x}\Omega^{2}\int_{C}\dtot t_{p}\ppr\sum_{\vec{x}\ppr}\sum_{c=1,2}
\hat{\mcal{O}}_{\vec{x},\vec{x}\ppr}^{-1;a=1,c}(t_{+},t_{p}\ppr)\;\;J_{\psi;\vec{x}\ppr}^{c}(t_{p}\ppr)\;\;\;.
\eeq
We can transform the above relation (\ref{159}) to energy momentum space which simplifies for a homogenous translation
invariant system with \(\ovv{u}=u(\Delta\vec{k}=\vec{0})\) and
\(J_{\psi;\vec{k}}^{c}(\omega_{p})=\frac{2\pi}{T_{0}}\;\;\delta(\omega)\;\;
\delta_{\vec{k},\vec{0}}\;\;\frac{j}{2}\), (\(j\in\mbox{\sf R}\)) (\ref{156})
\be \lb{160}
\psi_{BEC}(\vec{x},t_{+})=\mcal{N}_{x}\Omega T_{0}\int_{C}\frac{\dtot\omega_{p}\ppr}{(\frac{2\pi}{T_{0}})}
\sum_{\vec{k},\vec{k}_{1}}\sum_{c=1,2}
\exp\{\im(\vec{k}\cdot\vec{x}-\omega\ppr t_{+})\} \;\;
\hat{\mcal{O}}_{\vec{k},+;\vec{k}_{1},p}^{-1;a=1,c}(\omega\ppr)\;\;J_{\psi;\vec{k}_{1}}^{c}(\omega_{p}\ppr)
\ee
\beq\lb{161}
\lefteqn{\hspace*{-1.0cm}\psi_{BEC}
\stackrel{\mbox{\scriptsize static disorder}}{=}\sum_{p=\pm}\sum_{c=1,2}
\bigg\{\delta_{p\ppr,q\ppr}\;\eta_{p\ppr}\;\delta_{c\ppr,d\ppr}\;\hat{1}_{4\times 4}
\Big(\im\;\ve_{p\ppr}+\mu_{0}-\ovv{u}\Big) + } \\ \no &-&
\frac{R_{I}}{\sqrt{\mcal{N}_{x}}\;\hbar}\;\eta_{p\ppr}\;
\hat{\Sigma}_{\Delta\vec{k}\equiv\vec{0};p\ppr q\ppr}^{c\ppr d\ppr}(\omega\equiv0)\;\eta_{q\ppr}
\bigg\}_{+p}^{-1;a=1,c}\hspace*{-1.0cm}(\vec{k}\equiv\vec{0},\omega\equiv0) \;\;\;\;
\eta_{p}\;\Omega T_{0}\mcal{N}_{x}\frac{j}{2}
\eeq
\beq\lb{162}
\lefteqn{\hspace*{-1.0cm}\psi_{BEC}
\stackrel{\mbox{\scriptsize dynamic disorder}}{=}\sum_{p=\pm}\sum_{c=1,2}
\bigg\{\delta_{p\ppr,q\ppr}\;\eta_{p\ppr}\;\delta_{c\ppr,d\ppr}\;\hat{1}_{4\times 4}
\Big(\im\;\ve_{p\ppr}+\mu_{0}-\ovv{u}\Big) + } \\ \no
&-&\frac{1}{2}\;\eta_{p\ppr}\;
\hat{\Sigma}_{\Delta\vec{k}\equiv\vec{0};p\ppr q\ppr}^{c\ppr d\ppr}\;\eta_{q\ppr}
\bigg\}_{+p}^{-1;a=1,c}\hspace*{-1.0cm}(\vec{k}\equiv\vec{0},\omega\equiv0) \;\;\;\;
\eta_{p}\;\Omega T_{0}\mcal{N}_{x}\frac{j}{2}_{\mbox{.}}
\eeq
In the thermodynamic limit \(j\to 0\), a finite coherent BE-wavefunction \(\psi_{BEC}\) for a homogenous
system remains if an 'effective zero eigenvalue' appears in the denominators of (\ref{161},\ref{162}).
The order of magnitude of the BE-wavefunctions \(\psi_{BEC}\) (\ref{161},\ref{162}) can then be estimated as follows
\be \lb{163}
\psi_{BEC}\approx\;\;\;\frac{\mcal{N}_{x}\;\Omega\;T_{0}\;j}{\im\;\ve_{+}+(\mbox{'effective zero eigenvalue'})}
\to \mbox{finite value}\;\;\;\;.
\ee
Using the properties of non-equilibrium Green functions \cite{Klein1},
the $U(1)$ invariant density of non-condensed atoms
'\(\lim_{\delta t_{+}\to 0_{+}}\langle\Psi_{\vec{x}}^{+,b=1}(t_{+})\;\;
\Psi_{\vec{x}}^{a=1}(t_{+}+\delta t_{+})\rangle\)'
contains the density \(|\psi_{BEC}|^{2}\) of the coherent BEC wavefunctions
(\ref{160}-\ref{162})  which is {\it subtracted} from the total
density given by the first terms in (\ref{157},\ref{158})
\beq \lb{164}
\lefteqn{
\lim_{\delta t_{+}\to 0_{+}}\langle\Psi_{\vec{x}}^{+,b=1}(t_{+})\;\;\Psi_{\vec{x}}^{a=1}(t_{+}+\delta t_{+})\rangle
\stackrel{\mbox{\scriptsize static disorder}}{=}\im\hbar\Omega\mcal{N}_{x}\sum_{\vec{k}}
\int\frac{\dtot\omega}{(\frac{2\pi}{T_{0}})}\;\exp\{-\im\;\delta t_{+}\;\omega\} \;\times} \\ \no &&\hspace*{-0.91cm}
\bigg\{\delta_{p\ppr,q\ppr}\;\eta_{p\ppr}\;\delta_{c\ppr,d\ppr}\;\hat{1}_{4\times 4}
\bigg[\hbar\omega+\im\;\ve_{p\ppr}-\bigg(\frac{\hbar^{2}\;|\vec{k}|^{2}}{2m}-\mu_{0}+\ovv{u}\bigg)\bigg]
-\eta_{p\ppr}\;\frac{R_{I}\;
\hat{\Sigma}_{\Delta\vec{k}=\vec{0};p\ppr q\ppr}^{c\ppr d\ppr}(\omega)}{\sqrt{\mcal{N}_{x}}\;\hbar}
\;\eta_{q\ppr}\bigg\}_{++}^{-1;a=1,b=1}\hspace*{-1.27cm}(\vec{k},\omega) \hspace*{0.46cm} -
\Big|\psi_{BEC}\Big|^{2}
\eeq
\beq \lb{165}
\lefteqn{\hspace*{-0.28cm}
\lim_{\delta t_{+}\to 0_{+}}\langle\Psi_{\vec{x}}^{+,b=1}(t_{+})\;\;\Psi_{\vec{x}}^{a=1}(t_{+}+\delta t_{+})\rangle
\stackrel{\mbox{\scriptsize dynamic disorder}}{=}\im\hbar\Omega\mcal{N}_{x}\sum_{\vec{k}}
\int\frac{\dtot\omega}{(\frac{2\pi}{T_{0}})}\;\exp\{-\im\;\delta t_{+}\;\omega\} \;\times } \\ \no &&\hspace*{-1.18cm}
\bigg\{\delta_{p\ppr,q\ppr}\;\eta_{p\ppr}\;\delta_{c\ppr,d\ppr}\;\hat{1}_{4\times 4}
\bigg[\hbar\omega+\im\;\ve_{p\ppr}-\bigg(\frac{\hbar^{2}\;|\vec{k}|^{2}}{2m}-\mu_{0}+\ovv{u}\bigg)\bigg]
-\frac{1}{2}\;\eta_{p\ppr}\;
\hat{\Sigma}_{\Delta\vec{k}=\vec{0};p\ppr q\ppr}^{c\ppr d\ppr}\;\eta_{q\ppr}
\bigg\}_{++}^{-1;a=1,b=1}\hspace*{-1.27cm}(\vec{k},\omega) \hspace*{0.64cm} -\hspace*{0.28cm}
\Big|\psi_{BEC}\Big|^{2}_{\mbox{.}}
\eeq
The corresponding relations for the bosonic anomalous or pair condensate terms
\(\langle\psi_{\vec{x}}(t_{+})\;\psi_{\vec{x}}(t_{+}+\delta t_{+})\rangle=
\langle\Psi_{\vec{x}}^{+,b=2}(t_{+})\;\Psi_{\vec{x}}^{a=1}(t_{+}+\delta t_{+})\rangle\) can be taken
from Eqs. (\ref{141}) to (\ref{158}) by setting the index '\(b=1\)' in these equations to the value '\(b=2\)'.
This follows from the 'Nambu' doubling of fields where \(\Psi_{\vec{x}}^{+,b=2}(t_{+})\) is equivalent
to \(\psi_{\vec{x}}(t_{+})\). The limit process \(\lim_{\delta t_{+}\to 0_{+}}\) need not be considered
for the anomalous parts \(\langle\psi_{\vec{x}}(t_{+})\;\;\psi_{\vec{x}}(t_{+}+\delta t_{+})\rangle\)
whereas the limit process is of central importance for the density term (\ref{144},\ref{149})
\(\lim_{\delta t_{+}\to 0_{+}}\langle\psi_{\vec{x}}^{*}(t_{+})\;\;\psi_{\vec{x}}(t_{+}+\delta t_{+})\rangle\)
because one must not compute with the operator \(\hat{\psi}_{\vec{x}}(t)\) and its hermitian conjugate
\(\hat{\psi}_{\vec{x}}^{+}(t)\) at the same space time point in coherent state path integrals
of many body theory.

\section{Summary and conclusion for $\boldsymbol{d=2}$ spatial dimensions} \lb{s5}

It has already been mentioned that the derived mean field equations (\ref{87}-\ref{91}),
(\ref{127}-\ref{131}) for static and dynamic disorder in sections \ref{s32}, \ref{s33} are
particularly applicable for \(d=3\) spatial dimensions because of the singularity of the
density of states at \(\vec{k}\equiv\vec{0}\). Since the mean field approach is less
applicable in \(d=2\) dimensions, we briefly describe and point out an alternative method
\cite{Bm7,Bm2,Bm1} which extracts the Goldstone modes from a spontaneous symmetry breaking
in a nonlinear sigma model for the anomalous pair condensates. According to this procedure
of spontaneous symmetry breaking, the HST
transformation has to be considered in a different way with so-called 'hinge'-functions
\(\delta\hat{\Sigma}_{\vec{x};pq}^{aa}(t)\) (\ref{166},\ref{171}), \(\delta\sigma_{\vec{x}}^{aa}(t_{p})\)
(\ref{178}-\ref{182}) as subgroups of the 'larger' symmetry groups of the total disorder-self-energy and the
self-energy of the repulsive interaction. Using these modified HST's, a nonlinear sigma model can be derived
from spontaneous symmetry breaking and a gradient expansion for the Goldstone modes.
After introducing the block diagonal densities as 'hinge'-functions, they can be
eventually removed from the determinant and the part for the coherent-BE wavefunction
with the bilinear source field \(J_{\psi;\vec{x}}^{a}(t_{p})\). They remain in
Gaussian integrals and parts of the factorized invariant measure and can be eliminated
by integration with a coupling to the source matrix for the anomalous terms.
In order to acquire the HST for the disorder-self-energy of model II, we introduce
the diagonal self-energy \(\sigma_{R_{II}}^{(0)}(\vec{x},t)\) (\ref{166}) and
the 'hinge' parts \(\delta\hat{\Sigma}_{\vec{x};pq}^{11}(\vec{x},t)\),
\(\delta\hat{\Sigma}_{\vec{x};pq}^{22}(\vec{x},t)\) (\ref{168},\ref{169}) and also the terms
\(\delta\hat{\Sigma}_{\vec{x};pq}^{12}(\vec{x},t)\),
\(\delta\hat{\Sigma}_{\vec{x};pq}^{21}(\vec{x},t)\) (\ref{170}) with \(\delta c_{\vec{x};pq}(t)\)
(\ref{167}) for the pair condensates
\beq \hspace*{-1.0cm}\lb{166}
\sigma_{R_{II}}^{(0)}(\vec{x},t)&\in&\mbox{\sf R}  \hspace*{0.64cm}\mbox{'hinge' functions :}\;
\delta\hat{\Sigma}_{\vec{x};pq}^{11}(t),\;\delta\hat{\Sigma}_{\vec{x};pq}^{22}(t)   \\ \lb{167}
\delta c_{\vec{x};pq}(t)&\in&\mbox{\sf C};\hspace*{0.64cm}
\delta B_{\vec{x};++}(t),\;\;\delta B_{\vec{x};--}(t)\;\in\;\mbox{\sf R}\;;\;
\delta B_{\vec{x};+-}(t)\;\in\;\mbox{\sf C}   \\ \lb{168}
\delta\hat{\Sigma}_{\vec{x};pq}^{11}(t) &=&
\left(\bea{cc}
\delta B_{\vec{x};++}(t) & \delta B_{\vec{x};+-}(t) \\
\delta B_{\vec{x};+-}^{*}(t) & \delta B_{\vec{x};--}(t)
\eea\right)  \\ \lb{169}
\delta\hat{\Sigma}_{\vec{x};pq}^{22}(t) &=&
\left(\bea{cc}
\delta B_{\vec{x};++}(t) & \delta B_{\vec{x};+-}^{*}(t) \\
\delta B_{\vec{x};+-}(t) & \delta B_{\vec{x};--}(t)
\eea\right)\hspace*{0.55cm}
\Big(\delta\hat{\Sigma}_{\vec{x};pq}^{22}(t)\Big)^{T}=
\delta\hat{\Sigma}_{\vec{x};pq}^{11}(t)  \\ \lb{170}
\delta\hat{\Sigma}_{\vec{x};pq}^{12}(t) &=&
\left(\bea{cc}
\delta c_{\vec{x};++}(t) & \delta c_{\vec{x};+-}(t) \\
\delta c_{\vec{x};+-}(t) & \delta c_{\vec{x};--}(t)
\eea\right)\hspace*{0.55cm}
\Big(\delta\hat{\Sigma}_{\vec{x};pq}^{21}(t)\Big)^{+}=
\delta\hat{\Sigma}_{\vec{x};pq}^{12}(t)\;\;\;.
\eeq
One can generalize from the disorder model II to model I with a double time
dependence in the disorder-self-energy. In the case of a stationary state,
the self-energy of model I with static disorder obtains a single energy dependence
after Fourier transformation because of the reduced dependence to the difference
of the two times. Therefore, we can replace the single time dependence of the
disorder-self-energy in model II with dynamic disorder with the single frequency
dependence of model I with static disorder under restriction to stationary states.
The corresponding HST with 'hinge' functions is listed for the ensemble
averaged, non-hermitian interaction of dynamic disorder in the following relation
\beq \lb{171}
\lefteqn{\exp\bigg\{-\frac{R_{II}^{2}}{2\hbar^{2}}\int_{-\infty}^{\infty}\dtot t
\sum_{p,q=\pm}\sum_{\vec{x}}\Big(\psi_{\vec{x}}^{*}(t_{p})\;\eta_{p}\;
\psi_{\vec{x}}(t_{p})\Big)\;\;
\Big(\psi_{\vec{x}}^{*}(t_{q})\;\eta_{q}\;\psi_{\vec{x}}(t_{q})\Big)\bigg\}= } \\ \no &=&
\int\dtot[\sigma_{R_{II}}^{(0)}(\vec{x},t)]\;\;
\exp\bigg\{-\frac{1}{4}\frac{1}{R_{II}^{2}}\int_{-\infty}^{\infty}\dtot t\sum_{\vec{x}}
\sigma_{R_{II}}^{(0)}(\vec{x},t)\;\;\sigma_{R_{II}}^{(0)}(\vec{x},t)\bigg\}\;\; \times \\ \no &\times&
\int\dtot[\wtilde{\Sigma}_{\vec{x};pq}^{ab}(t)\;\wtilde{K}]\;\;
\exp\bigg\{-\frac{1}{8}\frac{1}{R_{II}^{2}}\int_{-\infty}^{\infty}\dtot t
\sum_{\vec{x}}\TRAB\Big[\delta\wtilde{\Sigma}_{\vec{x};pq}^{ab}(t)\;
\wtilde{K}\;\delta\wtilde{\Sigma}_{\vec{x};qp}^{ba}(t)\;\wtilde{K}\Big]\bigg\}
\\ \no &\times &\exp\Bigg\{-\frac{\im}{4\hbar}\int_{-\infty}^{\infty}\dtot t\sum_{\vec{x}}
\TRAB\Bigg[\Bigg(\bea{cc}
\hat{R}_{\vec{x};qp}^{11}(t) & \hat{R}_{\vec{x};qp}^{12}(t) \\
\hat{R}_{\vec{x};qp}^{21}(t) & \hat{R}_{\vec{x};qp}^{22}(t)
\eea\Bigg)\underbrace{\Bigg(\bea{cc} \hat{\eta} & 0 \\ 0 & \hat{\eta} \eea\Bigg)}_{\hat{K}} \\ \no &&
\Bigg(\bea{cc}
\hat{\Sigma}_{\vec{x};pq}^{11}(t) & \delta\hat{\Sigma}_{\vec{x};pq}^{12}(t) \\
\delta\hat{\Sigma}_{\vec{x};pq}^{21}(t) & -\hat{\Sigma}_{\vec{x};pq}^{22}(t)
\eea\Bigg)\underbrace{\Bigg(\bea{cc} \hat{\eta} & 0 \\ 0 & \hat{\eta} \eea\Bigg)}_{\hat{K}}
\Bigg]\Bigg\}_{\mbox{.}}
\eeq
Note the minus sign before \(\hat{\Sigma}_{\vec{x};pq}^{22}(t)\) (\ref{171},\ref{172},\ref{177})
and the tilde '\(\wtilde{\ph{\Sigma}}\)'
of \(\delta\wtilde{\Sigma}_{\vec{x};pq}^{a\neq b}(t)\) (\ref{172}) which refers to 'anti-hermitian'
anomalous parts \(\delta\wtilde{\Sigma}_{\vec{x};pq}^{ab}(t)=
\im\;\delta\hat{\Sigma}_{\vec{x};pq}^{ab}(t)\) (\(a\neq b\)) in this section \ref{s5}.
We have also included a second metric
\(\wtilde{K}\) (\ref{171},\ref{174}) apart from the metric \(\hat{K}_{pq}^{ab}=\hat{\eta}_{p}
\;\delta_{p,q}\;\delta_{a,b}\) (\ref{173}) for indefinite orthogonal symmetry concerning the two
branches of the contour time. The diagonal matrix \(\wtilde{K}\) (\ref{174}) is the appropriate metric
for the symplectic Lie algebra which changes the disorder-self-energy
\(\delta\wtilde{\Sigma}_{\vec{x};pq}^{ab}(t)\;\wtilde{K}\)
to an element of \(sp(4)\), thereby fulfilling
the exact commutation relations (with antihermitian coset parts !)
\beq \lb{172}
\delta\wtilde{\Sigma}_{\vec{x};pq}^{aa}(t)&=&
\delta\hat{\Sigma}_{\vec{x};pq}^{aa}(t) \hspace*{1.0cm}
\delta\wtilde{\Sigma}_{\vec{x};pq}^{ab}(t)=\im\;\;
\delta\hat{\Sigma}_{\vec{x};pq}^{ab}(t)\;\;\;\;\;\;(a\neq b) \\ \lb{173}
\hat{K}_{pq}^{ab}&=&\delta_{a,b}\;\;\delta_{p,q}\;\;\eta_{p} \\ \lb{174}
\wtilde{K}_{pq}^{ab}&=&\delta_{a,b}\;\;\delta_{p,q}\;\;\eta_{p}\;\;
\wtilde{\kappa}^{ab}\;;\hspace*{1.0cm}
\wtilde{\kappa}^{ab}=\delta_{a,b}\;\;
\Big\{\underbrace{+1}_{a=1}\;;\;\underbrace{-1}_{a=2}\Big\} \\ \lb{175}
\hat{R}_{\vec{x};pq}^{ab}(t)&=&\Psi_{\vec{x};p}^{a}(t)\otimes\Psi_{\vec{x};q}^{+b}(t) \\ \lb{176}
\hat{\Sigma}_{\vec{x};pq}^{11}(t) &=&\sigma_{R_{II}}^{(0)}(\vec{x},t)\;\eta_{p}\;\delta_{p,q}+
\delta\hat{\Sigma}_{\vec{x};pq}^{11}(t)  \\  \lb{177}
\hat{\Sigma}_{\vec{x};pq}^{22}(t)&=& -\sigma_{R_{II}}^{(0)}(\vec{x},t)\;\eta_{p}\;\delta_{p,q}+
\delta\hat{\Sigma}_{\vec{x};pq}^{22}(t)\;\;\;.
\eeq
In a similar manner the repulsive interaction term can be transformed with a diagonal
self-energy \(\sigma_{V_{0}}^{(0)}(\vec{x},t_{p})\), 'hinge' functions
\(\delta\sigma_{\vec{x}}^{11}(t_{p})=\delta\sigma_{\vec{x}}^{22}(t_{p})\) and anti-hermitian
anomalous terms \(\delta\wtilde{\sigma}_{\vec{x}}^{(a\neq b)}(t_{p})=\im\;
\delta\hat{\sigma}_{\vec{x}}^{(a\neq b)}(t_{p})\)
\beq \lb{178}
\lefteqn{\exp\bigg\{-\frac{\im}{\hbar}\int_{C}\dtot t_{p}\sum_{\vec{x}}V_{0}\;\;
\Big(\psi_{\vec{x}}^{*}(t_{p})\Big)^{2}\;\;\Big(\psi_{\vec{x}}(t_{p})\Big)^{2}\bigg\} = } \\ \no &=&
\int\dtot[\sigma_{V_{0}}^{(0)}(\vec{x},t_{p})]\;\;
\exp\bigg\{\frac{\im}{2\hbar}\frac{1}{V_{0}}\int_{C}\dtot t_{p}\sum_{\vec{x}}
\sigma_{V_{0}}^{(0)}(\vec{x},t_{p})\;\;\sigma_{V_{0}}^{(0)}(\vec{x},t_{p})\bigg\}  \\ \no &&
\int\dtot[\delta\wtilde{\sigma}_{\vec{x}}^{ab}(t_{p})\;\wtilde{\kappa}]\;\;
\exp\bigg\{\frac{\im}{4\hbar}\frac{1}{V_{0}}\int_{C}\dtot t_{p}\sum_{\vec{x}}\trab
\Big[\delta\wtilde{\sigma}_{\vec{x}}^{ab}(t_{p})\;\wtilde{\kappa}\;
\wtilde{\sigma}_{\vec{x}}^{ba}(t_{p})\;\wtilde{\kappa}\Big]\bigg\} \\ \no &\times&
\exp\Bigg\{-\frac{\im}{2\hbar}\int_{C}\dtot t_{p}\sum_{\vec{x}}\trab\Bigg[
\Bigg(\bea{cc} \hat{R}_{\vec{x};pp}^{11}(t) & \hat{R}_{\vec{x};pp}^{12}(t) \\
\hat{R}_{\vec{x};pp}^{21}(t) & \hat{R}_{\vec{x};pp}^{22}(t)
\eea\Bigg)\Bigg(\bea{cc}
\sigma_{\vec{x}}^{11}(t_{p}) & \delta\sigma_{\vec{x}}^{12}(t_{p}) \\
\delta\sigma_{\vec{x}}^{21}(t_{p}) & -\sigma_{\vec{x}}^{22}(t_{p})
\eea\Bigg)\Bigg]\Bigg\}
\eeq
\beq \lb{179}
\sigma_{\vec{x}}^{11}(t_{p})&=&\sigma_{V_{0}}^{(0)}(\vec{x},t_{p}) +
\delta\sigma_{\vec{x}}^{11}(t_{p})\;\;;\hspace*{0.64cm}
\delta\sigma_{\vec{x}}^{11}(t_{p})\;,\;\;\delta\sigma_{\vec{x}}^{22}(t_{p})\;\in\;\mbox{\sf R} \\ \lb{180}
\sigma_{\vec{x}}^{22}(t_{p})&=&-\sigma_{V_{0}}^{(0)}(\vec{x},t_{p}) +
\delta\sigma_{\vec{x}}^{22}(t_{p})\;\;;\hspace*{1.81cm}
\sigma_{V_{0}}^{(0)}(\vec{x},t_{p})\;\in\;\mbox{\sf R} \\ \lb{181}
\delta\wtilde{\sigma}_{\vec{x}}^{aa}(t_{p})&=&\delta\sigma_{\vec{x}}^{aa}(t_{p})\;\;;\hspace*{4.33cm}
\delta\wtilde{\sigma}_{\vec{x}}^{ab}(t_{p})=\im\;\delta\sigma_{\vec{x}}^{ab}(t_{p})\;(a\neq b) \\ \lb{182}
\delta\sigma_{\vec{x}}^{12}(t_{p})&\in&\mbox{\sf C}\;\;;\hspace*{0.46cm}
\Big(\delta\sigma_{\vec{x}}^{21}(t_{p})\Big)^{*}=\delta\sigma_{\vec{x}}^{12}(t_{p})\;\;;\hspace*{1.0cm}
\delta\sigma_{\vec{x}}^{11}(t_{p})=\delta\sigma_{\vec{x}}^{22}(t_{p})\;\;\;.
\eeq
In analogy to relations (\ref{64}-\ref{66}), (\ref{108},\ref{109}),
we continue by shifts of the total disorder-self-energy with \(\delta\hat{\sigma}_{\vec{x}}^{ab}(t_{p})\)
(\ref{183}) and of the self-energy \(\sigma_{V_{0}}^{(0)}(\vec{x},t_{p})\) of the repulsive interaction
with \(\sigma_{R_{II}}^{(0)}(\vec{x},t)\) (\ref{184}) and also include the shift with the source matrix
\(\hat{J}_{\psi\psi;\vec{x}}^{ab}(t_{p})\) (\ref{185}) for the creation of the bosonic pair condensates
\beq\lb{183}
\delta\hat{\Sigma}_{\vec{x};pq}^{ab}(t)&\to&\delta\hat{\Sigma}_{\vec{x};pq}^{ab}(t)-2\;\delta_{p,q}\;
\eta_{p}\;\delta\hat{\sigma}_{\vec{x}}^{ab}(t_{p}) \\ \lb{184}
\sigma_{V_{0}}^{(0)}(\vec{x},t_{p})&\to&\sigma_{V_{0}}^{(0)}(\vec{x},t_{p})-\frac{1}{2}\;\;
\sigma_{R_{II}}^{(0)}(\vec{x},t)  \\ \lb{185}
\delta\hat{\Sigma}_{\vec{x};pq}^{ab}(t)&\to&\delta\hat{\Sigma}_{\vec{x};pq}^{ab}(t)-2\;\delta_{p,q}\;
\eta_{p}\;\hat{J}_{\psi\psi;\vec{x}}^{ab}(t_{p})\;\;\;.
\eeq
After several transformations we finally achieve a coherent state path integral (\ref{186}) which only depends
on the anomalous terms, determined by the matrices \(\hat{T}_{pq}^{ab}(\vec{x},t)\) (\ref{190}-\ref{192})
of the coset part \(Sp(4)\backslash U(2)\), and on the self-energy \(\sigma_{V_{0}}^{(0)}(\vec{x},t_{p})\)
of the repulsive interaction. We list as starting point for a gradient expansion the relation (\ref{186})
where the block diagonal 'hinge' functions \(\delta\hat{\Sigma}_{\vec{x};pq}^{aa}(t)\) are still present in
Gaussian factors which are to be removed by integration after a change to the corresponding invariant measure
for \(Sp(4)\backslash U(2)\otimes U(2)\). The diagonal self-energy \(\sigma_{R_{II}}^{(0)}(\vec{x},t)\)
has already been absorbed by a shift into \(\sigma_{V_{0}}^{(0)}(\vec{x},t_{p})\) in the determinant
and in the bilinear term with \(J_{\psi;\vec{x}}^{a}(t_{p})\) so that its remaining in a Gaussian factor
has easily been eliminated by integration in \(\ovv{Z_{II}[\mcal{J}]}\) (\ref{186})
\beq\lb{186}
\lefteqn{\ovv{Z_{II}[\mcal{J}]}=\exp\bigg\{-\frac{1}{2\;R_{II}^{2}}
\sum_{p=\pm}\int_{-\infty}^{\infty}\dtot t\sum_{\vec{x}}
\Big(1-\mu_{p}^{(II)}\Big)\;
\trab\Big[\wtilde{J}_{\psi\psi;\vec{x}}^{ab}(t_{p})\;\wtilde{\kappa}\;
\wtilde{J}_{\psi\psi;\vec{x}}^{ba}(t_{p})\;\wtilde{\kappa}\Big]\bigg\} } \\ \no &&
\int\dtot[\sigma_{V_{0}}^{(0)}(\vec{x},t_{p})]\;\;
\exp\bigg\{-\frac{1}{4}\frac{R_{II}^{2}}{(\hbar\;V_{0})^{2}}\int_{-\infty}^{\infty}\dtot t\sum_{\vec{x}}
\Big(\sigma_{V_{0}}^{(0)}(\vec{x},t_{+})-\sigma_{V_{0}}^{(0)}(\vec{x},t_{-})\Big)^{2}\bigg\} \\ \no &\times&
\exp\bigg\{\frac{\im}{2\hbar}\frac{1}{V_{0}}\int_{C}\dtot t_{p}\sum_{\vec{x}}
\sigma_{V_{0}}^{(0)}(\vec{x},t_{p})\;\;\sigma_{V_{0}}^{(0)}(\vec{x},t_{p})\bigg\} \hspace*{0.46cm}
\int\dtot[\delta\wtilde{\Sigma}_{\vec{x};pq}^{ab}(t)\;\wtilde{K}]  \\ \no &&
\exp\bigg\{-\frac{1}{8}\frac{1}{R_{II}^{2}}\sum_{p,q=\pm}\int_{-\infty}^{\infty}\dtot t\sum_{\vec{x}}
\Big(1-\delta_{p,q}\;\mu_{p}^{(II)}\Big)\;
\trab\Big[\delta\wtilde{\Sigma}_{\vec{x};pq}^{ab}(t)\;\wtilde{K}\;
\delta\wtilde{\Sigma}_{\vec{x};qp}^{ba}(t)\;\wtilde{K}\Big]\bigg\}  \\ \no &\times&
\exp\bigg\{\frac{1}{2R_{II}^{2}}\sum_{p=\pm}\int_{-\infty}^{\infty}\dtot t\sum_{\vec{x}}\eta_{p}\;
\Big(1-\mu_{p}^{(II)}\Big)
\trab\Big[\delta\wtilde{\Sigma}_{\vec{x};pp}^{ab}(t)\;\wtilde{\kappa}\;
\wtilde{J}_{\psi\psi;\vec{x}}^{ba}(t_{p})\;\wtilde{\kappa}\Big]\bigg\}  \\ \no &\times&
\exp\bigg\{-\frac{1}{2}\int_{C}\frac{\dtot t_{p}}{\hbar}\eta_{p}\sum_{\vec{x}}\hbar\Omega\mcal{N}_{x}
\trab\ln\Big[\hat{\mcal{O}}_{\vec{x}\ppr,\vec{x}}^{ba}(t_{q}\ppr,t_{p})\Big]\bigg\}  \\ \no &\times&
\exp\bigg\{\frac{\im}{2}\frac{\Omega^{2}}{\hbar}\int_{-\infty}^{\infty}\dtot t\;\;\dtot t\ppr
\sum_{p\ppr,q\ppr=\pm}\;\;\sum_{a\ppr,b\ppr=1,2}\;\;\sum_{p,q=\pm}\;\;\sum_{a,b=1,2}\;\;
\sum_{\vec{x},\vec{x}\ppr}\hspace*{0.28cm}\mcal{N}_{x}
\\ \no && J_{\psi;\vec{x}\ppr}^{+b\ppr}(t_{q\ppr}\ppr)\;\hat{I}\;\wtilde{K}\;
\hat{T}_{q\ppr q}^{b\ppr b}(\vec{x}\ppr,t\ppr) \;\;
\hat{\mcal{O}}_{\vec{x}\ppr,\vec{x}}^{-1;ba}(t_{q}\ppr,t_{p})\;\;
\hat{T}_{pp\ppr}^{-1;aa\ppr}(\vec{x},t)\;\hat{I}\;J_{\psi;\vec{x}}^{a\ppr}(t_{p\ppr})\bigg\}
\eeq
\beq\lb{187}
\lefteqn{\hat{\mcal{O}}_{\vec{x}\ppr,\vec{x}}^{ba}(t_{q}\ppr,t_{p})= } \\ \no &=&
\delta_{a,b}\;\delta_{p,q}\;\delta_{\vec{x},\vec{x}\ppr}\;\delta(t_{p}-t_{q}\ppr)
\;\Big(\hat{H}_{p}^{a}(t_{p})+\sigma_{V_{0}}^{(0)}(\vec{x},t_{p})\Big)+
\Big(\hat{T}^{-1}\;\hat{I}\;\frac{\mcal{J}}{\mcal{N}_{x}}\;\hat{I}\;\wtilde{K}\;\hat{T}
\Big)_{\vec{x}\ppr,\vec{x}}^{ba}(t_{q}\ppr,t_{p}) +  \\ \no &+&\underbrace{\delta_{\vec{x},\vec{x}\ppr}\;
\delta(t-t\ppr)\;
\Big(\hat{T}_{qp\ppr}^{-1;ba\ppr}(\vec{x}\ppr,t\ppr)\;\hat{H}_{p\ppr}^{a\ppr}(t)\;
\hat{T}_{p\ppr p}^{a\ppr a}(\vec{x},t)-
\delta_{a,b}\;\delta_{p,q}\;\hat{H}_{p}^{a}(t)\Big)}_{\delta\hat{\mcal{H}}(\hat{T}^{-1},\hat{T})}
\eeq
\beq\lb{188}
\hat{H}_{p}^{a=1}(t_{p}) &=&
\hat{h}_{p}(t_{p})=-\im\hbar\frac{\pp}{\pp t_{p}}-\im\;\ve_{p}-\frac{\hbar^{2}}{2m}\Delta+u(\vec{x})-\mu_{0}
\\ \lb{189} \hat{H}_{p}^{a=2}(t_{p}) &=& \hat{h}_{p}^{T}(t_{p})=+\im\hbar\frac{\pp}{\pp t_{p}}-\im\;\ve_{p}-
\frac{\hbar^{2}}{2m}\Delta+u(\vec{x})-\mu_{0}\;\;\;.
\eeq
The matrix \(\hat{T}_{pq}^{ab}(\vec{x},t)\) (\ref{190}) in \(\hat{\mcal{O}}_{\vec{x}\ppr,\vec{x}}^{ba}(t_{q}\ppr,t_{p})\)
(\ref{187},\ref{186}) contains the pair condensates with matrices
\(\hat{Y}_{pq}^{ab}(\vec{x},t)\) (\ref{191}), \(\hat{X}_{pq}(\vec{x},t)\) (\ref{192}) as the coset part
\(Sp(4)\backslash U(2)\) of \(Sp(4)\)
\beq \lb{190}
\hat{T}_{pq}^{ab}(\vec{x},t)&=&\Big(\exp\Big\{-\hat{Y}_{p\ppr q\ppr}^{a\ppr b\ppr}(\vec{x},t)\Big\}\Big)_{pq}^{ab}
\\ \lb{191}
\hat{Y}_{pq}^{ab}(\vec{x},t)&=&
\left(\bea{cc} \Big(0\Big)_{pq}^{11} & \Big(\hat{X}_{pq}(\vec{x},t)\Big)^{12} \\
-\Big(\eta_{p}\;\hat{X}_{pq}^{+}(\vec{x},t)\;\eta_{q}\Big)^{21} & \Big(0\Big)_{pq}^{22}
\eea\right)_{pq}^{ab}
\eeq
\beq\lb{192}
\hat{X}_{pq}(\vec{x},t)&=&\left(\bea{cc}
-\delta c_{D;++}(\vec{x},t) & \delta c_{D;+-}(\vec{x},t)  \\
-\delta c_{D;+-}(\vec{x},t) & \delta c_{D;--}(\vec{x},t)
\eea\right)_{\mbox{.}}
\eeq
One can extract the Goldstone modes of a spontaneous symmetry breaking \(Sp(4)\backslash U(2)\otimes U(2)\)
in a gradient expansion of the operator \(\delta\hat{\mcal{H}}(\hat{T}^{-1},\hat{T})\)
with the matrix \(\hat{T}_{pq}^{ab}(\vec{x},t)\) (\ref{190}) in
\(\hat{\mcal{O}}_{\vec{x}\ppr,\vec{x}}^{ba}(t_{q}\ppr,t_{p})\) (\ref{187}).
The relevant parameter for classifying the various terms of the gradient expansion is
the number \(\mcal{N}_{x}\) of discrete space points. Furthermore, one obtains special properties of
the coefficients multiplying the traces of the gradients with the matrix \(\hat{T}_{pq}^{ab}(\vec{x},t)\)
in \(d=2\) spatial dimensions. Apart from the conformal invariance of the nonlinear sigma model
in \(d=2\) \cite{Ketov,Fran}, the coefficients reduce to one point functions in the spatial isotropic case which
allow computations by saddle point approximations for the coefficients containing the self-energy
\(\sigma_{V_{0}}^{(0)}(\vec{x},t_{p})\).

The one dimensional case with white noise disorder can be preferably treated by transfer matrices
of ensemble averaged generating functions because large fluctuations about mean field solutions may occur
\cite{Bm4}-\cite{Bm6}. One can also try to extend the transfer matrix approach to \(d=2\) spatial dimensions by
approximating and restricting to the lowest momentum modes perpendicular to the transfer direction \cite{Bm6}.
\vspace*{0.37cm}

\noindent {\bf Acknowledgement}
I would like to thank R. Graham and A. Pelster for giving me this problem of disordered bosons
and the discussions about it in the course of time. I would also like to thank the theoretical physics
group of the University Duisburg-Essen for giving me the opportunity for this contribution to the
'Sonderforschungsbereich' SFB/TR12 'Symmetries and Universality in Mesoscopic Systems'.

\end{document}